\documentclass[twocolumn]{aastex62}

\usepackage{amsmath}

\usepackage{bm}

\newcommand{\ASmod}[1]{{{#1}}}
\newcommand{\ASmodmath}[1]{{#1}}
\newcommand{\ASmodd}[1]{{{#1}}}

\def\st{\sin\theta}
\def\4p{\frac{1}{4\pi}}
\def\p{\partial}

\def\dpt2{\frac{1}{r \sin \theta}\frac{\partial}{\partial \theta}}

\def\nab{\mbox{\boldmath $\nabla$}}
\def\omm{\mbox{\boldmath $\omega$}}
\def\rb{\hat{\rho}\,}
\def\tb{\hat{T}}
\def\sb{\hat{S}}
\def\vv{{\bf{v}}}
\def\vrb{\overline{v_r}} % mean velocities
\def\vtb{\overline{v_{\theta}}}
\def\vpb{\overline{v_{\phi}}}

\def\vrp{{v'_r}} % fluct velocities

\newcommand{\BB}{{\bf B}}
\newcommand{\JJ}{{\bf J}}

\def\brb{\overline{B_r}} % mean B fields
\def\btb{\overline{B_{\theta}}}
\def\bpb{\overline{B_{\phi}}}
\def\brp{{B'_r}} % fluct B fields
\def\btp{{B'_{\theta}}}
\def\bpp{{B'_{\phi}}}

\def\vr2{\overline{v^2_r}}  % all kind of stresses
\def\vt2{\overline{v^2_{\theta}}}
\def\vp2{\overline{v^2_{\phi}}}

\def\vrvrp{\overline{{v'}_r^2}}
\def\vrvt{\overline{v_r v_{\theta}}}
\def\vrvtp{\overline{{v'}_r {v'}_{\theta}}}
\def\vrvp{\overline{v_r v_{\phi}}}
\def\vrvpp{\overline{{v'}_r {v'}_{\phi}}}
\def\vtvt{\overline{v_{\theta}^2}}
\def\vtvtp{\overline{{v'}_{\theta}^2}}
\def\vtvp{\overline{v_{\theta} v_{\phi}}}
\def\vtvpp{\overline{{v'}_{\theta} {v'}_{\phi}}}
\def\vpvp{\overline{v_{\phi}^2}}

\def\br2{\overline{B^2_r}}  
\def\bt2{\overline{B^2_{\theta}}}
\def\bp2{\overline{B^2_{\phi}}}
\def\bb2{\overline{B^2}}
\def\brbrp{\overline{{B'}_r^2}}

\def\brbt{\overline{B_r B_{\theta}}}
\def\brbtp{\overline{{B'}_r B'_{\theta}}}
\def\brbp{\overline{B_r B_{\phi}}}
\def\brbpp{\overline{{B'}_r {B'}_{\phi}}}
\def\btbt{\overline{B_{\theta}^2}}
\def\btbtp{\overline{{B'}_{\theta}^2}}
\def\btbp{\overline{B_{\theta} B_{\phi}}}
\def\btbpp{\overline{{B'}_{\theta} {B'}_{\phi}}}
\def\bpbpp{\overline{{B'}_{\phi}^2}}
\def\bpbp{\overline{B_{\phi}^2}}
\def\brvt{\overline{B_r v_{\theta}}}
\def\brvtp{\overline{{B'}_r {v'}_{\theta}}}
\def\brvp{\overline{B_r v_{\phi}}}
\def\brvpp{\overline{{B'}_r {v'}_{\phi}}}
\def\btvr{\overline{B_{\theta} v_r}}
\def\btvrp{\overline{{B'}_{\theta} {v'}_r}}

\def\btvp{\overline{B_{\theta} v_{\phi}}}
\def\btvpp{\overline{{B'}_{\theta} {v'}_{\phi}}}
\def\bpvr{\overline{B_{\phi} v_r}}
\def\bpvrp{\overline{{B'}_{\phi} {v'}_r}}
\def\bpvt{\overline{B_{\phi} v_{\theta}}}
\def\bpvtp{\overline{{B'}_{\phi} {v'}_{\theta}}}

\graphicspath{{Figures/}}

\received{June 30, 2021}
\revised{November 30, 2021}
\accepted{December 23, 2021}
\shorttitle{Dynamo action in G and K stars}
\shortauthors{A.S. Brun et al.}
\begin{document}

%\title{Dynamo Action, Energy Budget and Transfers in Sun-like Stars}
%\title{From Gravitational to Magnetic Energies: Energy Transfers in Convective Dynamo of Sun-like Stars}
\title{Powering Stellar Magnetism: Energy Transfers in Cyclic Dynamos of Sun-like Stars}

\correspondingauthor{Allan Sacha Brun}
\author[0000-0002-1729-8267]{Allan Sacha Brun}
\affil{D\'epartement d'Astrophysique/AIM, CEA/IRFU, CNRS/INSU, Univ. Paris-Saclay \& Univ. de Paris, 91191 Gif-sur-Yvette, France}
\email{sacha.brun@cea.fr}

\author[0000-0002-9630-6463]{Antoine Strugarek}
\affil{D\'epartement d'Astrophysique/AIM, CEA/IRFU, CNRS/INSU, Univ. Paris-Saclay \& Univ. de Paris, 91191 Gif-sur-Yvette, France}

\author[0000-0002-7422-1127]{Quentin Noraz}
\affil{D\'epartement d'Astrophysique/AIM, CEA/IRFU, CNRS/INSU, Univ. Paris-Saclay \& Univ. de Paris, 91191 Gif-sur-Yvette, France}

\author[0000-0002-2137-2896]{Barbara Perri}
\affil{Centre for mathematical Plasma Astrophysics, KU Leuven, Celestijnenlaan 200b-box 2400, 3001 Leuven, Belgium} 
\affil{D\'epartement d'Astrophysique/AIM, CEA/IRFU, CNRS/INSU, Univ. Paris-Saclay \& Univ. de Paris, 91191 Gif-sur-Yvette, France}

\author[0000-0002-6114-0539]{Jacobo Varela}
\affil{Universidad Carlos III de Madrid, Legan\'es, 28911, Spain}
\affil{D\'epartement d'Astrophysique/AIM, CEA/IRFU, CNRS/INSU, Univ. Paris-Saclay \& Univ. de Paris, 91191 Gif-sur-Yvette, France}

\author[0000-0003-4714-1743]{Kyle Augustson}
\affil{D\'epartement d'Astrophysique/AIM, CEA/IRFU, CNRS/INSU, Univ. Paris-Saclay \& Univ. de Paris, 91191 Gif-sur-Yvette, France}

\author[0000-0003-1618-3924]{Paul Charbonneau}
\affiliation{D\'epartement de physique, Universit\'e de Montr\'eal, C.P. 6128 Succ. Centre-Ville, Montr\'eal, QC H3C-3J7, Canada}

\author[0000-0002-3125-4463]{Juri Toomre}
\affiliation{JILA and Department of Astrophysical and Planetary Sciences, University of Colorado, Boulder, CO 80309-O440, USA}




\begin{abstract}
We  use  the  ASH  code  to  model  the  convective  dynamo  of  solar-type  stars.   Based  on  a  series  of 15  3-D  MHD  simulations  spanning  4  bins  in  rotation  and  mass, we show what mechanisms are at work in these stellar dynamos with and without magnetic cycles and  how  global  stellar  parameters affect the outcome. We also derive scaling laws for the differential rotation and magnetic field based on these simulations.  We find a weaker trend between differential rotation and stellar rotation rate, ($\Delta \Omega \propto (|\Omega|/\Omega_{\odot})^{0.46}$) in the MHD solutions than in their HD counterpart  $(|\Omega|/\Omega_{\odot})^{0.66}$), yielding a  better agreement  with  the  observational trends based on power laws. We  find  that  for a fluid Rossby number between  $0.15 \lesssim Ro_{\rm f} \lesssim 0.65$  the  solutions  possess  long  magnetic cycle,  if $Ro_{\rm f} \lesssim 0.42$ a short cycle  and if $Ro_{\rm f} \gtrsim 1$ (anti-solar-like differential  rotation) a  statistically  steady  state.   We show that short-cycle dynamos follow the classical Parker-Yoshimura rule whereas the long-cycle period ones do not. 
We  also  find  an efficient energy transfer between reservoirs leading to the conversion of several percent of the star’s luminosity into magnetic energy that could provide enough free energy to sustain intense eruptive behavior at the star's surface. We further demonstrate that the Rossby number dependency of the large-scale surface magnetic field in the simulation ($B_{\rm L,surf} \sim Ro_{\rm f}^{-1.26}$) agrees better with observations ($B_{V} \sim Ro_{\rm s}^{-1.4\pm 0.1}$) and differs from dynamo scaling based on the global magnetic energy ($B_{bulk} \sim Ro_{\rm f}^{-0.5}$).
\end{abstract}


\section{Introduction} \label{sec:intro}

\ASmod{Sun-like stars go through various magnetic activity phases in their lives. From young very active TTauri stars rotating much faster than
our Sun to old stars that are less active, it is key to understand how convection, rotation, turbulence, magnetism and surface activity evolve and feedback on one another 
over secular time}. Of particular interest is the generation of magnetic field via dynamo action, because it is both as the source of key temporal variabilities like the Schwabe 11-yr or Gleissberg 90yr magnetic cycles in the Sun and at the heart of a complex
feedback loop between stellar magnetism and rotation via wind braking and the loss of mass and angular momentum by the star \citep{2015ApJ...799L..23M,2017LRSP...14....4B,2021arXiv210315748V}.
It is also key in providing the free energy reservoir needed to power eruptive events such as flares or CME's \citep{2013PASJ...65...49S,2015ApJ...802...53A,2017PASJ...69...41M}. In this work, we seek to assess how solar-like stars with different masses and rotation rates can power their magnetism by means of dynamo action in their convective envelopes.

Various activity indicators have been derived observationally over the last 50 years using for instance photometric and spectroscopic variability \citep{1995ApJ...438..269B,2009AandA...501..703O,2017PhDT.........3E,2018A&A...616A.108B}, and more recently through asteroseismology \citep{2019LRSP...16....4G} to connect the spectral class and age of a star to its dynamical properties and activity level. 
Turning specifically to solar type stars, spectropolarimetric studies have revealed several interesting properties \citep{2014MNRAS.444.3517M}. In \cite{2014MNRAS.441.2361V}, it was shown that the large-scale magnetic field is following a scaling law with the stellar Rossby number $\left\langle B_V\right\rangle \propto Ro_s^{-1.38 \pm 0.14}$ for stars with $Ro_s > 0.1$ \ASmod{(here the stellar Rossby number is defined as the ratio between the rotation period and the convective turnover time)}. \ASmod{More recently, \citet{2019ApJ...876..118S} have revisited this trend and found a similar result with $\left\langle B_V\right\rangle \propto Ro_s^{-1.40 \pm 0.10}$.} It was also proposed in \cite{2008MNRAS.388...80P} and later by \cite{2015MNRAS.453.4301S} that toroidal magnetic field dominates over poloidal field for fast rotators. It was further shown that no significant collapse of the large-scale field with respect to higher multipole moments was observed as the star evolved and are found less active \citep{2016MNRAS.455L..52V}. Recent work by \cite{2021ApJ...910..110L} including more evolved stars seems to help constrain better the rotation-activity relationship, confirming that considering the Rossby number is better than the rotation period alone. In \cite{2018ApJ...852...46K} the possibility that larger metallicity increases the activity level of solar analogs was also  proposed.

Moreover, long observational studies based on Ca II H\&K chromospheric observations have shown that magnetic activity of solar-like stars \citep{1978ApJ...226..379W,1990MmSAI..61..559S,1999ApJ...514..402P,2008LRSP....5....2H,1538-3881-133-3-862} can be found to be either irregular with no obvious cyclic activity or to possess activity cycles with short magnetic cycle periods  \citep{2010ApJ...723.1583M,2018MNRAS.479.5266J}  or long (decadal) ones \citep{1984ApJ...279..763N,1995ApJ...438..269B}
as in the Sun. Such studies have further indicated the existence of a relation between Rossby number and magnetic cycle periods, its exact nature being still debated given the relatively small numbers of truly confirmed cyclic magnetic stars \citep{2041-8205-790-2-L23,2017PhDT.........3E}. 

A puzzling property regarding stellar magnetic cycles has been the existence (or not) of active and inactive branches of stellar activity, as proposed by \cite{1999ApJ...524..295S} and \cite{2007ApJ...657..486B}. Recently it has been argued that stars may be transiting from one state to the other as they evolve and that such distinct activity branches do not exist. Instead, activity would be decreasing while rotation would be almost unchanged beyond a certain age or stellar internal dynamo state. A key quantity to characterize this activity state transition is again the Rossby number. \cite{2017SoPh..292..126M} proposed that once their Rossby number becomes large, stars stop braking through their stellar wind, hence departing  from the classical Skumanich law $\Omega(t) \propto t^{-0.5}$ and gyrochronology trend \citep{1972ApJ...171..565S,2003ApJ...586..464B,2007ApJ...669.1167B}. This is still highly debated in the community as some observers find stars older than the Sun still following Skumanich's law \citep[][]{2015Natur.517..589M,2016ApJ...823...16B,2018A&A...619A..73L,2019MNRAS.485L..68L,2020ApJ...898..173D} while others do not \citep{2016ApJ...826L...2M,2019ApJ...871...39M,2021NatAs.tmp...71H}.
The disagreement could also be due to the observation techniques (photometric versus  chromospheric studies for instance) and observational data set (Kepler data vs long-term monitoring of individual stars) used, since each have rotation rates and ages determination that differ sometime significantly  \citep{2016A&A...594L...3L,2020ApJ...898..173D}. Another alternative would be that stars temporarily stop spinning down before starting again \citep{2020A&A...636A..76S,2020ApJ...904..140C} or that the coronal temperature drops, yielding a smaller mass loss for older stars \citep{2018MNRAS.476.2465O}.
Thus, understanding what happens from a theoretical point of view to stellar dynamo and magnetic field geometry for large Rossby numbers is crucial in helping to interpret the most recent observations. This is one of the goals of this study.
Given the close link between surface activity and stellar magnetism, a key aspect to characterize is the amount of magnetic energy made available in a given solar-like star by dynamo action. We know that flare intensity is linked to the magnetic energy made available to the magnetic structures. 
It is thus crucial to better characterize energy transfers in solar-type star dynamos for a wide range of Rossby numbers.

\ASmod{Characterizing the differential rotation realised at the base and in the convective envelope of solar-type stars is central to the understanding of their magnetic field generation, activity level and rotation, as it is directly linked to the $\Omega$-effect (e.g. stretching of the poloidal magnetic field lines by large scale shear).} Hence, the role of the differential rotation (DR) in driving the star's magnetic activity level and field properties should be clarified \citep{1996ApJ...466..384D}.
Doppler imaging \citep{1997MNRAS.291....1D,2005MNRAS.357L...1B}, asteroseismology \citep{2004SoPh..220..169G,2013AandA...560A...4R,2014AandA...572A..34G}, classical spot models \citep{2014AandA...564A..50L} and short-term Fourier-transform \citep{2014MNRAS.441.2744V} are methods to infer differential rotation. The combination of all these observations on stellar rotation and magnetism helps constrain the trends linking rotation with stellar differential rotation and magnetic activity. Various analyses of stellar differential rotation revealed different dependencies between DR and star's rotation ($\Delta \Omega \propto \Omega^n$), with $n$ varying between 0.15 and 0.7 \citep{2005MNRAS.357L...1B,2006AandA...446..267R,2013AandA...560A...4R}. There is no clear consensus in the community for now, some authors are even advocating that such laws should be derived per spectral stellar classes and that the confusion comes from mixing together F and K stars \citep{2016MNRAS.461..497B}. \cite{2011IAUS..273...61S,2018ApJ...855L..22B} also propose that the dependency of the differential rotation with the rotation rate may not be monotonic, with a break near Rossby equals unity.
By contrast, a more systematic and stronger dependency is observed with the star's temperature ($\Delta \Omega \propto T_{\rm eff}^{8.92}$ \citealt{2005MNRAS.357L...1B,2013AandA...560A...4R} and $\Delta \Omega \propto T_{\rm eff}^{8.6}$ \citealt{2007AN....328.1030C}). Hence, we expect large-scale shear to vary both in amplitude and profile (as a function of latitude and depth), as the global stellar parameters change. \ASmod{Some recent studies have confirmed this is happening in solar-type stars by inverting seismically their profile \citep{2018Sci...361.1231B}, pointing to a possible anti-solar differential rotation state (e.g. slow equator/fast poles) which was possibly already guessed in F stars \citep{2007AN....328.1034R} and advocated to exist in numerical simulations \citep[][see below]{2011AN....332..897M,2014MNRAS.438L..76G,2015SSRv..tmp...30B}.}

Considering the large number of global stellar parameters probed by these different observational studies, it is expected that the excitation of various types of convective dynamos may occur \citep{1994lspd.conf...59W,1998ASPC..154.1349T,2017LRSP...14....4B,2018ApJ...855L..22B,2020LRSP...17....4C}. In order to quantify the influence of key parameters such as rotation and mass in characterizing the dynamo and magnetic level achieved in solar-like stars and given the intrinsic nonlinear mechanisms at work in stellar dynamos, multi-D numerical simulations have been developed over the years in an attempt to provide more quantitative answers.

Some studies have used the 2.5D mean field dynamo approach to do so, extending solar mean field dynamo models to other stellar spectral types \citep[][and references therein]{2006A&A...446.1027C,2010A&A...509A..32J,2011A&A...530A..48K,2018A&A...615A..38K}. While these studies are very helpful, most of them lack the full nonlinearity and genuine parametric dependence of 3D magnetohydrodynamic (MHD) simulations. Recent developments by \cite{2021MNRAS.502.2565P} are starting to overcome these limits and have extended the work of \cite{2006ApJ...647..662R} on the Sun to solar-type stars with various rotation rates.
Nevertheless, with the arrival of more powerful supercomputers, other authors have used instead global 3D MHD simulations to model differential rotation and stellar magnetism in the convection zone of solar-like stars  \citep{1982ApJ...256..316G,2000ApJ...532..593M,2004ApJ...614.1073B,2006ApJ...641..618M,2008ApJ...689.1354B,2010ApJ...715L.133G,2010ApJ...711..424B,2011ApJ...742...79B,2011AandA...531A.162K,2014AandA...570A..43K,2014MNRAS.438L..76G,2015ApJ...809..149A,2015AandA...576A..26K}. These studies pointed out the large magnetic temporal variability and the critical effect of stellar rotation and mass on magnetic field generation through dynamo mechanism, leading in some parameter regimes to configurations with cyclic activity \citep{Gilman2,1983ApJS...53..243G,1985ApJ...291..300G,0004-637X-735-1-46,2011ApJ...731...69B,2013ApJ...762...73N,2013GApFD.107..244K,2013ApJ...777..153A,2015ApJ...809..149A,2016ApJ...826..138B,2016ApJ...819..104G,2017Sci...357..185S,2018ApJ...863...35S,2018A&A...616A..72W,2018A&A...616A.160V,2019ApJ...886...21V,2019ApJ...880....6G,2020ApJ...892..106M}. Several studies pointed out the positive effect of a stable region underneath the convection zone \citep{1993ApJ...408..707P} on the efficient storage of intense toroidal field and the  lengthening of the stellar dynamo cycle period \citep{1985GApFD..31..137G,2006ApJ...648L.157B,2015ApJ...813...95L,2016ApJ...826..138B,2016ApJ...819..104G,2019ApJ...880....6G,2019GApFD.113..149K,2020ApJ...893..107B}. Over the last decade significant progress has been made in successfully simulating large-scale mean flows and stellar activity cycle using different numerical codes and methods \citep{2011Icar..216..120J}. This is quite reassuring that a global consensus is growing on the nature of solar-like star dynamos. It is common knowledge that there are still key transitions in Rossby number (at low and high values of this parameter) that need to be understood further, as well as what is the exact type of convective dynamos realized in solar-like stars as their global parameters are varied. 
This study continues this effort by doing an even broader systematic parametric study of solar-like star dynamos coupled to a stably stratified layer below than what have been published so far. \ASmod{It extends the work published in \cite{2016AdSpR..58.1507V} and \cite{2017ApJ...836..192B} with the MHD anelastic spherical harmonic code (ASH) \citep{2004ApJ...614.1073B}.} In particular, we wish to better characterize energy transfers and how much of a star's energy (luminosity) is converted into magnetic energy by nonlinear global convective dynamos over a wide range of Rossby numbers, generalizing to solar-like stars the work by \cite{1966PApGe..64..145S} and \cite{2006ApJ...647..662R}.

In the following sections, we analyze how differential rotation and magnetism feedback on one another \citep{2004SoPh..220..333B,2014ApJ...789...35F} as well as  how kinetic and magnetic energies flow within a stellar magnetized rotating convective envelope, using 15 convective dynamo MHD simulations for model stars with different masses and rotation rates (hence Rossby numbers) in order to achieve this goal. In section \S \ref{sec:numericalSetup} we present the equations and model setup. In section \S \ref{sec:overview} we make a quick overview of one of the dynamo solutions emphasizing the main properties of a cyclic solution. In \S \ref{sec:LargeScaleFlows} we discuss the various DR profiles obtained in our parametric studies, expanding \cite{2016AdSpR..58.1507V} to include 15 models. We discuss angular momentum and various scaling laws of the differential rotation contrast $\Delta \Omega$. In \S \ref{sec:MagneticProperties} we analyze our dynamo solutions for various key properties as a function of the Rossby number, such as their activity level, the amount of magnetic flux generated by the dynamo, the existence or not of an activity cycle and torsional oscillations, how the cycle period for cyclic solutions changes, what is the relative contribution of dipolar and quadrupolar magnetic fields in the overall dynamo generated magnetic field, and interpret our simulations in terms of mean field $\alpha$-$\Omega$ classification. We further expand our data set with the 17 simulations published previously  \cite{2017Sci...357..185S,2018ApJ...863...35S} with the Eulag-MHD code \citep{2013JCoPh.236..608S}, in order to improve the statistics. In \S \ref{sec:EnergyContent} we perform an extensive study of energy transfer between various reservoirs in stellar dynamos, assessing how much magnetic energy is accessible to stars like our Sun to power eruptive events. We compute all MHD transfers between kinetic and energy reservoirs for the large-scale flows and magnetic fields. In \S \ref{sec:AstrophysicalImplications} we reflect on our findings in an astrophysical context, comparing our results with recent observational results and then conclude.

\section{Numerical setup}
\label{sec:numericalSetup}

In this section we present the main features of the ASH code, describing the boundary and initial conditions of the numerical models and our choice of global parameters.

\subsection{Set of equations solved}
We perform 3D MHD simulations of convective dynamo action coupled to a stable radiative interior \ASmod{where the anelastic MHD equations are solved for the motions of a conductive plasma in a rotating sphere \citep{2011Icar..216..120J}}.
The anelastic approximation captures the effects of density
stratification without having to resolve sound waves, which would severely limit the time step \citep{2012ApJ...756..109B}. In the MHD context, the anelastic approximation filters out fast magneto-acoustic waves but retains Alfv\'en waves. 

The code ASH uses a pseudo-spectral method \citep{Clune}.  The velocity ($\mathbf{v}$), magnetic ($\mathbf{B}$), and thermodynamic variables (entropy $S$, pressure $P$) are expanded in spherical harmonics $Y_{\ell m}(\theta,\phi)$ for their horizontal structure and in Chebyshev polynomials $T_n(r)$ for their radial structure \citep{2004ApJ...614.1073B}. 
The density ($\rho$), entropy, pressure and temperature ($T$) are linearized about the spherically symmetric background values, denoted by the symbol ( $\hat{}$ ).  The equations solved by ASH are \citep{2004ApJ...614.1073B}:
\begin{eqnarray}
\boldsymbol{\nabla} \cdot \rb\vv = 0  \mbox{\,\, , \,\, } \boldsymbol{\nabla} \cdot \mathbf{B} &=&0 \\
\rb \left(\frac{\p {\bf v}}{\p t}+({\bf v}\cdot\nab){\bf v}+2{\bf \Omega_*}\times{\bf v}\right) 
 &=& -\nab P + \rho {\bf g} \nonumber \\ 
 + \frac{1}{4\pi} (\nab\times{\bf B})&\times&{\bf B} + \nab\cdot\mbox{\boldmath $\cal D$} \\
 %-[\nab\hat{P}-\rb{\bf g}] \\
 \rb \tb \frac{\partial S}{\partial t} = -\rb \tb \vv \cdot {\nab}(\sb +S)  
 - \boldsymbol{\nabla} \cdot \mathbf{q} &+& \Phi_d + \rb \epsilon \\
\frac{\partial \mathbf{B}}{\partial t} = \boldsymbol{\nabla} \times [\vv \times \mathbf{B} &-& \eta \boldsymbol{\nabla} \times \mathbf{B}] , 
\end{eqnarray}
with the velocity field $\mathbf{v} = v_{r} \hat{\bf e}_r +  v_{\theta} \hat{\bf e}_{\theta} +  v_{\varphi}\hat{\bf e}_{\varphi}$, the magnetic field $\mathbf{B} = B_{r}\hat{\bf e}_r +  B_{\theta}\hat{\bf e}_{\theta} +  B_{\varphi}\hat{\bf e}_{\varphi}$, the angular velocity in the rotation frame $\mathbf{\Omega}_*  = \Omega_{*} \hat{\bf e}_z $, $\hat{\bf e}_z$ the unit vector along the rotation axis, $g$ the magnitude of the gravitational acceleration. 
A volumetric heating term $\rb \epsilon$ is also taken
into account to approximate generation of energy by nuclear
reactions in the stellar core. The nuclear reactions are modeled very simply by
assuming that $\epsilon = \epsilon_0 \tb^{n_c}$. By enforcing that the integrated
luminosity of the star matches its known surface value, we can
determine $\epsilon_0$ and $n_c$ as listed in Table 7 of \cite{2017ApJ...836..192B}. 
Note that only \ASmod{the low-mass star series of models (e.g. 0.5 and 0.7 $M_{\odot}$) require} that heating source term, since their computational domain includes a portion of the nuclear energy generation core.

The diffusion tensor $\boldsymbol{D}$ and the dissipative term $\Phi_d$ are defined as:
$$ D_{ij} = 2\rb \nu \left[ e_{ij} - \frac{1}{3}\boldsymbol{\nabla} \cdot \vv \delta_{ij} \right],$$
$$ \Phi_d = 2\rb \nu \left[ e_{ij} e_{ij} - \frac{1}{3} (\boldsymbol{\nabla} \cdot \mathbf{v})^{2} \right] + \frac{4\pi\eta}{c^{2}}\mathbf{J}^{2},$$
with $e_{ij}$ the stress tensor and $\mathbf{J} = c/4\pi\boldsymbol{\nabla}\times\mathbf{B}$ the current density. The energy flux $\mathbf{q}$ is the sum of a radiation flux and of a turbulent entropy diffusion flux:
$$ \mathbf{q} = \kappa_{r} \rb c_{p} \boldsymbol{\nabla} (\tb + T) + \kappa\rb\tb \boldsymbol{\nabla} S + \kappa_{0}\rb\tb \frac{\partial \sb}{\partial r} \hat{\bf e}_r , $$ 
with $\nu$, $\kappa$ and $\eta$ the effective eddy diffusivities of the momentum, heat and magnetic field transport, $\kappa_{r}$ the atomic radiation diffusion coefficient, $\kappa_{0}$ the effective thermal diffusivity acting only on the spherically symmetric ($l = 0$) entropy gradient and $c_{p}$ the specific heat at constant pressure. 

\begin{figure}[!ht]
\begin{center}
\includegraphics[width=1.0\linewidth]{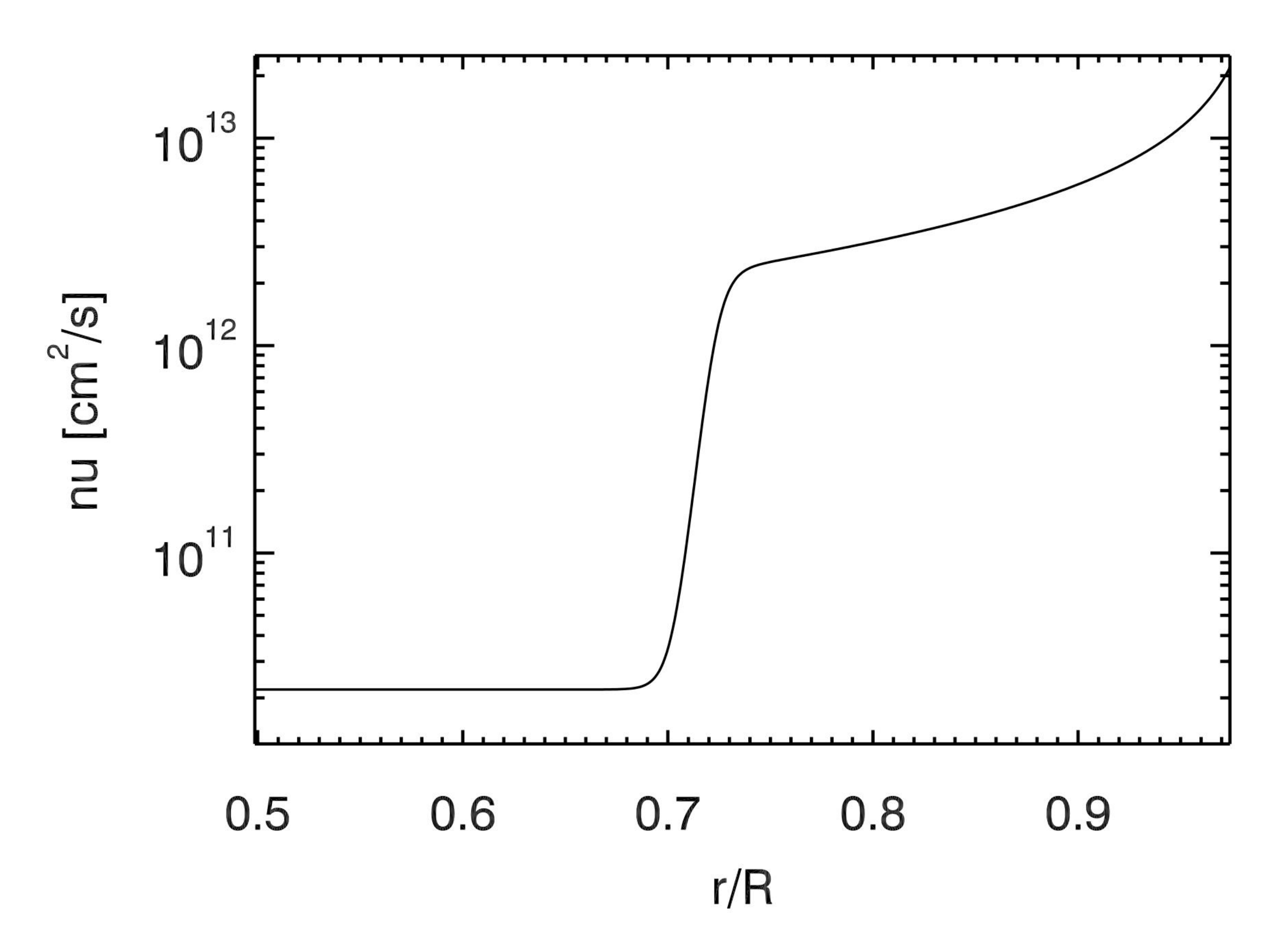}
\end{center}
\caption{Typical radial profile of kinematic viscosity $\nu$ used in this study, here for case M11R3m. Profiles of $\kappa$ and $\eta$ are the same, but their amplitude depends respectively for each cases on the chosen Prandtl and magnetic Prandtl numbers (see Table \ref{tab:tableAdimParams}).}
\label{diff1D}
\end{figure}

Due to limitations in computing resources, current numerical simulations cannot capture all scales of solar convective motions and magnetic fields from global to atomic dissipation
scales.  The simulations described in this study resolve nonlinear interactions
among a large range of scales but motions and magnetic field still exist in solar-like stars on scales smaller than our grid resolution.  Hence, our
models should be considered as large-eddy simulations (LES) with
parameterization to account for subgrid-scale (SGS) motions. The effective eddy diffusivities $\nu$, $\kappa$, and $\eta$ represent
momentum, heat, and magnetic field transport by motions which are not
resolved by the simulation.  They are allowed to vary with radius but
are independent of latitude, longitude, and time for a given 
simulation.  In the simulations
reported here, $\nu$, $\kappa$, and $\eta$ have the following profile:

\begin{equation}
  \nu(r) = \nu_{bot} + \nu_{top} f_{step}(r),\nonumber
\end{equation}

\noindent where 
\begin{eqnarray}
  f_{step}(r)&=(\rb/\rb_{top})^{\alpha}[1 - \beta]f(r), \nonumber \\
  f(r)&=0.5(\tanh((r-r_t)/\sigma_t)+1), \nonumber \\
  \beta &= \nu_{bot}/\nu_{top}=10^{-3}, \nonumber
\end{eqnarray}

\noindent and with $\nu_{top}$ in {\rm cm$^2$ s$^{-1}$} and $r_t$ and $\sigma_t$ in {\rm cm}, as provided in Table 7 of \cite{2017ApJ...836..192B}, $\alpha$ is -0.5 for all cases.  All models assumed a Prandtl number $Pr=\nu/\kappa$ of 0.25, so that $\kappa$ can be
directly obtained from the amplitude and profile of $\nu$. The magnetic Prandtl number $P_m=\nu/\eta$ is equal to 1 or 2 depending on the case considered (see Table \ref{tab:tableAdimParams}), so that $\eta$ can as well be deduced from $\nu$. These tapered profiles are chosen in order to take into
account the much smaller sub-grid scale transport expected in the stably stratified radiative interior.  A representative profile is shown in Figure \ref{diff1D}. Their amplitudes are adapted for each rotation rates and stellar masses considered in order to achieve the best turbulent convective dynamo states while retaining a reasonable numerical resolution and computing effort (still, each model has used of the order of 8 to 10 million cpu hours spread over several years).

The diffusivity $\kappa_0$ is set such as to have the unresolved eddy flux carrying the solar flux outward 
at the top of the domain (see Figure \ref{Model}).  It drops off exponentially with depth in order to avoid a large inward heat flux in the stable zone
(see \citealt{2000ApJ...532..593M}). Of course there is some arbitrariness in choosing the exact shape and amplitude of our diffusivity profiles, and we do our best to limit
their influence on the results reported here.

The mass flux and magnetic vector fields are maintained divergenceless by a streamfunction formalism \citep{2004ApJ...614.1073B}:
\begin{eqnarray}
{\rb\bf v}=\nab\times\nab\times (W \hat{\bf e}_r) +  \nab\times (Z \hat{\bf e}_r), \\ 
{\bf B}=\nab\times\nab\times (C \hat{\bf e}_r) +  \nab\times (A \hat{\bf e}_r) ~~~. 
\end{eqnarray}

A perfect ideal gas equation is used for the mean state and the fluctuations are linearized:
\[ \hat{P} = (\gamma - 1) c_{p} \rb \tb / \gamma \]
\[ \rho / \rb = P / \hat{P} - T / \tb = P / \gamma \hat{P} - S / c_{p} \]
with $\gamma = 5/3$ the adiabatic exponent.

The anelastic MHD system of equations requires 12 boundary conditions. 
We use an impenetrable and stress-free boundary conditions at the top and bottom of the domain, i.e.: 
$$ v_{r} = \frac{\partial}{\partial r}\left( \frac{v_{\theta}}{r} \right) = \frac{\partial}{\partial r}\left( \frac{v_{\phi}}{r} \right) = 0.$$
Magnetic boundary conditions are perfectly conducting at the lower radial boundary and the magnetic field matches to a potential field in the upper boundary:
\hspace{5pt} $ B_{r}|_{r_{\rm bot}} = \frac{\partial}{\partial r} \left(\frac{B_{\theta}}{r} \right)|_{r_{\rm bot}} = \frac{\partial}{\partial r} \left(\frac{B_{\varphi}}{r} \right) |_{r_{\rm bot}} = 0 $ and $ B_{r}|_{r_{\rm top}} = \nabla \Psi \Rightarrow \Delta \Psi = 0 $,
with $r_{\rm top}$, $r_{\rm bot}$ respectively the radius of the top and bottom of the numerical domain and $r_{bcz}$ that of the base of the convective layer (cf. Table \ref{tablespectraltype}). 

\begin{figure*}[!ht]
\centering
\includegraphics[width=1.0\linewidth]{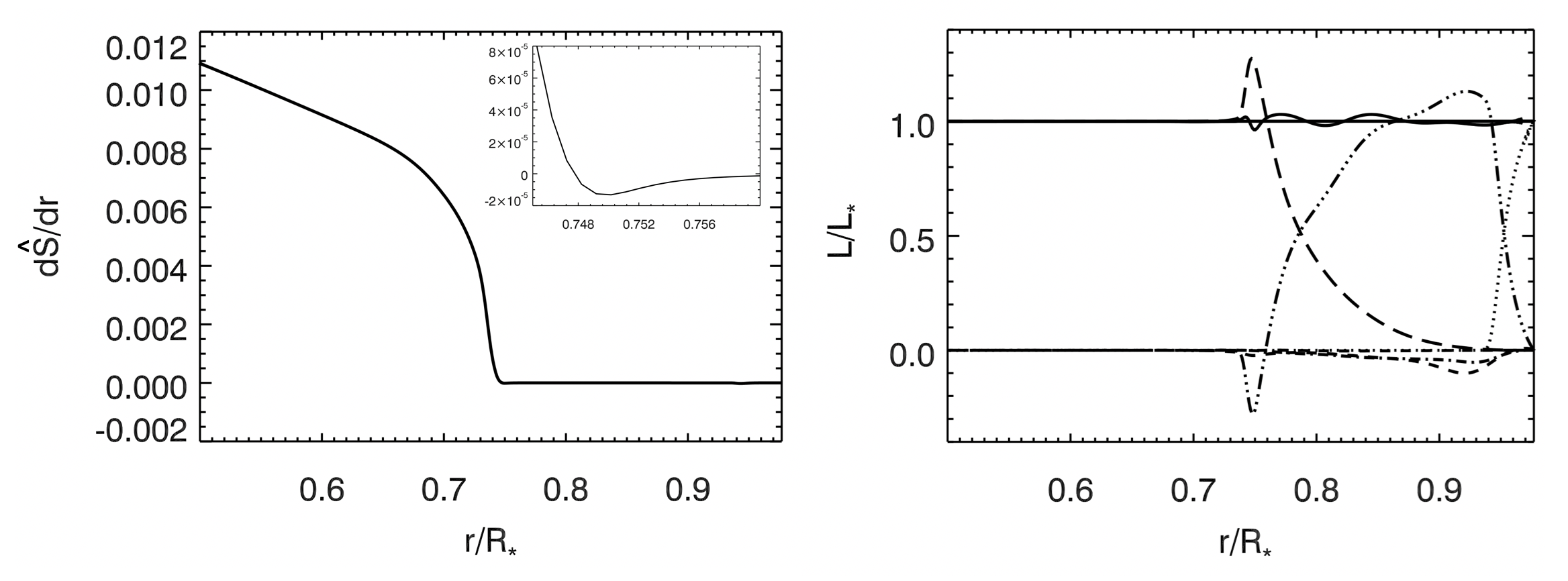}
\caption{(a) \ASmod{Radial dependence of the mean entropy gradient for case M11R3m. The region of the tachocline is shown in the inset figure}. (b) Time and horizontally averaged radial energy fluxes as luminosities (normalized to the star luminosity) for case M11R3m. The solid line is the total flux, the long-dashed line the radiative flux, the dash-triple-dotted line the enthalpy flux, the dotted line the conductive entropy flux, the thick dash-dotted line the kinetic energy, the dashed line the viscous diffusion flux and the thick dashed line the Poynting flux.}  
\end{figure*}\label{Model}

Finally, we maintain the entropy flux at the top and bottom. Keeping the values of $d\hat{S}/dr|_{r_{top},r_{bot}}$ fixed at all times in the simulations further implies that the fluctuating $dS/dr$ is set to zero at both boundary conditions.

\subsection{Model structure and initialization}

The simulation is focused on the bulk convection zone, avoiding regions too close to the stellar surface. We include a stably stratified layer below the convective envelope, hence providing a realistic bottom boundary condition for the fields and flow that are allowed to be pumped down and to penetrate into the radiative interior. The code uses a realistic background stratification for the profiles of entropy ($\hat{S}$), density ($\rb$), temperature ($\tb$) derived from a one-dimensional solar structure model CESAM \citep{1997AandAS..124..597M,2002A&A...391..725B}. Our starting point are the G and K star rotating convective 3D models published in \cite{2017ApJ...836..192B} (see also \citealt{2011AN....332..897M,2015SSRv..tmp...30B} and Table \ref{tablespectraltype}).

\begin{table*}[!ht]
\begin{center}
\caption{Global properties on the main sequence of the 4 stars used in our ASH dynamo models}\label{tablespectraltype}
\vspace{0.2cm}
%\begin{tabular}{||p{1.8cm}*{1}{||c} cccc ||}
\begin{tabular}{ccccccccccccc}
\tableline
\tableline
\\ [-1.5ex]
 Mass & Radius & $L_{*}$ & $T_{\rm eff}$ & Sp. T. & $M_{bcz}$ & $r_{bcz}$ & $\bar{T}_{bcz}$ & $\rb_{bcz}$ & $\Delta_{cz} \rb$ & $\Delta_{f} \rb$ & $r_{\rm bot}$& $r_{\rm top}$\\ [0.5ex]
 $(M_{\odot})$ & $(R_{\odot})$ &  $(L_{\odot})$ & $(K)$ & & $(M_{\odot},M_*)$ & $(R_{\odot},R_*)$  & $(K)$ & $(g cm^{-3})$ & - & - & $(R_*)$ & $(R_*)$\\ [0.8ex] 
\tableline
\tableline
\\ [-1.5ex]
 0.5 & 0.44 & 0.046 & 4030 & K7 & 0.18, 0.36 & 0.25,0.56 & $4.3\times 10^6$ & 14.0 & 42 & 193 & 0.13 & 0.95 \\
 0.7  & 0.64 & 0.15 & 4500 & K4/K5 & 0.079, 0.11 & 0.42,0.66 & $3.0\times 10^6$ & 2.1 & 50 & 605 & 0.32 & 0.97 \\
 0.9  & 0.85 & 0.55 & 5390 & G8 & 0.042, 0.046 & 0.59,0.69 & $2.6\times 10^6$ & 0.51 & 67 & 1013 & 0.38 & 0.97 \\
 1.1 & 1.23 & 1.79 & 6030 & G0 & 0.011, 0.010 & 0.92,0.75 & $1.6\times 10^6$ & 0.048 & 81 & 830 & 0.5 & 0.97 \\ [0.5ex]
\hline
 \tableline
 \tableline
\end{tabular}
\end{center}
All the listed values were computed with the CESAM stellar evolution code \citep{1997AandAS..124..597M}.
We adopt $M_{\odot}=1.989\times 10^{33}\, g$, $R_{\odot}=6.9599\times 10^{10}\, cm$, and $L_{\odot}=3.846\times 10^{33}\, erg\cdot s^{-1}$.
The density ratios $\Delta_{cz} \rb$ and $\Delta_{f} \rb$ are evaluated by forming the ratio between the value of the 
density respectively at the base of the convection and the top of the domain and at the bottom and the top of the domain.
\end{table*}

The MHD models are initialized from their equivalent progenitor hydrodynamical models in which a small magnetic field perturbation is introduced in the convective envelope (many orders smaller than the final magnetic field observed in the simulation). 
In that hydrodynamic study we published 15 simulations covering 4 mass bins and 4 rotation rates. We have models for stellar masses 0.5, 0.7, 0.9, 1.1 $M_{\odot}$ and rotation rates ranging from 1/8 to 5 times the solar rotation rates.
In keeping with the naming nomenclature of \cite{2017ApJ...836..192B}, we name our model such as to indicate the mass of the star and its rotation rate. The models are named MAxrm, where 'A' is the mass of the star and 'r' the rotation rate of the star (in solar rotation rate). The index 'x' indicates slow/anti-solar (x = s) and prograde (x = R) differential rotation models (\ASmod{except model M11R1m that is also anti-solar}) and m stands for magnetism, to distinguish between the hydrodynamic progenitor published in 2017 and their MHD dynamo counterparts considered in this study.

\ASmod{The models have a numerical resolution of} ($N_{r}$, $N_{\theta}$, $N_{\phi}$) 769 x 256 x 512 except for few cases in the M09m \& M11m series rotating at $\Omega_{*} = 3 \mbox{ or } 5\Omega_{\odot}$ where $N_{\theta}$ is 512 and $N_{\phi}$ is 1024. 

In Fig. \ref{Model} we show in (a) an example \ASmod{of the radial dependence of the entropy gradient}, and in (b) the temporal and azimuthally averaged radial energy fluxes balance as luminosities for the model with 1.1 solar mass and 3 times the rotation rate of the Sun. We note the sharp increase of the stratification at the base of the convective envelope that is coherent with the stiff radiative interior found in main sequence solar-like stars. Such a realistic interface, as opposed to an impenetrable wall, allows the convective motions to overshoot beyond the radius where the entropy gradient changes sign. By doing so they generate internal gravity waves and pump magnetic field. Since we are in this study mostly interested with the magnetic state of our simulation, we refer the reader to the following multidimensional studies of internal gravity waves generation in solar-like stars \citep{2006ApJ...653..756R,2011ApJ...742...79B,2014AandA...565A..42A,2015A&A...581A.112A}. Turning to the radial flux balance, we note that the enthalpy flux (dash-triple-dotted line) dominates energy transport in most of the convective envelope. The diffusive fluxes (radiative in the bottom half of the computational domain (long-dash) and unresolved near the top (dotted line)) carry the stellar luminosity at each end of the domain. We note an inward kinetic energy flux (dash dot line) reaching about 10\% of the star's luminosity, as is common to find in stratified convection simulations. The Poynting and viscous fluxes account for less than 1\% of the radial energy balance. Finally, we note the negative enthalpy flux near the base of the convective envelope, that is compensated by a local increase of the radiative flux, such as to reach a satisfactory radial energy balance and thermal equilibrium.

\begin{table*}[!ht]
\begin{center}

\vspace{0.2cm}
\begin{tabular}{l|lllllllllllll}
\hline
{} & $\Omega_*$ & $\tilde{V}_r$ & $\tilde{V}_\theta$ & $\tilde{V}_\phi$ & $\Delta \Omega$ & $\tilde{B}_r$ & $\tilde{B}_\theta$ & $\tilde{B}_\phi$  &  $\tau_c$ & $\tau_\nu$ & $\tau_\kappa$  & $\tau_\eta$  \\
{} & $[\Omega_{\odot}]$ & [m/s] & [m/s] & [m/s] & [nHz] & [G] & [G] & [G] &  [days] & [years] & [years] & [years]    \\
{} & {} & {} & {} & {}  & MHD (HD) & {} & {} & {} & {} &{} & {} & {}    \\
\hline

M05Sm  & 1/8 & 13.52 &  12.19 &  29.35 &  -23  (-24) & 13.54 & 15.01 & 24.87  & 102.23 & 15.78 &  3.94 & 15.78 \\
M05R1m & 1  & 7.27 &   7.39 &  29.72 &  112  (129) & 15.92 & 15.66 & 40.92  & 190.03 & 37.33 &  9.33 & 74.65 \\
M05R3m & 3  & 6.21 &   6.80 &  56.85 &  200  (85)  & 10.32 &  9.34 & 25.91  & 222.49 & 64.65 & 16.16 & 64.65 \\
M05R5m & 5  & 6.95 &   4.69 &   6.59 &  9    (146) & 39.36 & 49.61 & 70.71  & 198.79 & 64.65 & 16.16 & 64.65 \\
M07Sm  & 1/4  & 25.44 &  18.14 &  30.35 &  -53  (-32) & 11.85 & 11.05 & 17.72  &  62.82 &  4.50 &  1.13 &  4.50 \\
M07R1m & 1 & 16.34 &  14.48 &  44.72 &  111  (120) &  5.28 &  5.00 &  8.12  &  97.82 &  8.22 &  2.06 & 16.45 \\
M07R3m & 3  & 14.74 &  11.21 &  38.41 &  68   (187) & 29.62 & 33.21 & 67.96  & 108.46 & 14.24 &  3.56 & 28.49 \\
M07R5m & 5 & 13.42 &  11.55 &  14.34 &  -2   (223) & 35.85 & 42.33 & 54.71  & 119.11 & 18.39 &  4.60 & 18.39 \\
M09Sm  & 1/2 & 53.51 &  36.80 &  48.98 &  -36  (-25) &  1.70 &  1.66 &  1.68  &  35.83 &  2.72 &  0.68 &  2.72 \\
M09R1m & 1 & 38.74 &  35.32 &  68.55 &  102  (108) &  2.32 &  2.44 &  3.17  &  49.50 &  3.86 &  0.97 &  7.72 \\
M09R3m & 3 & 30.61 &  32.42 & 148.70 &  265  (288) &  1.07 &  1.07 &  1.93  &  62.64 &  6.67 &  1.67 &  6.67 \\
M09R5m & 5  & 27.94 &  19.74 &  56.43 &  76   (338) & 20.33 & 19.82 & 47.02  &  68.62 &  7.18 &  1.80 &  7.18 \\
M11R1m & 1 & 130.77 & 93.56 & 140.61 &  -102 (-131)& 12.69 & 11.81 & 13.06  &  16.67 &  1.46 &  0.37 &  2.93 \\
M11R3m & 3 & 90.23 &  81.57 & 272.53 &  278  (291) &  4.49 &  4.66 &  6.83  &  24.17 &  2.54 &  0.63 &  2.54 \\
M11R5m & 5 & 88.50 &  61.74 &  88.73 &  109  (435) & 18.63 & 18.48 & 32.93  &  24.63 &  2.62 &  0.65 &  3.27 \\
\hline
\end{tabular}
\caption{Models dimensional characteristics, averaged over a small interval of $0.01 R_\star$ at the middle of the convective envelopes (unless stated otherwise). Characteristic velocities, differential rotation, magnetic fields, and time scales are listed. The differential rotation is taken between latitude 60$^\circ$ and the equator at the surface of the models (see \S \ref{sec:DR}). Likewise, the total magnetic flux is computed at the surface of the models, and averaged over at least one magnetic cycle for the cyclic cases (see \S \ref{sec:TotFlux}). The {\rm rms} velocity and magnetic field are $\tilde{v} = \left( \tilde{V}_r^2 + \tilde{V}_\theta^2 + \tilde{V}_\phi^2\right)$ and $\tilde{B} = \left( \tilde{B}_r^2 + \tilde{B}_\theta^2 + \tilde{B}_\phi^2\right)$. Here $\tau_{c} = D / \tilde{v_{r}}$ is the overturning convection time, and the dissipation time scales are defined as $\tau_{x} = D^{2} /x$ with $x \in [\nu,\kappa,\eta]$, where $D = r_{top} - r_{bcz}$ the thickness of the convective layer that differs for each mass bin. }
 \label{tab:tableDimParams}
\end{center}
\end{table*}

\begin{table*}[!ht]
\begin{center}

\vspace{0.2cm}
\begin{tabular}{l|llllllllllll}
\hline
{} &    $Re$ &  $R_m$ &    $Pe$ & $P_m$ &   $Ra$  &   $Ra^\star/Ra_c$  &     $Ta$   & $Ro_{\rm f}$ & $Ro_{\rm c}$ & $Ro_{\rm s}$ &  $Ek$ &  $\Lambda$   \\
{} &  & &&  & [$10^6$] &  & [$10^6$] & & & & [$10^{-3}$] & [$10^{-3}$] \\
\hline

M05Sm  &  56.34 &  56.34 & 14.08 & 1 & 0.01 &   67.64 &    0.21 & 1.74 & 0.38 & 3.96 & 4.37 &  4.57 \\
M05R1m &  71.71 & 143.42 & 17.93 & 2 & 0.89 &   16.83 &   37.49 & 0.33 & 0.31 & 0.70 & 0.33 &  1.88 \\
M05R3m & 106.08 & 106.08 & 26.52 & 1 & 7.15 &    8.58 & 1012.05 & 0.15 & 0.17 & 0.23 & 0.06 &  0.14 \\
M05R5m & 106.08 & 106.08 & 29.68 & 1 & 7.60 &    5.69 & 2811.24 & 0.07 & 0.10 & 0.14 & 0.04 &  4.59 \\
M07Sm  &  26.16 &  26.16 &  6.54 & 1 & 0.01 &   73.32 &    0.05 & 1.24 & 0.72 & 2.98 & 9.03 &  7.18 \\
M07R1m &  30.69 &  61.38 &  7.67 & 2 & 0.11 &   28.75 &    1.82 & 0.42 & 0.50 & 0.89 & 1.48 &  0.39 \\
M07R3m &  47.94 &  95.88 & 11.98 & 2 & 0.64 &   14.78 &   49.12 & 0.16 & 0.23 & 0.30 & 0.29 &  8.40 \\
M07R5m &  56.36 &  56.36 & 14.09 & 1 & 1.59 &    8.91 &  227.44 & 0.09 & 0.17 & 0.18 & 0.13 &  8.68 \\
M09Sm  &  27.69 &  27.69 &  6.92 & 1 & 0.01 &   54.09 &    0.05 & 1.28 & 0.74 & 3.02 & 8.98 &  0.12 \\
M09R1m &  28.48 &  56.96 &  7.12 & 2 & 0.04 &   24.71 &    0.40 & 0.68 & 0.66 & 1.51 & 3.16 &  0.14 \\
M09R3m &  38.90 &  38.90 &  9.73 & 1 & 0.33 &   11.02 &   10.79 & 0.27 & 0.35 & 0.50 & 0.61 &  0.01 \\
M09R5m &  38.21 &  38.21 &  9.55 & 1 & 0.36 &    7.00 &   34.70 & 0.10 & 0.20 & 0.30 & 0.34 &  5.33 \\
M11R1m &  32.06 &  64.12 &  8.02 & 2 & 0.01 &   47.17 &    0.06 & 1.38 & 0.78 & 3.30 & 8.33 & 11.36 \\
M11R3m &  38.31 &  38.31 &  9.58 & 1 & 0.10 &   17.16 &    1.56 & 0.54 & 0.50 & 1.10 & 1.60 &  0.51 \\
M11R5m &  38.81 &  38.81 &  9.70 & 1 & 0.13 &    6.87 &    4.62 & 0.27 & 0.34 & 0.66 & 0.93 & 13.06 \\

\hline
\end{tabular}
\caption{Model characteristics non-dimensional numbers, averaged over a small interval of $0.01 R_\star$ at the middle of the convective envelopes. $Re = \tilde{v} D / \nu$ is the Reynolds number. The Prandtl number $Pr = \nu / \kappa=1/4$ in all cases. $P_m = \nu / \eta$ is the magnetic Prandtl number, $R_m = Re P_m$ is the magnetic Reynolds number, and $Pe = Re Pr$ is the P\'eclet number. $Ra = (-\partial \rb/ \partial S)\Delta Sg D^{3} / \rb \nu \kappa$ is the Rayleigh number, and $Ra^\star/Ra_c$ is the modified Rayleigh number as computed by \citet{2020ApJ...893...83T}. $Ta = 4\Omega_*^{2} D^{4}/\nu^{2}$ is the Taylor number. We also list three Rossby numbers: the fluid Rossby number  $Ro_{\rm f} = \tilde{\omega}/2\Omega_\star$, the convective Rossby number $Ro_{\rm c} = \sqrt{Ra / Ta Pr}$, and the stellar Rossby number $Ro_{\rm s} = P_{\rm rot}/\tau_c^{\rm CS}$. \ASmodd{The latter is useful for comparison from observationnally-derived Rossby numbers. For $Ro_{\rm s}$ we have therefore considered the empirical convective turnover time derived by \citet{2011ApJ...741...54C} which is $\tau_c^{\rm CS} = 314.24{\rm exp}\left[ - \frac{T_{\rm eff}}{1952.5\, {\rm K}} - \left(\frac{T_{\rm eff}}{6250\, {\rm K}}\right)^{18}\right]$ + 0.002 days. We note that it correlates well with our fluid Rossby number and find $Ro_{\rm s} \simeq 2.26 Ro_{\rm f}$.} \ASmod{The Ekman number is defined as $Ek = \nu/(\Omega_\star D^2$), and the Elsasser number as} $\Lambda = \tilde{B}^2/8\pi \rb D \tilde{v} \Omega_\star$.
 \label{tab:tableAdimParams}
}
\end{center}
\end{table*}

\vspace{0.5cm}
\subsection{G and K stars parametric study}

As indicated above, we initiate each of the 15 dynamo simulations from mature, relaxed hydrodynamics convective states and introduce a seed magnetic field in the convective envelope only. These hydrodynamical progenitors have been run long enough to reach a statistically stationary state in the convection zone and a well established rotation profile. They possess a genuinely established tachocline, defined as the transition between differential rotation in the outer convective envelope to solid-body rotation in their stable radiative interior, leading to regions with strong shear \citep{1992AandA...265..106S}. The tachocline plays an important role in the dynamo process of magnetic field generation in solar-like stars as reported in simulations performed by several authors \citep{1985GApFD..31..137G,2006ApJ...648L.157B,0004-637X-735-1-46,2013ApJ...778...11M,2015ApJ...813...95L,2016ApJ...819..104G}.

The main parameters of the models are listed in Tables \ref{tab:tableDimParams} and \ref{tab:tableAdimParams}. 
The density scale heights between the top and the base of the convection zone and  between the top and the bottom of the model are defined as $N_{\rb_{bcz}} = {\rm ln} (\rb_{out}/\rb_{bcz})$ and $N_{\rb{tot}} = {\rm ln} (\rb_{out}/\rb_{in})$. For the M05 model $N_{\rb_{bcz}} = 3.25$ and $N_{\rb_{tot}} = 4.70$, M07 model $N_{\rb_{bcz}} = 3.48$ and $N_{\rb_{tot}} = 5.78$, M09 model $N_{\rb_{bcz}} = 3.31$ and $N_{\rb_{tot}} = 5.99$, M11 model  $N_{\rb_{bcz}} = 3.28$ and $N_{\rb_{tot}} = 5.60$. The convective flows at the middle of the convective envelope vary from 5 m/s up to about 300 m/s in our sample, and the convective turnover time from 7 (case M11R1m) to 222 (case M05R3m) days. The surface differential rotation between the equator and latitude $60^\circ$ varies from -102 to +278 nHz in our sample, and we will study its maintenance in detail in \S\ref{sec:DR}. In the middle of the convection zone, the root-mean-squared magnetic field typically varies from 1 G to 70 G in our models. It is found to be maximum close to the bottom of the convective envelope, where the large-scale shear efficiently powers the dynamo, and the magnetic field can be stored in the tachocline close to the convective-radiative boundary.

 Our sample of simulations were designed to operate in a relatively homogeneous turbulent Reynolds number regime, as seen in the first column of Table \ref{tab:tableAdimParams}. The supercriticality degree can be characterized by the Rayleigh number achieved in our models, compare to a critical Rayleigh number for the onset of convection. Such a modified Rayleigh number was proposed by \citet{2020ApJ...893...83T} and is listed in column 6 of Table \ref{tab:tableAdimParams}. All our models exhibit a Rayleigh number at least five time larger than the critical Rayleigh number. We have run these models as long as we could while maintaining a reasonable numerical cost to achieve the large parameter study presented here, and while computing the models showing a magnetic cycle over several decades. In this study we use the Rossby number as a measure of the influence of rotation on the flows maintaining the differential rotation as well as to power dynamos. Several Rossby number definitions have been proposed in the community, and we have computed the fluid Rossby number $Ro_{\rm f}$, the convective Rossby number $Ro_{\rm c}$, and the stellar Rossby number $Ro_{\rm s}$ in Table \ref{tab:tableAdimParams}. We refer the reader to \citet{2017ApJ...836..192B} and their Appendix for a more in-depth discussion of these various definitions of the Rossby number. We will focus here on the fluid Rossby number $Ro_{\rm f}$ and note that the other two are related to $Ro_{\rm f}$ through a nearly linear relationship. The fluid Rossby number decreases with rotation rate and increases with mass, and varies from 0.07 to 2.35 in our sample of models. This range nicely covers the transition from solar-like to anti-solar differential rotation regimes (the transition is nearly at $Ro_{\rm f} \simeq 1$), and our smallest Rossby number (namely model M05R5m) is close to the expected fast-rotators saturation regime. \\
 We now briefly present one representative cyclic dynamo solution before entering a more detailed analysis of our dynamo simulations ensemble in sections \ref{sec:LargeScaleFlows} to \ref{sec:EnergyContent}.

\begin{figure}[!h]
\centering
\includegraphics[width=0.75\linewidth,angle=-90]{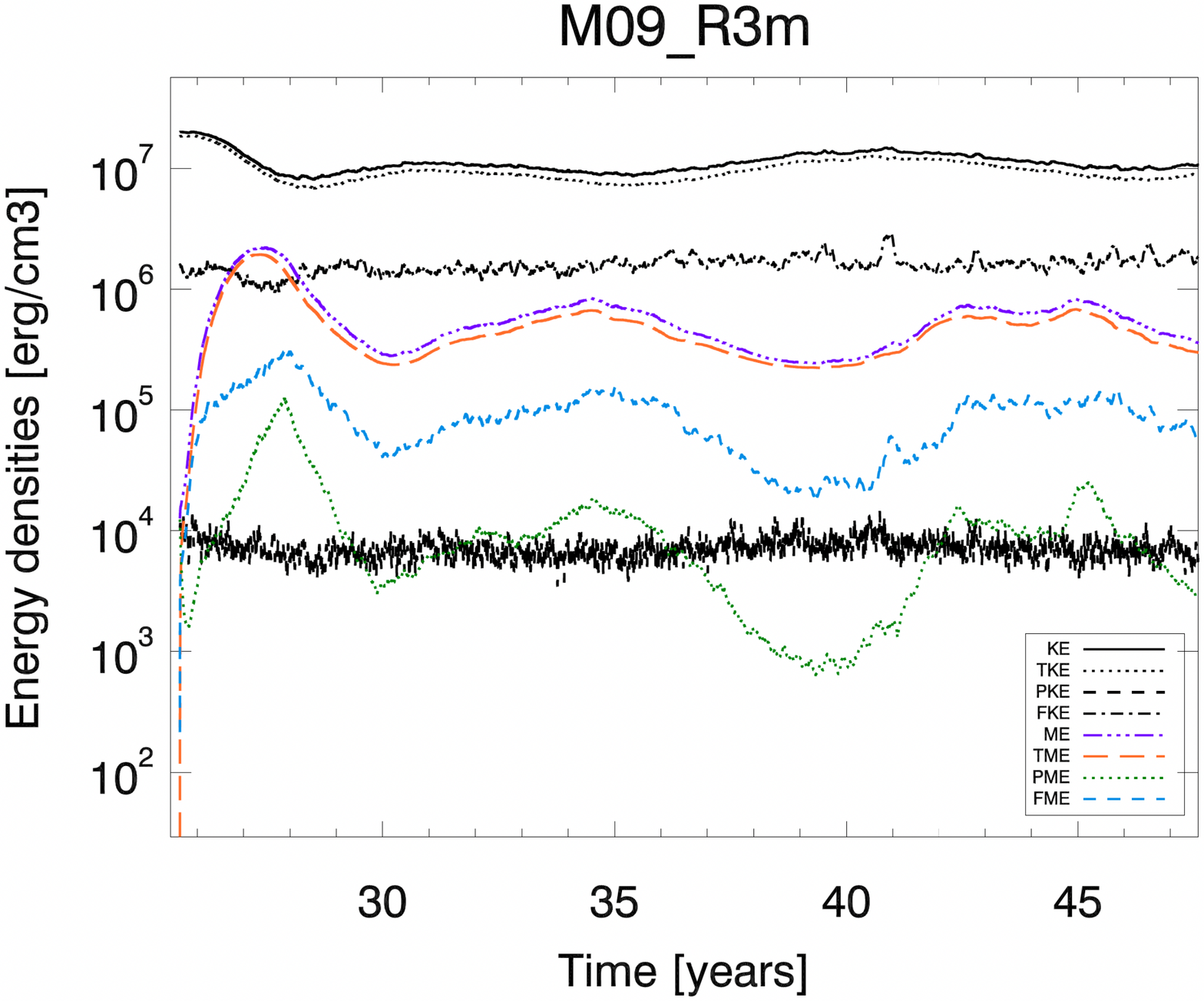}
\caption{Temporal evolution in case M09R3m of kinetic (KE) and magnetic (ME) energies. We also show their axisymmetric toroidal TKE, TME and poloidal PKE and PME components and their fluctuating components FKE and FME. We note the rise over about 500 days of ME just after having introduced a weak seed field. \ASmod{Then follows a modulation of ME with a 9 year period. Case M09R3m is indeed} one of our cyclic cases (see also Figure \ref{fig:SummaryDynamoM09R3}).}
\end{figure}\label{fig:KEME}

\section{Overview of one cyclic dynamo case}
\label{sec:overview}

\begin{figure*}[!h]
\centering
\includegraphics[width=0.9\linewidth]{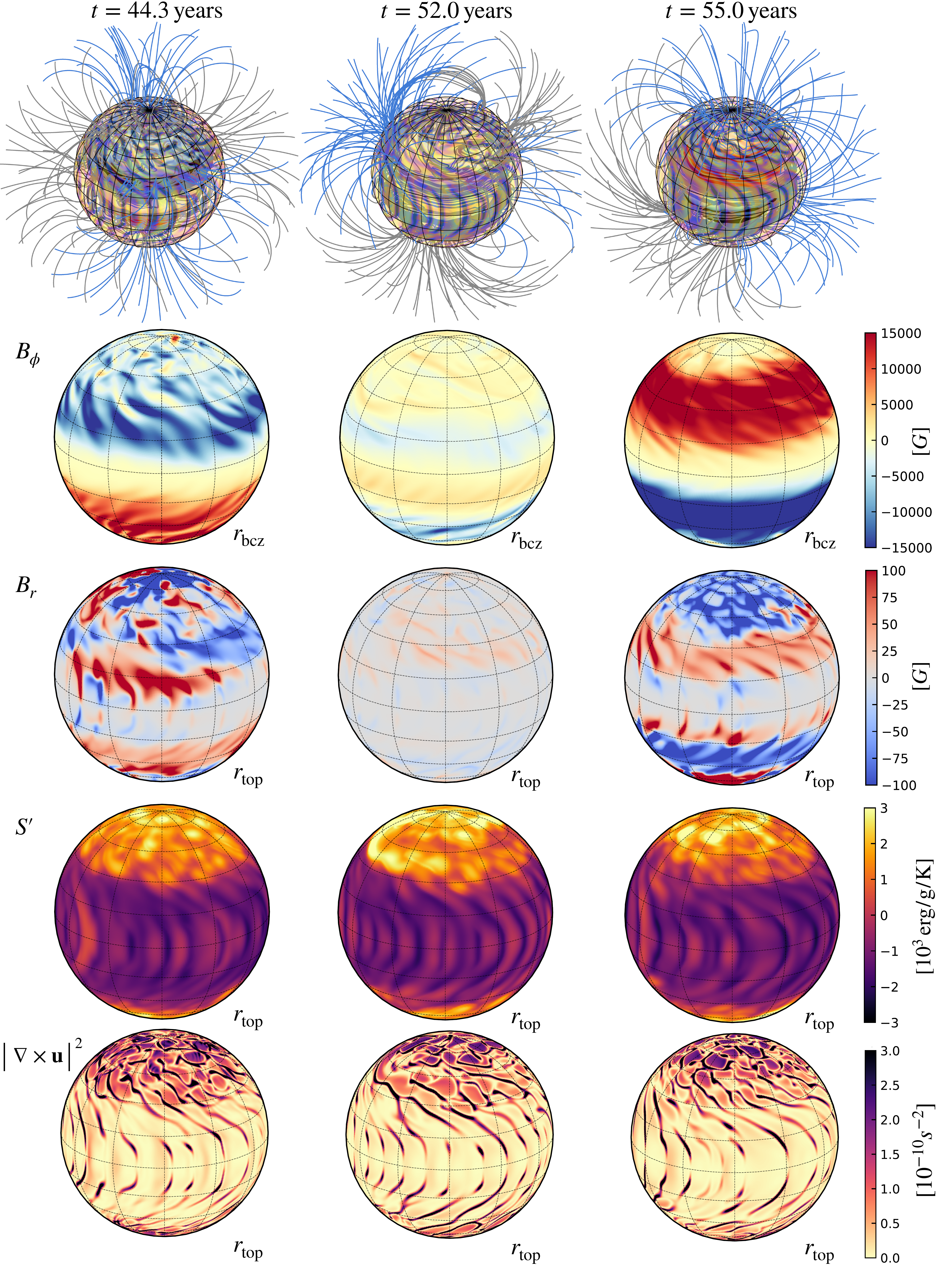}
\caption{Temporal evolution of a magnetic cycle for case M09R3m, taken at three different instances during one reversal. The upper row shows a 3D potential extrapolation of the modelled magnetic field, with blue lines denoting field lines oriented outward and black lines oriented inward. Behind a semi-transparent representation, the radial velocity close to the surface, while deeper below the magnetic wreaths are shown by red (oriented eastward) and blue (oriented westward) lines. The second row shows the azimuthal field at the bottom of the convection zone, and the third row the radial field at the top of the domain. The fourth row shows the entropy fluctuations at the top of the domain, and the lowest row the enstrophy at the same depth. \label{fig:SummaryDynamoM09R3}}
\end{figure*}

\ASmod{To illustrate the richness of the dynamo solutions discussed in this study, it is key to show how the subtle nonlinear interplay between convection,
rotation, and turbulence leads to the generation of time dependent complex magnetic fields}. \ASmod{All 15 models discussed in detail in this study successfully generate and maintain a dynamo-generated magnetic field against Ohmic dissipation}. 

We defer the systematic comparison between all 15 models to the next sections and focus here on the representative case M09R3m. Indeed, M09R3m is in an intermediate Rossby number regime ($R_{of}=0.27$) and therefore lies in the middle of our sample of models. The temporal evolution of kinetic and magnetic energies is shown in Fig. \ref{fig:KEME}. The magnetic energies first rise very fast to then saturate after about 1000 days in this case, and exhibit long-term oscillations over a decadal timescale reminiscent of a solar-like magnetic cycle. All components of the magnetic energy (toroidal, poloidal, fluctuating) oscillate in phase in this model. The mean toroidal kinetic energy also presents oscillation of the same amplitude, albeit anti-correlated with the magnetic ones. These energy trends are similar to the ones found in the magnetic cycles obtained with the EULAG code in \citet{2018ApJ...863...35S} (see their Fig. 3) and points toward a similar dynamo mechanism involving a strong feedback of the magnetic field on the differential rotation within the convective envelope. We will perform a detailed analysis of this mechanism in \S\ref{sec:MagneticProperties}. 
Here, we first illustrate the dynamics of the dynamo achieved in model M09R3m in Fig. \ref{fig:SummaryDynamoM09R3}. The top row shows the 3D structure of our model by means of a potential field extrapolation outside our computational domain, at three different instances covering a magnetic reversal. We see that the field at the South Pole changes from blue to black, showing the polarity reversal. The strong toroidal field at the base of the convective envelope can be seen through transparency. In the leftmost panel, this deep wreath is mainly blue (westward oriented). Its polarity is reversed in the rightmost panel (red, eastward oriented), showing that the polarity reversal takes places over the full convective domain. The subsequent rows show spherical slices of $B_\phi$ at the base of the convective envelope (second row), $B_r$ at the top of the domain (third row), entropy fluctuations ($S'$) at the top of the domain (fourth row), and enstrophy $\left|\nabla\times \vv  \right|^2$ (last row). We recover in the second row the magnetic wreath located at the base of the convective envelope and at mid-latitude that changes polarity as the cycle progresses. The toroidal field reaches high values up to $1.5\times 10^4 $G, with a strong temporal variation, as seen in the middle panel during the reversal. The surface radial field (third row) reaches values of about $100 G$ and exhibits a complex topology, mixing dipolar and quadrupolar symmetries. We see again here that both fields oscillate in phase and reach a minimum in the midst of the magnetic reversal (middle panels). Finally, the two last rows show the thermal (entropy) fluctuations and the vortical motions (enstrophy) in our simulations. The first striking aspect is that these two quantities vary very little along the magnetic cycle. Indeed, the magnetic field modifies the large-scale motions and the average convective state in our models. Yet the magnetic cycles (when present) show little imprint on the convective flows themselves and mainly act on the mean flows (see \S 
\ref{sec:DR}). The specific entropy fluctuations have two distinctive features. \ASmod{Firstly, a mean pole-to-equator contrast is well established in the model, with higher entropy fluctuations at the poles. Such contrast is expected} in models with a solar-like differential rotation \citep{2002ApJ...570..865B} and can be generally related to the pressure field required to drive the observed meridional flow. \ASmod{Secondly, patterns} in $S'$ are imprinted  by the non axisymmetric convective motions themselves, which are also recognizable in the enstrophy in the lower panel. The enstrophy is concentrated at the boundaries of the so-called banana cells at low latitude \citep{2000ApJ...532..593M}, and is distributed between convective cell centers and boundaries at high latitude.     

We now turn to the detailed analysis of the large-scale flows (\S\ref{sec:LargeScaleFlows}), magnetic properties (\S\ref{sec:MagneticProperties}), and energetic balances (\S\ref{sec:EnergyContent}) achieved in the 15 models. The 
reader mostly interested in the astrophysical consequences of our study may consider going directly to \S\ref{sec:AstrophysicalImplications} for a summary.

\begin{figure*}[!th]
\centering
\includegraphics[width=\linewidth]{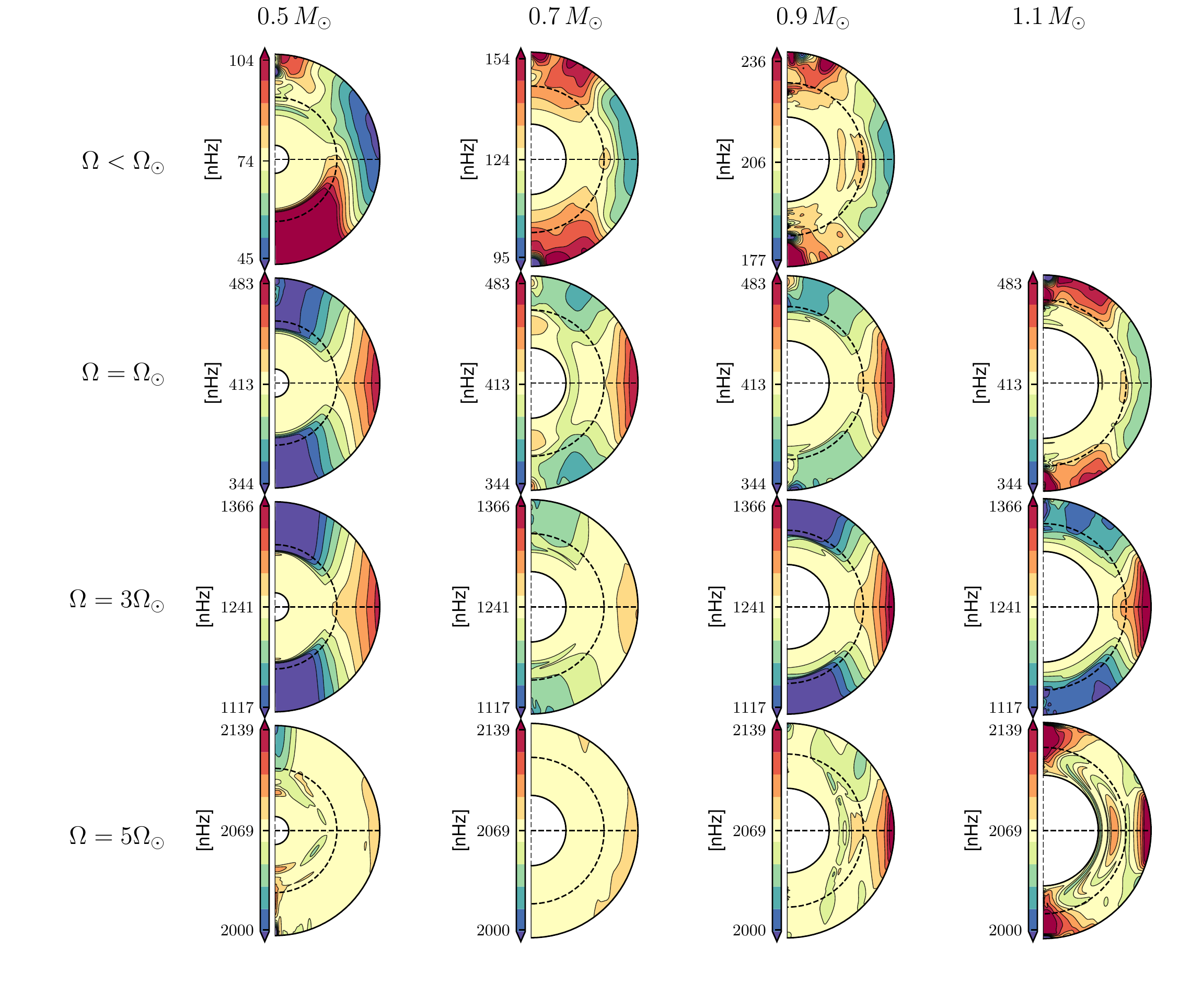}
\caption{Temporal and longitudinal averages of the angular velocity profiles \ASmod{over ten convective overturning times (10 $\tau_{c}$) in our suite of models}. Prograde flows are in reddish tones and retrograde ones in blueish tones. In each panel, the dashed semicircle represents the base of the convective envelope and the dashed horizontal line the equator.}  
\end{figure*}\label{RP}

\begin{figure*}[!th]
\centering
\includegraphics[width=0.95\linewidth]{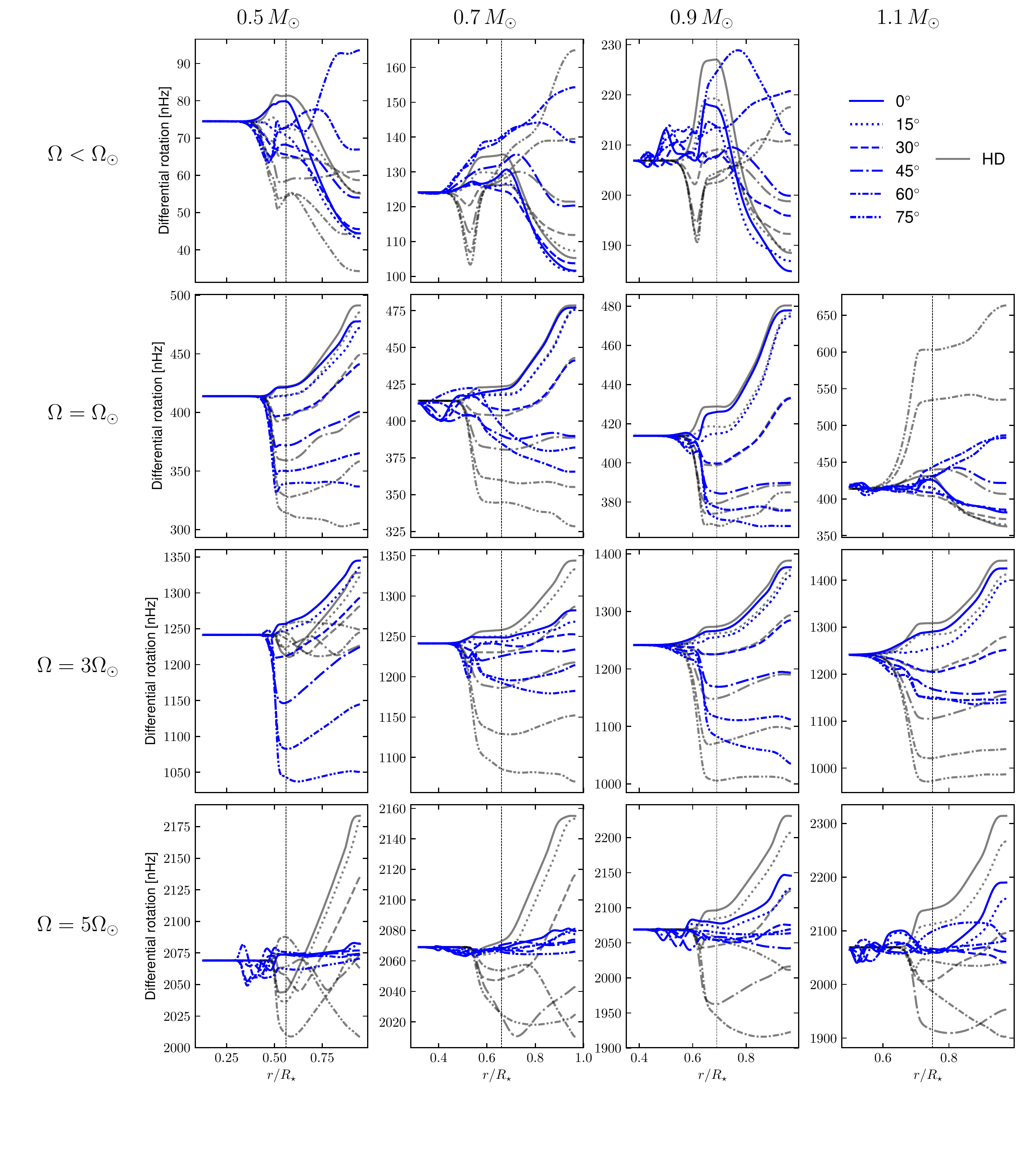}
\caption{Radial cuts of the temporal and longitudinal averaged angular velocity from the equator to the latitude 75$^{\circ}$ each 15$^{\circ}$ (black lines are for the hydrodynamical cases and the blue lines the MHD/dynamo cases). The solid vertical line in each panel shows the bottom of the convective layer.}  
\end{figure*}\label{DRprof}

\section{\ASmod{Large-scale flows in the models}}
\label{sec:LargeScaleFlows}

In this section we analyze the differential rotation profiles of the models including both a stable subadiabatic layer and magnetic field self-consistently generated by dynamo action. The aim of the study is to compare the differential rotation profiles of the hydrodynamical and MHD models. We confirm our preliminary results \citep{2016AdSpR..58.1507V} and that of others \citep{2015AandA...576A..26K,2016ApJ...819..104G,2018A&A...616A.160V} that the presence of magnetic fields leads to different trends for the differential rotation with stellar rotation rate and mass when compared to their hydrodynamical counterpart.
We further discuss how the meridional circulation is impacted by the presence of a magnetic field and discuss the main mechanisms acting to redistribute angular momentum within the convective shell. We also observe torsional oscillations in our set of dynamo simulations but delay their discussion to section \ref{sec:TO}.

\subsection{Differential rotation profiles as a function of Rossby number}
\label{sec:DR}

We analyze the differential rotation of the simulations that results from the angular momentum redistribution occurring mostly in the convection zone. The panels of Figure \ref{RP} show a meridional cut of the axisymmetric differential rotation averaged over 10 overturning convective times, defined as $\tau_{c} = \int_{r_{bcz}}^{r_{top}} dr / \tilde{v_{r}}$ (see Table \ref{tab:tableDimParams}).

We observe that for the simulations M05Sm, M07Sm, M09Sm and M11R1m there is an anti-solar differential rotation, with the poles rotating faster than the equator, like their hydrodynamical counterparts (see Figure \ref{DRprof} and also \citealt{2017ApJ...836..192B}). The cases rotating at an intermediate rotation rate show a solar-like differential rotation. Finally, the cases rotating the fastest (R5 series) show almost no differential rotation (in particular for cases M05Sm and M07Sm). This constitutes a big difference with their hydrodynamical counterpart cases. The magnetic field here had a major impact, with almost solid body rotation imposed throughout the convective envelope. There is little asymmetry in the profiles between the northern and southern  hemispheres, as expected when the average is performed over an interval long enough with respect to the convective overturning time (except for M05Sm for which the rotational constraint is the weakest and the longitudinal average less meaningful). Figure \ref{DRprof} also displays radial cuts of the rotation for the MHD cases (blue lines) and hydrodynamic progenitor cases (gray lines). In cases rotating 1, 3 and 5 times the solar rotation rate (bottom three rows), the velocity range in latitude (different styles of line) is generally reduced in the presence of magnetic field. This effect is observed to be stronger as the rotation rate increases. Conversely, the effect of magnetic field is mild for the slowly-rotating cases (upper row), except on the slowly-rotating case M11R1m which still shows some degree of magnetic feedback on its differential rotation. As one may expect, the radial gradient of the differential rotation nearby the tachocline is generally weaker in all the MHD simulations compared with hydrodynamic progenitors. This points to a magnetic feedback of the dynamo field on differential rotation itself, a feedback that is observed to strengthen as the rotation rate increases.

We have calculated the surface latitudinal differential rotation $\Delta\Omega$ for each model, defined as the difference between the equator and 60$^{o}$ latitude. A positive value thus denotes a solar-like differential rotation, and a negative value an anti-solar differential rotation. We report these values for the magnetic cases as well as the hydrodynamic progenitors in Table \ref{tab:tableDimParams} (fourth column).

The differential rotation of our sample spans a range between -102 and +278 nHz, with some fast-rotating models presenting an extremely weak DR, like M05R5m with $\Delta\Omega = 9$ nHz. We find that the absolute differential rotation generally weakens in MHD models compared to their hydrodynamic progenitors, as expected from the radial profiles shown in Fig. \ref{DRprof}. This is particularly striking for fast rotators such as M09R5m that goes from 338 nHz in hydro to 76 nHz in MHD. 

We investigate in Fig. \ref{DR_fit} the differential rotation trends with respect to the rotation rate (top panel) and rotation period (bottom panel). The differential rotation of the hydrodynamic progenitors and of the MHD cases are respectively shown in small semi-transparent and large opaque symbols. The shape of the symbol labels the rotation of the model, and the color the mass of the modelled star, as indicated in the legend. In the bottom panel, we compare the model differential rotation to the differential rotation in the Kepler sample obtained by \citet{2015A&A...583A..65R} (shown as black dots). The dotted lines correspond to their estimated observational detection limits. We first note that the absolute value of our differential rotations agree well with the observed values. In addition, the differential rotation range in our sample increases as the rotation period decreases, like what is observed in the {\it Kepler} satellite sample. Several of our models nevertheless lie outside the observed values: the three anti-solar differential rotations on the right (triangles), and two of our fast-rotating models. Several reasons can explain this discrepancy. Slowly-rotating stars could produce very few starspots, or even no starspots at all (see for instance \citealt{2019ApJ...872..128V}), making their differential rotation impossible to detect with photometry. Another possibility is that they lie outside the presently detectable limit with the Kepler data, due to their long rotational period (up to about 200 days for our most slowly rotating model). Finally, the two fast rotating models (M05R5m and M07R5m) show very weak differential rotations due to magnetic feedback, which are outside the detection limits of {\it Kepler} (dotted black lines). 

The top panel of Fig. \ref{DR_fit} shows the differential rotation trend with the rotation rate. Using only the hydrodynamic progenitors, we previously showed that the differential rotation scales as $\Omega^{0.66}$ \citep{2017ApJ...836..192B}. Blindly trying to fit such a power-law to the MHD sample, we find that the exponent reduces to $\Omega^{0.46}$. This weaker dependency is expected due to the magnetic feedback on the differential rotation through the Lorentz force. It also agrees better with the observational trends, which are still quite uncertain and were found to vary from 0.2 (\citealt{2016MNRAS.461..497B} for G stars), 0.3 (\citealt{2013AandA...560A...4R} for cool stars) to even 0.7 (\citealt{1996ApJ...466..384D} for F-K stars). Looking more closely at our sample on the top panel of Fig. \ref{DR_fit}, it clearly appears that a power-law fit is a poor representation of the differential rotation in our sample. Rather, we see that $\Delta \Omega$ increases with $\Omega$ for slow rotators, while it dramatically drops for fast rotators due to the magnetic feedback. Following \citet{2011IAUS..273...61S}, we recast in Fig. \ref{DR_fit2} the differential rotation trend in terms of relative differential rotation $\Delta \Omega/\Omega$ with respect to the fluid Rossby number $R_{of}$ (for the different definitions of Rossby number used in this work, see the caption of Table \ref{tab:tableAdimParams} or the appendix A of  \citealt{2017ApJ...836..192B}). We find a trend that is very similar to the observational trend reported by \citet{2011IAUS..273...61S} (shown by the dashed line in Fig. \ref{DR_fit2}): $\Delta\Omega/\Omega$ is roughly constant for inverse Rossby numbers lower than a certain threshold (here $Ro_f^{-1} \lesssim 5$), and it drops for fast rotators as $\Delta\Omega/\Omega \propto Ro_f^{p}$. \citet{2011IAUS..273...61S} proposed that $p=2$, but here our sample agree with a somewhat large range $p\in [2,6]$. Additional models with even higher turbulence level are required to confirm the exact amplitude of the drop in differential rotation contrast found in the fast rotating cases. Finally, our sample also shows some hint of an increase of $\Delta\Omega/\Omega$ at large Rossby numbers, which is outside the observable constraints for now. It would be interesting to search observationally for candidate solar-like stars possibly possessing such anti-solar rotation states.

In \citet{2017ApJ...836..192B}, we have proposed that the differential rotation could follow two power-laws with respect to the Rossby number and the stellar mass. Here, we find that the differential rotation is weakened at high Rossby number, and therefore we do not recover a simple power-law trend, as we saw in Fig. \ref{DR_fit}. We can nevertheless attempt to fit such a combined power-law on a sub-sample of our models, excluding the fast rotating case but retaining the slow rotators. We obtain in this way 
\ASmodd{\begin{eqnarray}
    &\Delta\Omega \ASmodmath{= 107} Ro_{\rm f}^{-0.73 \pm 0.13} \frac{M_\star}{M_\odot}^{1.93 \pm 0.42} \, nHz \; {\rm  (HD)} \, , \\
    %&\Delta\Omega \ASmodddmath{= 84} Ro_{\rm f}^{-0.36 \pm 0.19} \frac{M_\star}{M_\odot}^{0.83 \pm 0.64} \, nHz \; {\rm (MHD)} \, .
    &\Delta\Omega \ASmodmath{= 84 Ro_{\rm f}^{-0.40 \pm 0.20} \frac{M_\star}{M_\odot}^{0.78 \pm 0.62}} \, nHz \; {\rm (MHD)} \, .
\end{eqnarray}}
In the MHD case, we find again that the differential rotation is less sensitive to both the Rossby number and the stellar mass. The power-law fit is nevertheless questionable here, as the range covered by our Rossby numbers and masses is quite small. We have nevertheless included the results of the fit here to compare with the purely hydrodynamic case \citep{2017ApJ...836..192B}. We can conclude here that the clear trend in stellar mass and effective temperature found in the hydrodynamic study \citep{2017ApJ...836..192B} is less significant when magnetism is taken into account, but overall we see a better agreement with observations of the dynamo models compared to their hydrodynamical progenitors.

\begin{figure}[!h]
\centering
\includegraphics[width=0.95\linewidth]{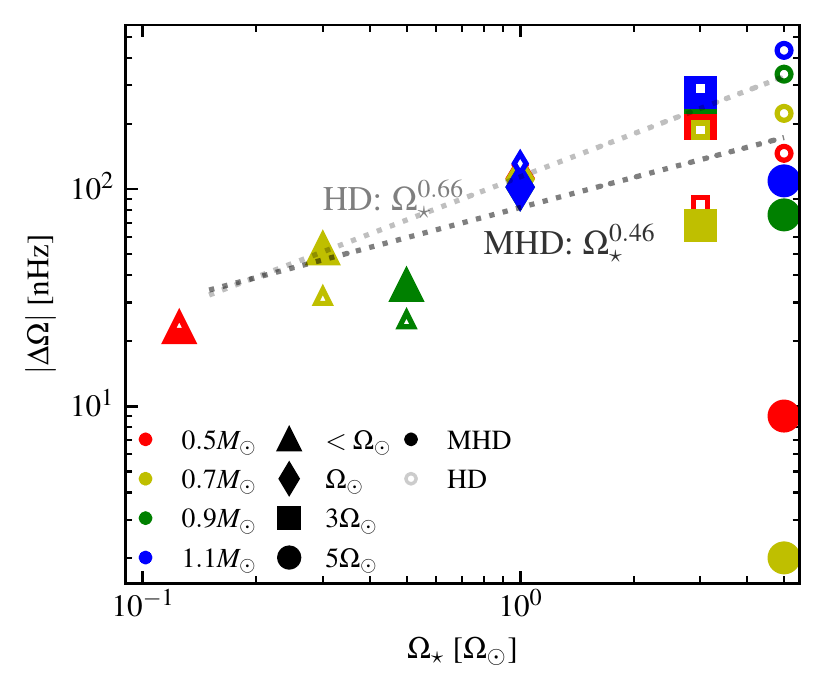}
\includegraphics[width=0.95\linewidth]{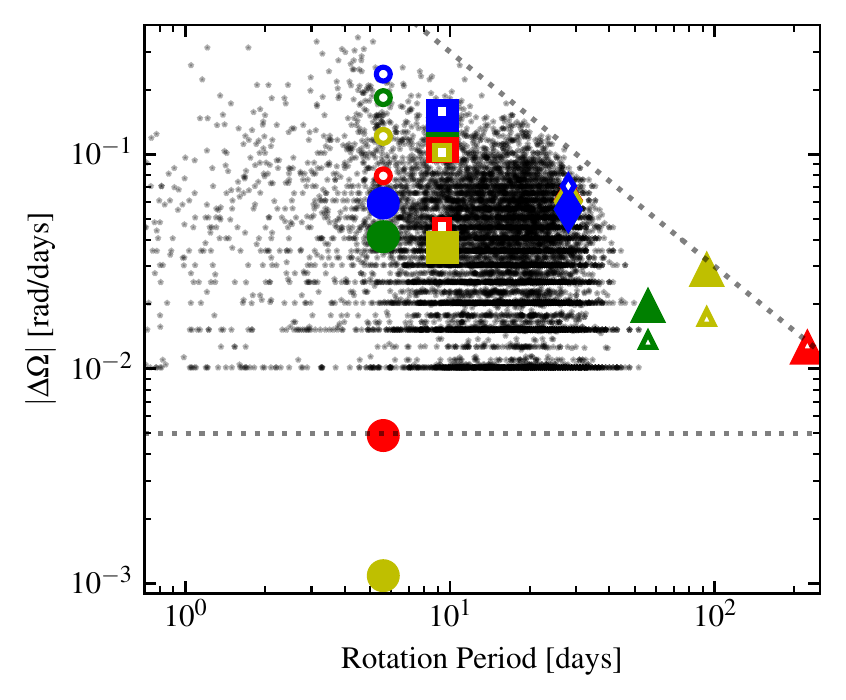}
\caption{Absolute value of the differential rotation between the equator and 60$^{o}$ latitude versus rotation rate (top panel) and rotation period (bottom panel). The symbols denote the rotation rate of the model and the color the mass of the modelled star, as shown in the legend. \ASmod{MHD models are shown by the large plain symbols, and the hydrodynamic progenitors by the smaller open ones.} On the bottom panel, the differential rotation in the Kepler sample obtained by \citet{2015A&A...583A..65R} is shown as black dots, and the observational detection limit by the two dotted black lines.} 
\end{figure}\label{DR_fit}

\begin{figure}[!h]
\centering
\includegraphics[width=0.95\linewidth]{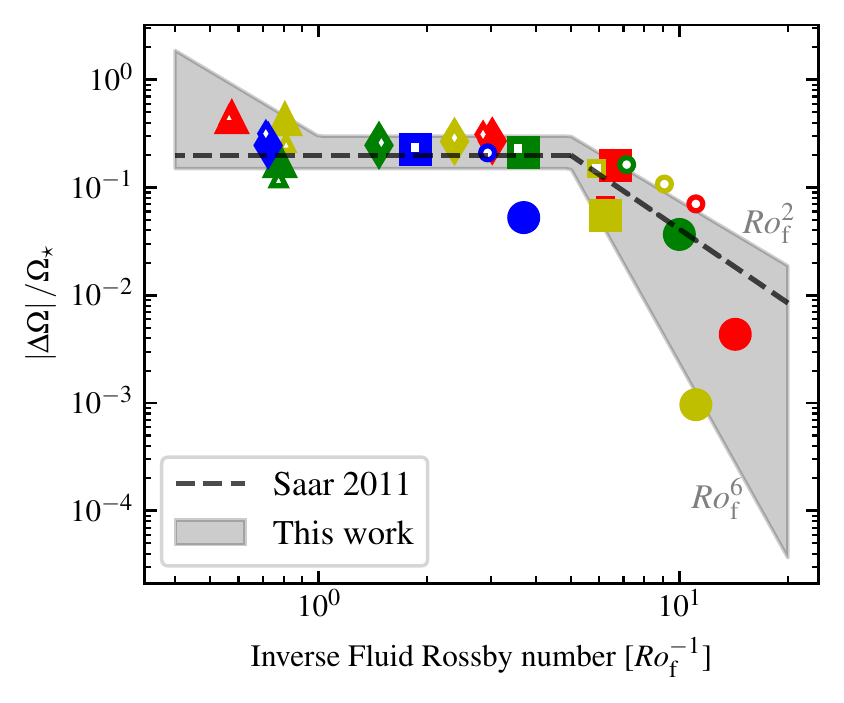}
\caption{Relative differential rotation between the equator and 60$^{o}$ latitude as a function of the fluid Rossby number. The symbols shape and color are the same as in Fig. \ref{DR_fit}. The trend found in the MHD sample is highlighted by the grey area, and the observational trend reported by \citet{2011IAUS..273...61S} is shown by the dashed black line.} 
\end{figure}\label{DR_fit2}

The MHD simulations therefore show that the magnetic field changes the angular momentum redistribution, especially for fast rotating stars. 
In the next section, we perform a detailed analysis of this balance for 4 representative models.

\subsection{Angular momentum transfer}

We can better understand how the differential rotation profiles are achieved in the dynamo models by identifying the main physical processes responsible for redistributing angular momentum within rotating convective shells. Our choice of stress-free and potential-field boundary conditions at the top and stress-free and perfect conductor boundary conditions at the bottom of the computational domain have the advantage that no net external torque is applied, and thus angular momentum is 
conserved.  We can assess the transport of angular momentum by considering the mean radial (${\cal F}_r$) and latitudinal
(${\cal F}_{\theta}$) angular momentum fluxes, applying the
procedure used in \citet{2004ApJ...614.1073B}. Starting from the 
$\phi$-component of the momentum equation expressed in conservative
form and averaged in time and longitude:

\begin{equation}
\frac{1}{r^2} \frac{\p(r^2 {\cal F}_r)}{\p r}+\frac{1}{r \sin\theta}
\frac{\p(\sin \theta {\cal F}_{\theta})}{\p
\theta}=0,
\end{equation}
involving the mean  radial angular momentum flux
\begin{eqnarray}
{\cal F}_r=\rb r\sin\theta [&-&\nu r\frac{\p}{\p
r}\left(\frac{\hat{v}}{r}\right)+\widehat{v_{r}^{'}
v_{\phi}^{'}}+\hat{v}_r(\hat{v}_{\phi}+\Omega r\sin\theta) \nonumber \\ 
&-&\frac{1}{4\pi\rb}\widehat{B_{r}^{'}
B_{\phi}^{'}}-\frac{1}{4\pi\rb}\hat{B}_r\hat{B}_{\phi}], \end{eqnarray}
and the mean latitudinal angular momentum flux
\begin{eqnarray}
{\cal F}_{\theta}&=&\rb r\sin\theta[-\nu
\frac{\sin\theta}{r}\frac{\p}{\p
\theta}\left(\frac{\hat{v}_{\phi}}{\sin\theta}\right)+\widehat
{v_{\theta}^{'} v_{\phi}^{'}} \\ &+&\hat{v}_{\theta}(\hat{v}_{\phi}+\Omega
r\sin\theta) -\frac{1}{4\pi\rb}\widehat{B_{\theta}^{'}
B_{\phi}^{'}}-\frac{1}{4\pi\rb}\hat{B}_{\theta}\hat{B}_{\phi}].\nonumber
\end{eqnarray}

In these equations, the terms on the
right-hand-side represent for both fluxes contributions respectively from viscous
diffusion (which we denote as ${\cal F}_r^{\rm VD}$ and ${\cal
F}_\theta^{\rm VD}$), Reynolds stresses (${\cal F}_r^{\rm RS}$ and ${\cal
F}_\theta^{\rm RS}$), meridional circulation (${\cal F}_r^{\rm MC}$ and ${\cal
F}_\theta^{\rm MC}$), Maxwell stresses (${\cal F}_r^{\rm MS}$ and ${\cal
F}_\theta^{\rm MS}$) and large-scale magnetic torques (${\cal F}_r^{\rm MT}$
and ${\cal F}_\theta^{\rm MT}$). The Reynolds stresses are linked to 
correlations of the fluctuating velocity components coming from
organized tilts within the convective structures, especially in the
downflow plumes. Likewise, the Maxwell stresses are associated with correlations of the fluctuating magnetic field components due to the twist and tilt of the dynamo generated magnetic structures.

\begin{figure*}[!th]
\centering
\includegraphics[width=0.45\linewidth]{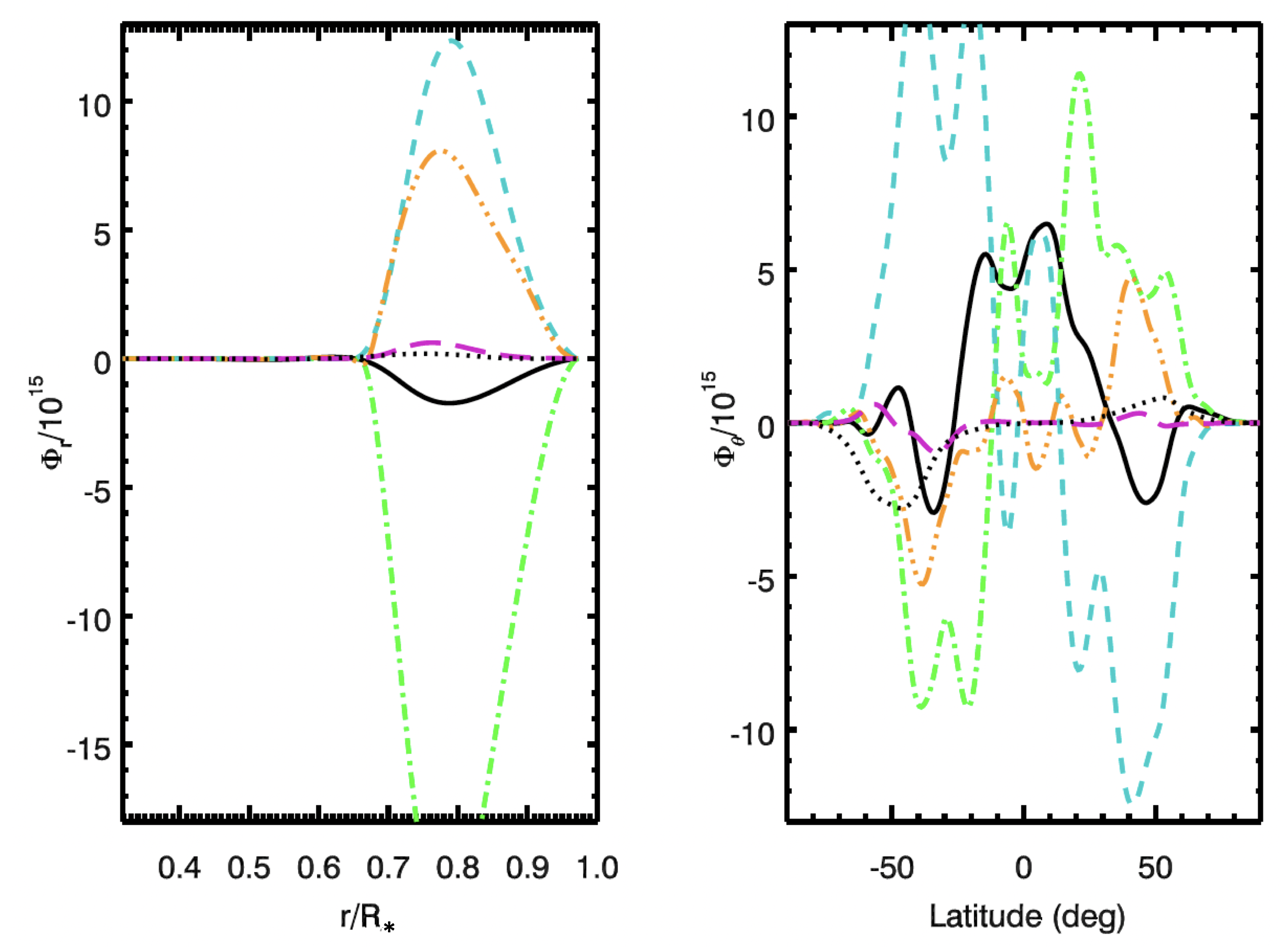}
\includegraphics[width=0.46\linewidth]{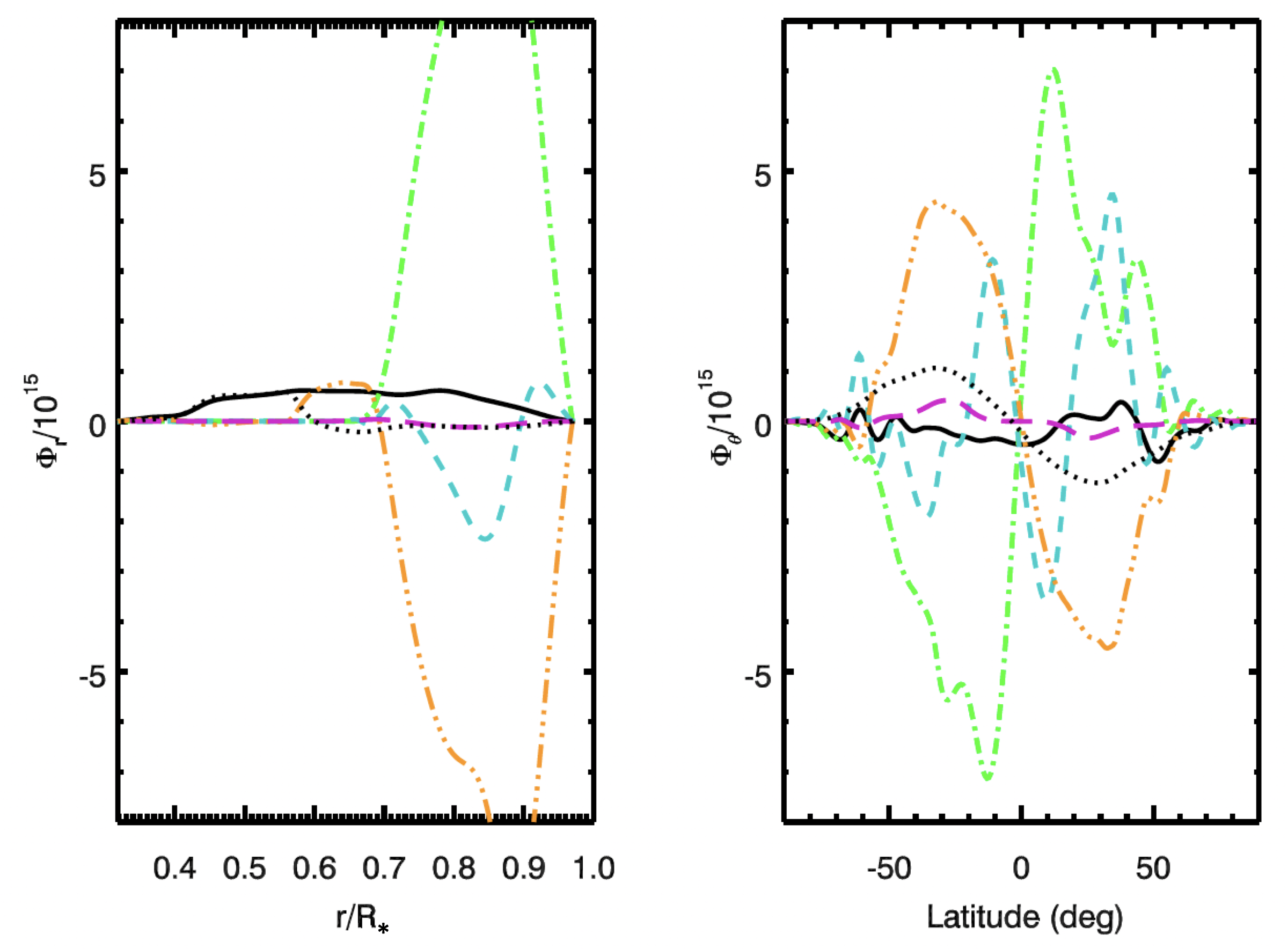}
\includegraphics[width=0.45\linewidth]{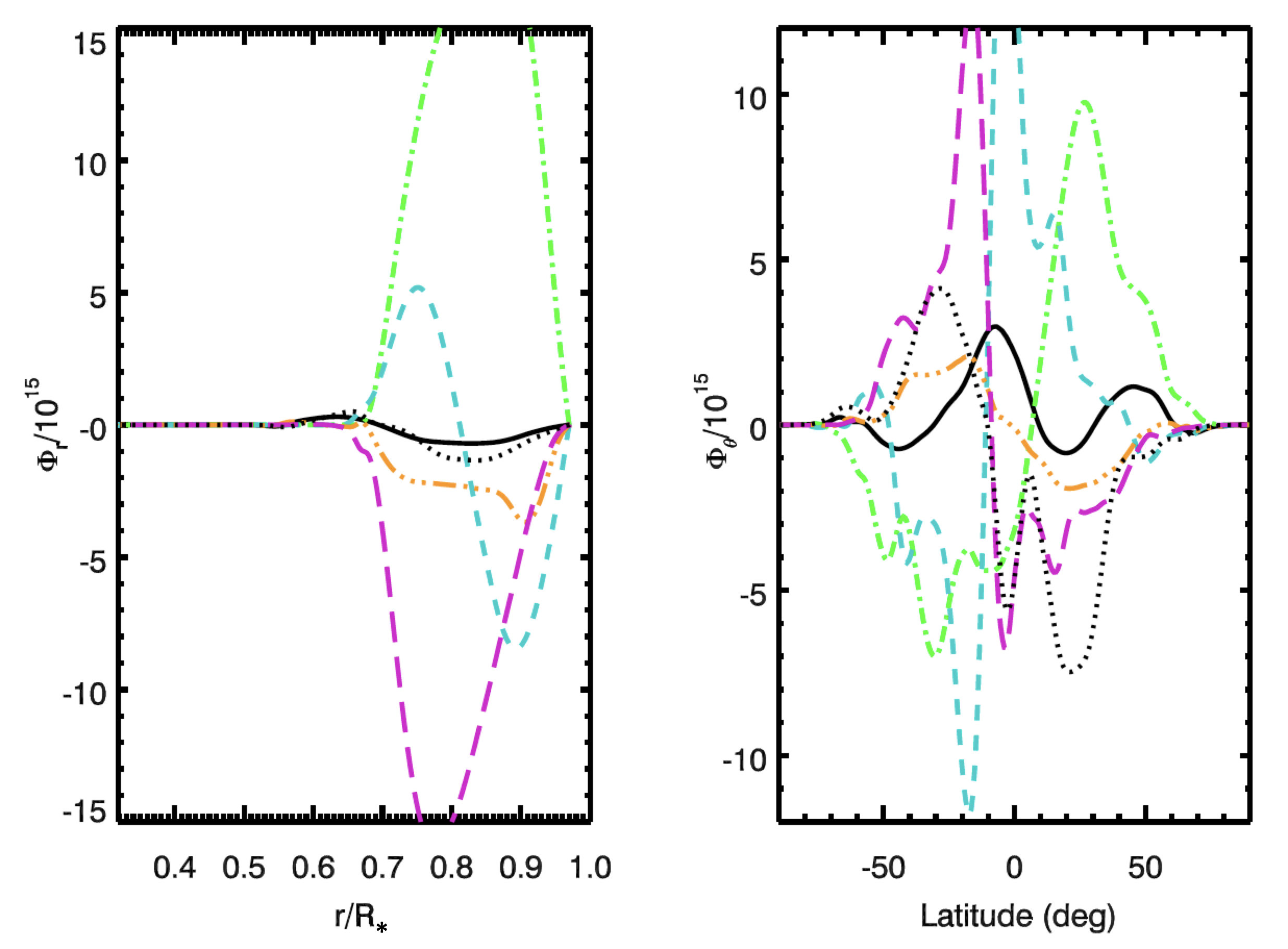}
\includegraphics[width=0.46\linewidth]{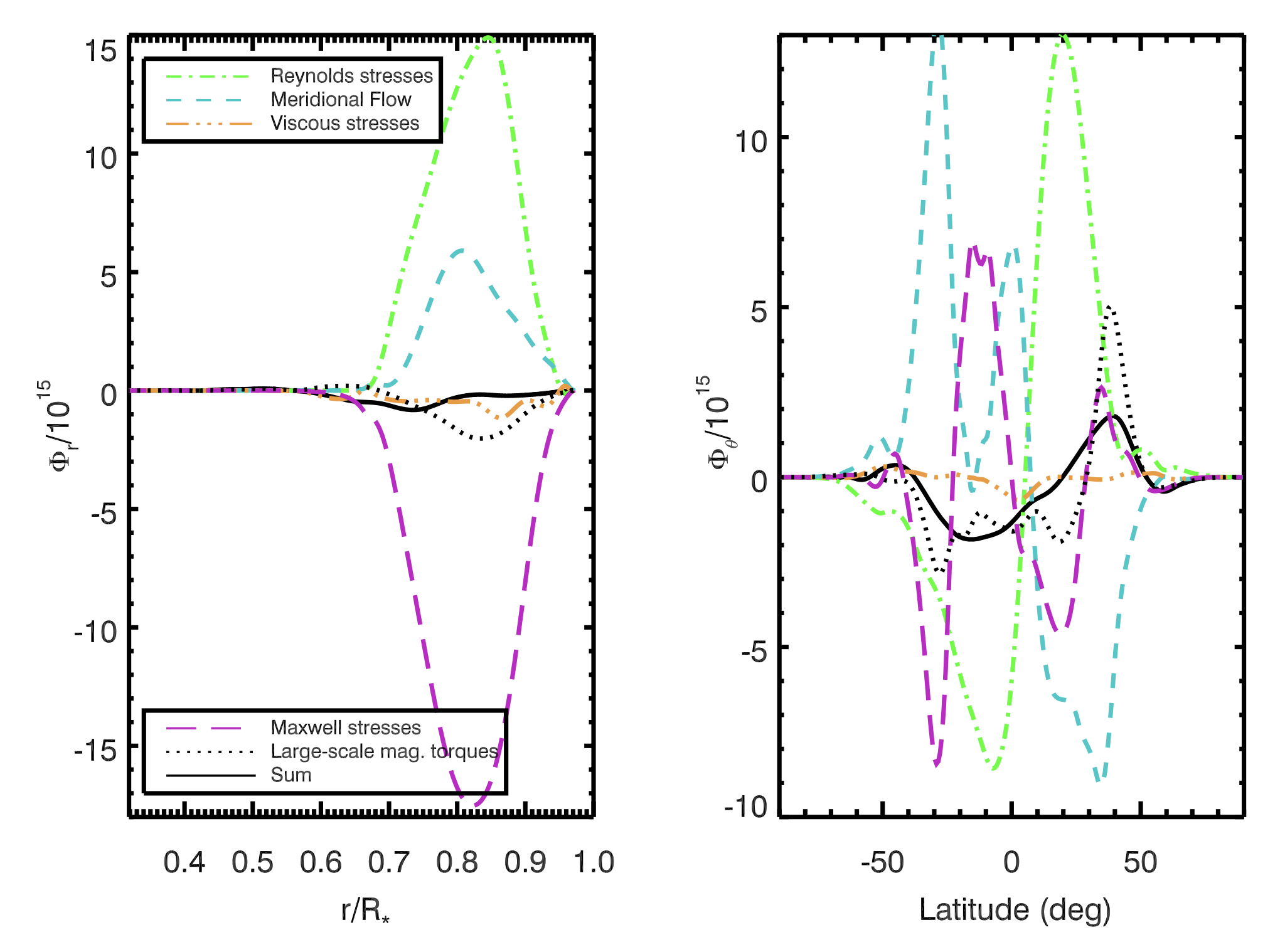}
\caption{Angular momentum transport in M07 case series. We display cases M07Sm (left) and M07R1m (right) in the top row and M07R3m (left) and M07R5m (right) in the bottom row. For each model radial (left) and latitudinal (right) angular momentum balance is shown, with Reynolds stresses contribution shown as green dash-dotted lines, viscous stresses as dash-triple-dotted yellow lines, meridional circulation as dashed cyan lines, Maxwell stresses as long dashed magenta lines, the large scale magnetic torques as black dotted line and the sum of all contributions as solid black line. We note that a very good angular momentum balance is achieved in most models in both directions with the sum being close to zero.}
\end{figure*}\label{amom}

In Figure \ref{amom} we show the components of  ${\cal F}_r$ and
${\cal F}_{\theta}$ for M07 case series, having
integrated over co-latitude and radius as follows: 
\begin{eqnarray}
\Phi_r(r)=\int_0^{\pi} {\cal F}_r(r,\theta) \, r^2 \sin\theta
\, d\theta \; \mbox{ , } \nonumber \\ \Phi_{\theta}(\theta)=\int_{r_{bot}}^{r_{top}} {\cal
F}_{\theta}(r,\theta) \, r \sin\theta \, dr \, .
\end{eqnarray}
Thus $\Phi_r$ represents the net angular momentum flux through
horizontal shells at different radii and $\Phi_\theta$ represents the net
flux through cones at different latitudes. This representation is helpful in assessing
the direction and amplitude of angular momentum transport within
the computational domain by each component of ${\cal F}_r$ and ${\cal F}_{\theta}$.

For each of the four cases we display $\Phi_r$ on the left panel and $\Phi_{\theta}$ on the right panel both normalized by $R_*^2$.
Turning to the radial angular momentum transfer, \ASmod{we first note a very good overall radial balance. We find that the} Reynolds stresses (green dash-dotted curves) transport angular momentum outward in all the low Rossby number models. By contrast, M07S the slowly rotating case has the Reynolds stresses transporting angular momentum inward. The viscous diffusion and Maxwell stresses oppose this transport, tending to rigidify the rotation state in the radial direction. The meridional circulation has one large cell per hemisphere for the M07Sm case (see \S\ref{sec:MC}). It opposes the Reynolds stresses, but as the rotation rate increases and the Maxwell stresses gain in amplitude, it changes in profiles and direction to yield a radial balance of angular momentum, from the angular momentum equation. Note that the mean large-scale magnetic torques (black dotted line) have little influence in the overall radial angular momentum balance.

Considering now $\Phi_{\theta}$, we can assess the balance of latitudinal angular momentum transport. We first notice that the Reynolds stresses (green curves) are systematically equatorward in both hemisphere (positive in the northern hemisphere and negative in the southern one). Since most cases exhibit a very good latitudinal balance, as demonstrated by the solid black curve, these Reynolds stresses must be nicely counterbalanced. A quick survey of the right panels for all four models indicates that many contributors act depending on the Rossby number of the simulations. For the slowly rotating case (upper left corner), we see that it is mostly the meridional circulation (cyan dashed curve) that does most of the work (we defer the reader to section \ref{sec:MC} for a discussion of the meridional circulation patterns in the various dynamo cases). By contrast, magnetic terms do not play much role in the case of M07Sm. For M07R1m (right top corner) it is now the viscous diffusion that plays that role of opposing the Reynolds stresses. For that case the meridional circulation is not doing much, but we do see a 20\% contribution of the large scale magnetic torques, the Maxwell stresses being still weak. As the Rossby number decreases and the dynamo action becomes more intense, we see that the magnetic terms start influencing the latitudinal angular momentum transport more and more, tending to oppose the Reynolds stresses. It is particularly noticeable for the Maxwell stresses. They are the dominant player for M07R3m case (bottom left corner), helped by the large scale magnetic torques. In that case the meridional circulation is somewhat helping the Reynolds stresses, notably at low latitudes near the equator. 

\begin{figure*}[!th]
\centering
\includegraphics[width=\linewidth]{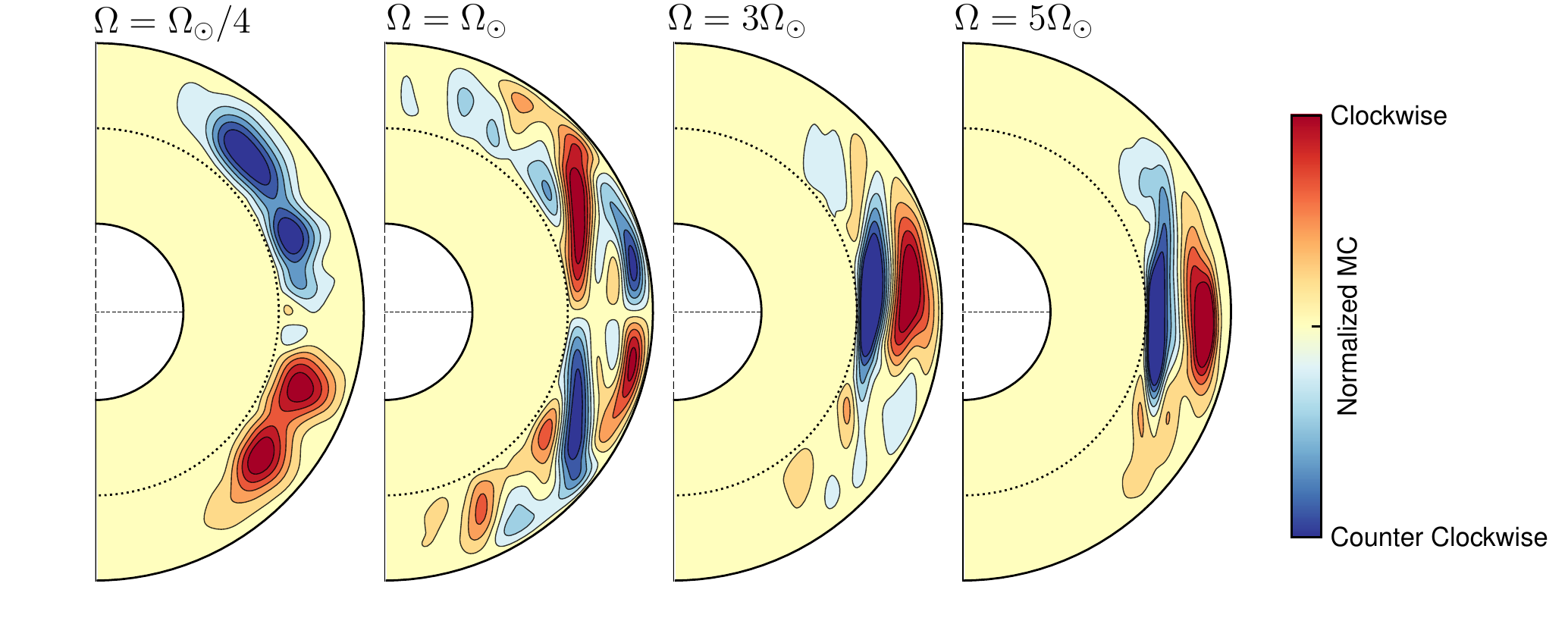}
\caption{Meridional flows in M07m case series. The contours are normalized and denote clockwise (red) and anti-clockwise (blue) circulations.}
\end{figure*}\label{MCM07}

For the M07R5m case, the story becomes less clear, except for the Reynolds stresses all terms fluctuate and sometimes oppose or reinforce the turbulent stresses. Maxwell stresses still play an important role as does the meridional circulation. In that model, the differential rotation has been so significantly quenched by dynamo action, that it is not surprising the trends are less clear and systematic. \ASmod{In summary, in most cases the transport of angular momentum by Reynolds stresses are opposed by a combination of meridional circulation, viscous stresses and Maxwell stresses.}

\subsection{Meridional circulation profiles}
\label{sec:MC}

The meridional flow patterns are also affected by the presence of magnetism in our set of models, especially for the fast rotating cases. We immediately note that the meridional circulation is indirectly modified by magnetism (as will be made clear in \S \ref{sec:Etransfers}). Indeed, magnetic stresses play a negligible role in setting the meridional flows in our models, and the differences we observe compared to the hydrodynamical counterparts originate from changes in the differential rotation (see e.g. \citealt{2014A&A...568A.113P}).

%\AS{I will add some text for MC here.}
%\ASB{Quid Meridional Circulation, need to show 3 representative profiles}

We illustrate the meridional flow pattern achieved in the M07m set of simulations in Fig. \ref{MCM07}. The slow rotating case (first panel) is very similar to its hydrodynamic progenitor, with a well-defined circulation cell in each hemisphere. Both cells circulate from the equator to the pole at the surface, and from the pole to the equator at the base of the convective envelope. The second model rotating at the solar rate (second panel) is also similar to its hydrodynamical progenitor and shows a more complex circulation profile. These are consistent with previous numerical experiments by e.g. \citet{2015AandA...576A..26K}. It consists of stacked cells elongated along the rotation axis outside the tangent cylinder, and two counter-rotating cells in each hemisphere at high latitude. Finally, the fast-rotating models (third and fourth panels) exhibit a peculiar meridional circulation pattern concentrated at the equator, with two stacked trans-equatorial cells (see \textit{e.g.} \citealt{2009EL.....8519001S}). These profiles can be understood as follows. In these models, the differential rotation is strongly quenched by magnetic feedback as seen in the previous section. In particular, the radial shear of differential rotation vanishes at the equator as seen in Fig. \ref{DRprof}. As a result, gyroscoping pumping \citep{2006ApJ...641..618M,2007sota.conf..183M,2015ApJ...804...67F} dramatically weakens along the equator and the resulting meridional circulation is both very weak (this can be seen in the drop of meridional flow kinetic energy in Table \ref{tab:tableEnergies}) and mainly driven by the remaining latitudinal shear. This leads to two meridional cells crossing the equator, as seen in the last panels of Fig. \ref{MCM07}. Having presented the large-scale flows achieved in the simulations, we now turn to discussing their magnetic properties.

\begin{table*}[!ht]
\begin{center}

\vspace{0.2cm}
\begin{tabular}{l|lllcllllll}
\hline
{} &    Long Cycle &   Short Cycle &  $\delta_t \Omega$ & Active lat.  & $\Phi_{\rm tot}$ & $B_{\rm rms}$ & $B_{r, \rm{dip}}$ & $B_{L,{\rm surf}}$ & $f_{\rm dip}$  & $f_{\rm quad}$  \\
{} &       (y/[yrs]/n) &      (y/[yrs]/n)  & [nHz]      &   [$^\circ$ - $^\circ$]      & [$10^{24}$ Mx]$^{\rm max}_{\rm min}$ & [G]$^{\rm max}_{\rm min}$ & [G]$^{\rm max}_{\rm min}$ & [G]$^{\rm max}_{\rm min}$ &                &                 \\
\hline

M05Sm  & n              & n             &  2.4 & [43 - 45] & $3.0_{ 2.6}^{ 3.3}  $ & $793_{683}^{884}   $ & $191_{156}^{217}$ & $357_{296}^{407}$ & 0.21 & 0.22  \\
M05R1m & $13.6 \pm 5.7$ & $1.2 \pm 0.6$ &  2.1 & [31 - 31] & $1.9_{ 1.3}^{ 2.5}  $ & $575_{343}^{805}   $ & $58_{17}^{99}   $ & $158_{88}^{234}$ & 0.05 & 0.25  \\
M05R3m & $\ASmodmath{21.4 \pm 9.4}$              & $0.5 \pm 0.2$ &  2.1 & [44 - 45] & $\ASmodmath{2.1_{1.0}^{ 3.4}  }$ & $\ASmodmath{552_{297}^{830}}$ & $\ASmodmath{38_{5}^{93}}$ & $\ASmodmath{146_{69}^{215}}$ & 0.28 & 0.37  \\
M05R5m & n              & $1.8 \pm 1.0$ &  3.0 & [49 - 55] & $8.6_{ 5.7}^{ 9.1}  $ & $1698_{1131}^{1851}$ & $196_{51}^{236} $ & $819_{426}^{929}$ & 0.19 & 0.35  \\
M07Sm  & n              & n             &  5.4 & [51 - 53] & $4.8_{ 4.5}^{ 5.2}  $ & $575_{524}^{648}   $ & $136_{130}^{142}$ & $306_{279}^{340}$ & 0.29 & 0.38  \\
M07R1m & $\ASmodmath{6.2 \pm 1.1}$          & $1.4 \pm 1.3$ &  3.7 & [26 - 35] & $8.6_{ 5.8}^{11.7}  $ & $949_{616}^{1325}  $ & $200_{100}^{266}$ & $439_{261}^{601}$ & 0.38 & 0.48  \\
M07R3m & y              & $2.5 \pm 0.8$ &  4.1 & [21 - 22] & $8.7_{ 3.2}^{13.0}  $ & $972_{127}^{1951}  $ & $157_{40}^{251} $ & $340_{198}^{573}$ & 0.12 & 0.40  \\
M07R5m & n              & $1.0 \pm 0.7$ &  1.2 & [55 - 58] & $18.4_{ 15.3}^{20.2}$ & $1597_{1320}^{1879}$ & $181_{52}^{351} $ & $925_{699}^{1072}$ & 0.35 & 0.48  \\
M09Sm  & n              & n             & 13.1 & [72 - 73] & $1.8_{ 1.7}^{ 1.8}  $ & $109_{97}^{123}    $ & $53_{49}^{55}   $ & $60_{55}^{66}$ & 0.60 & 0.17  \\
M09R1m & n              & n             &  9.0 & [20 - 21] & $1.1_{ 0.9}^{ 1.5}  $ & $68_{58}^{100}     $ & $11_{9}^{16}    $ & $23_{18}^{33}$ & 0.23 & 0.29  \\
M09R3m & $9.9 \pm 1.8$  & $0.9 \pm 0.6$ &  9.0 & [24 - 26] & $2.2_{ 0.3}^{ 4.0}  $ & $133_{15}^{261}    $ & $10_{0.9}^{23}  $ & $47_{6}^{91}$ & 0.16 & 0.27  \\
M09R5m & n              & $1.3 \pm 0.7$ &  9.5 & [30 - 35] & $13.4_{ 10.7}^{18.7}$ & $657_{485}^{970}   $ & $274_{221}^{400}$ & $392_{284}^{608}$ & 0.35 & 0.44  \\
M11R1m & n              & n             & 12.4 & [46 - 47] & $14.5_{ 13.2}^{15.8}$ & $589_{544}^{650}   $ & $10_{0.8}^{23}  $ & $184_{161}^{197}$ & 0.06 & 0.31  \\
M11R3m & $4.9 \pm 0.9$  & n             & 11.5 & [20 - 21] & $2.7_{ 0.7}^{7.1}   $ & $86_{20}^{226}     $ & $11_{0.7}^{31}  $ & $55_{13}^{160}$ & 0.22 & 0.29  \\
M11R5m & .              & .             & 39.0 & [52 - 77] & $53.1_{ 41.6}^{57.7}$ & $1208_{980}^{1329} $ & $713_{528}^{875}$ & $809_{596}^{986}$ & 0.51 & 0.11  \\
\hline
\end{tabular}
\caption{
Magnetic properties of the modelled dynamos. The first column indicates the presence or absence of a long (decadal), deeply-seated magnetic cycle. When the time-series were long-enough to identify unambiguously a cycle period, its value is given with error bars. Otherwise, the existence of such a cycle is indicated by a yes ('y'), and its absence by a no ('n'). The second column shows the same for the short magnetic cycle that we identify in the upper convection zone near the equator. We do not indicate this information for model M11R5m that was not run long enough to determines the existence or absence of magnetic cycles. The third column indicates the amplitude of the torsional oscillations at the surface in nHz (see \S\ref{sec:TO}). The fourth column shows the  active latitudinal band at the bottom of the convection zone, \ASmod{based on the azimuthally-averaged and temporally-varying azimuthal field straddling the base of the convection zone}. The fourth column shows the total magnetic flux at the surface, in units of $10^{24}$ Mx, with minimum and maximum as subscript and superscript (see \S\ref{sec:TotFlux}). The three next columns show the root-mean-squared surface field in Gauss, the surface dipole in Gauss, and the surface large-scale radial field $B_{L,{\rm surf}}$ (taken for $l<5$) in G with the same layout (see \S\ref{sec:AstrophysicalImplications}). Finally, the last two columns show the fractions of the large-scale dipole ($f_{\rm dip}$) and quadrupole ($f_{\rm quad}$), as defined in \S\ref{sec:fdipfquad}. 
 \label{tab:MagCyParams}
}
\end{center}
\end{table*}

\section{Magnetic properties}
\label{sec:MagneticProperties}
In this section, we discuss in more details various aspects of our dynamo simulations, such as their type, their temporal variability, the amount of magnetic flux they generate and the distribution in space and size of their magnetic fields.

\begin{figure*}[!th]
\centering
\includegraphics[width=\linewidth]{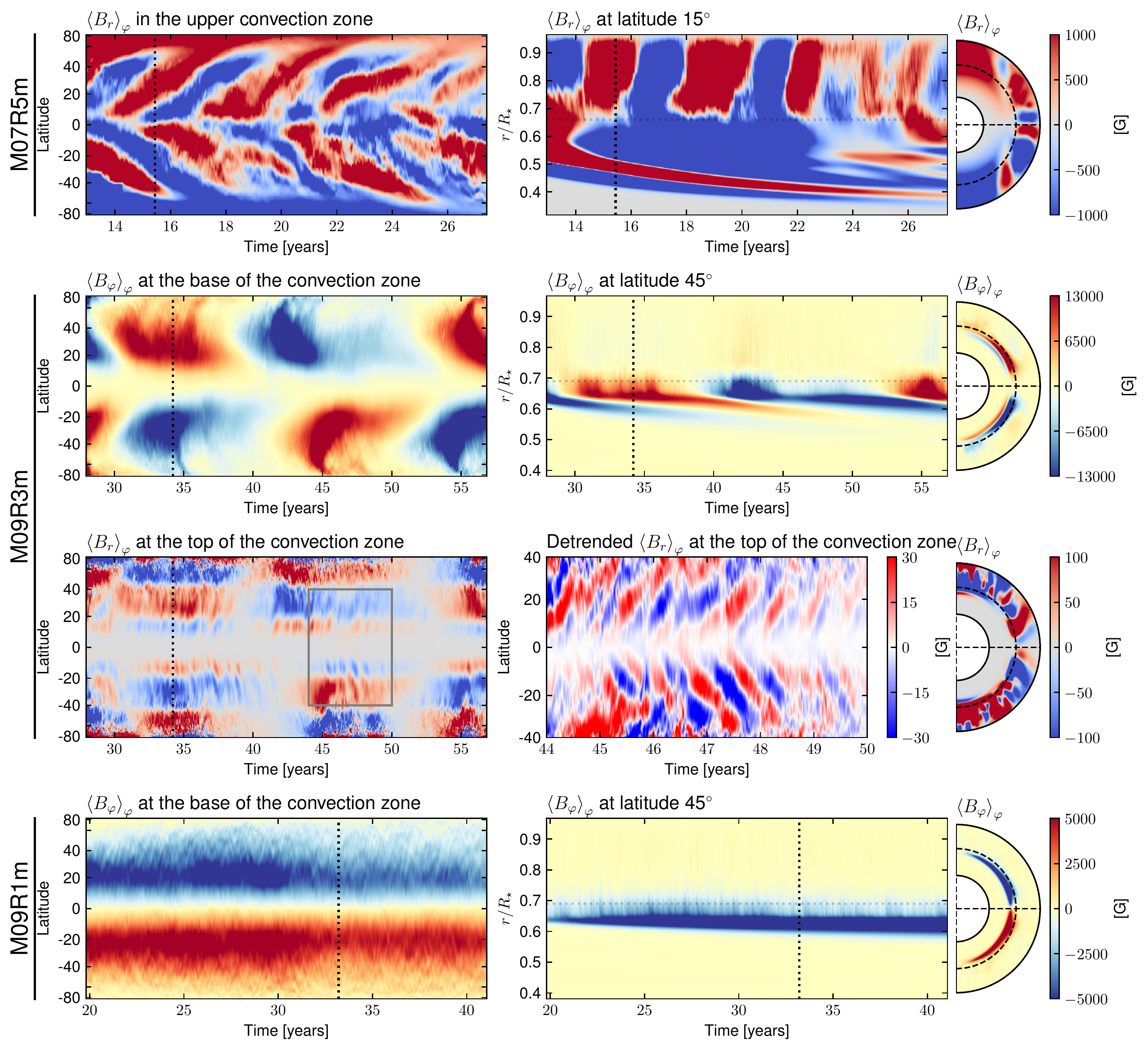}
\caption{Various dynamo states achieved in our sample our models, as illustrated by M07R5m (top panels), M09R3m (middle panels) and M09R1m (lower panels). In the top panel we show time-latitude, time-radius, and instantaneous meridional plane of $B_r$ (red denotes positive values and blue negative values)\ASmod{, with sampling times indicated by a vertical dashed line}. These illustrate the short magnetic cycles achieved by our models. The four middle panels illustrate both the short and long cycles achieved in model M09R3m. The first panels show latitude-time (at the base of the convection zone) and radius-time (at mid-latitude) diagrams of the mean azimuthal magnetic field that reverses on a decadal timescale. The two panels below show the mean radial field at the top, which also shows the same cyclicity. Once the long cycle is filtered, the short cycle appears in the zoomed panel on the right at a particular epoch and around the equator. The lowest panels show the mean azimuthal field for model M09R1m and illustrate a dynamo with no cycles but which sustains strong stable wreaths at the base of the convective envelope. \label{fig:DynamoTlTr}}
\end{figure*}

\subsection{Properties of dynamo solutions: long cycles, short cycles, and steady dynamos}
\label{sec:DynamoSolutions}

We find three dynamo states in our sample of 15 MHD models: long (decadal) magnetic cycles, short (yearly) magnetic cycles, and stable magnetic wreaths concentrated close to the bottom of the convection zones. These three states are illustrated in Fig. \ref{fig:DynamoTlTr} with models M07R5m, M09R3m and M09R1m.

Let us first focus on the decadal cycles, as the one found for M09R3m (see middle left panels in Fig. \ref{fig:DynamoTlTr}). In this model, we find that the global magnetic field of the star reverses with a period of 10 years (see first column in Table \ref{tab:MagCyParams}). The averaged azimuthal field at the bottom of the convection zone presents a solar-like butterfly diagram, with both a polar and an equatorial branches. The magnetic field is generally consistent with dipolar symmetry, with azimuthal field of inverse polarities in each hemisphere. We also see some departures from hemispheric symmetry (for instance around t=42 years). The azimuthal field is found to be concentrated at the base of the convective envelope and in the tachocline, where the radial shear of $\Omega$ is maximized, as shown in the time-radius and meridional diagrams. \ASmod{It develops over a relatively large latitudinal extent, as shown by the active latitudinal band reported in the fourth column of Table \ref{tab:MagCyParams}. We find this band to be centered at higher latitudes the slower the model rotates for low and intermediate Rossby numbers. Conversely, this activity band moves to high latitudes for models with anti-solar differential rotation.} The averaged radial magnetic field at the surface is also found to reverse with the same timescale. At the surface, the migration branches are nevertheless not as clear as deep within the convection zone in this case. \citet{2017Sci...357..185S,2018ApJ...863...35S} also found similar deeply-seated cycles using the EULAG code, as well as \citet{2015ApJ...809..149A} using the ASH code. The cyclic behavior in their results originate from the non-linear magnetic feedback of the large-scale Lorentz force onto the differential rotation. This weakens the source of mean toroidal field that decreases and reverses, while the associated poloidal field closely follows due to the sign inversion of the electromotive force. We find the same mechanism in this new sample of simulation with the ASH code. Indeed, the DRKE cyclic variation observed in Fig. \ref{fig:KEME} compensates the magnetic energy cyclic variation, pointing toward a magnetic cycle determined by the Lorentz force feedback. This fascinating dynamo regime sustaining a long decadal magnetic cycle, because of the existence of a subtle nonlinear feedback loop between the large-scale shear and the toroidal magnetic field, is therefore confirmed by the present study using a different numerical code than \citet{2017Sci...357..185S}. We stress that its existence can be unveiled here only because we consider fully nonlinear convective dynamos, with a self-consistent differential rotation maintenance and magnetic field generation.

\begin{figure}[!h]
\centering
\includegraphics[width=0.95\linewidth]{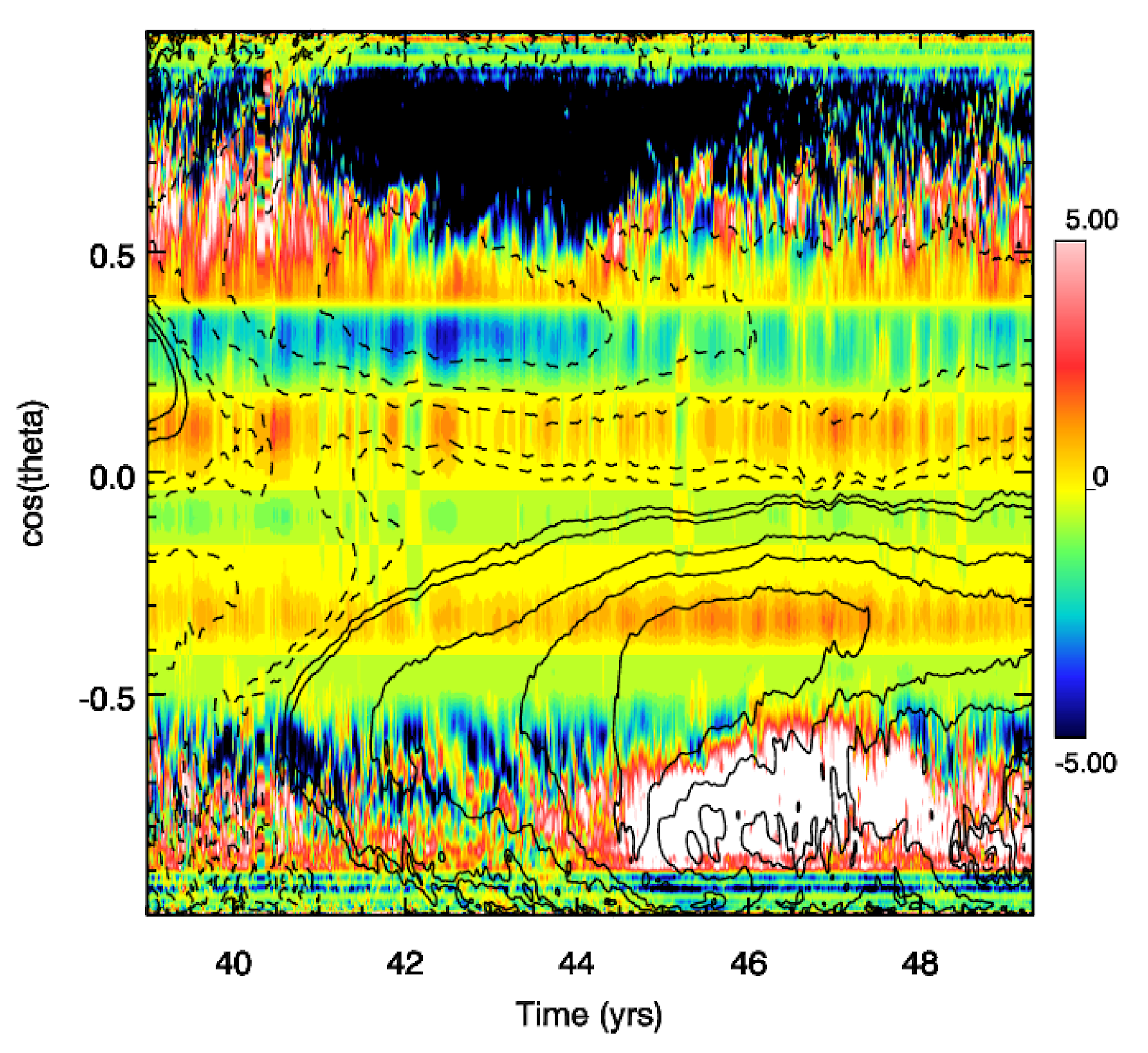}
\includegraphics[width=0.95\linewidth]{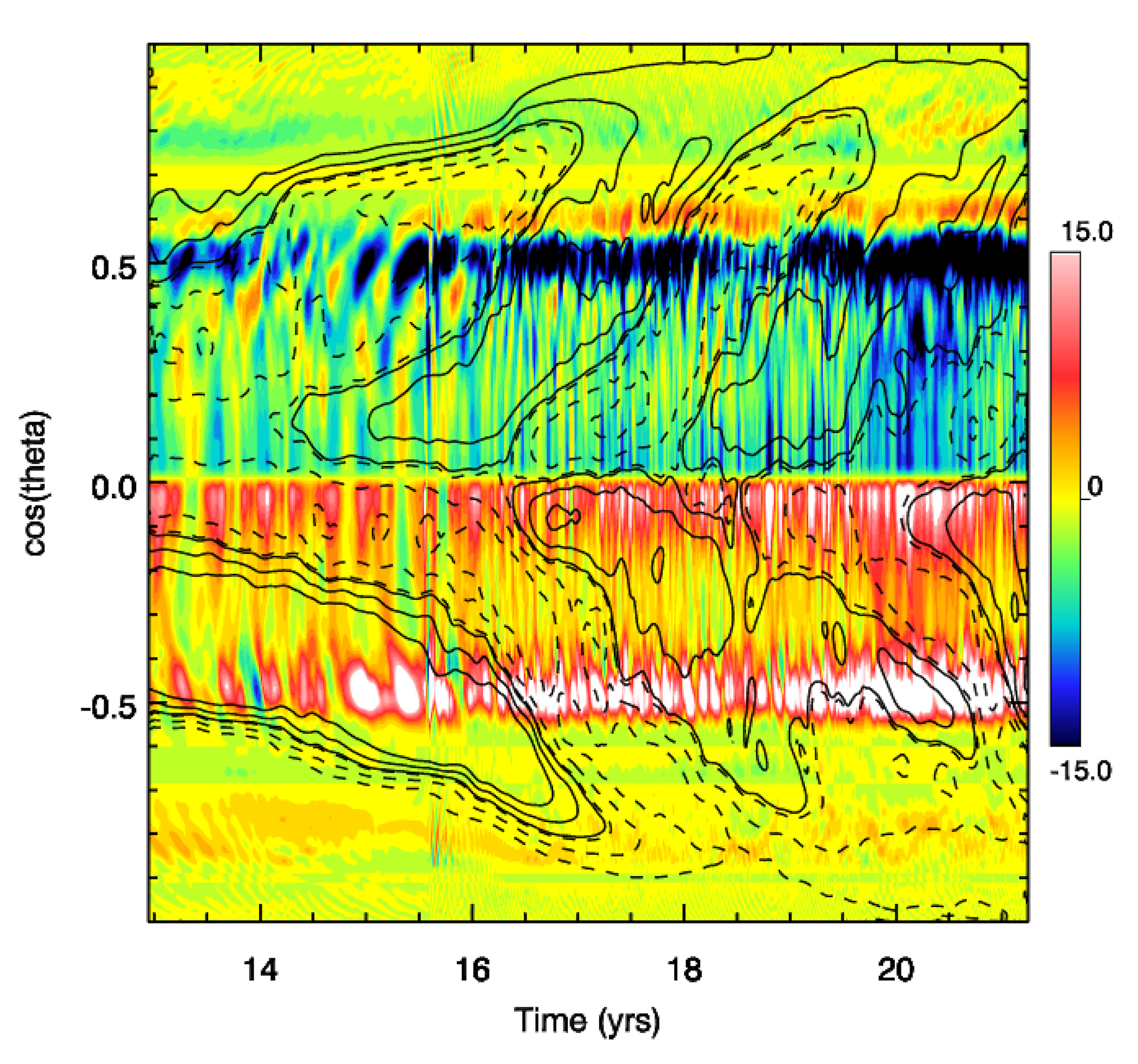}
\caption{Parker-Yoshimura rule $S_{\theta}$ (see equation \ref{eq:PYrule}) as deduced from the SVD extraction of the equivalent mean-field $\alpha$ tensor. At the top, it is shown for model M09R3m at the base of the convective envelope as a function of the cosine of the colatitude and time (color contours in {\rm m/s}). Contours of $B_\phi$ are overlaid in black contours (solid denotes positive values, dashed negative values; covering $\pm [250,10^4]$ G). At the bottom, the same is shown for model M07R5m in the mid-upper part of the convective envelope. In this case, the black lines label contours of $B_r$ (covering $\pm [75,1500]$ G). \label{fig:PYdiag}}
\end{figure}

Still, we have attempted to \ASmod{interpret our simulations through mean-field dynamo theory} by inverting the $\alpha$ tensor and its antisymmetric part $\gamma$ through the means of singular-value decomposition (SVD) technique \citep[see][]{2015ApJ...809..149A,2016AdSpR..58.1522S}. The details of this procedure are given in Appendix \ref{sec:AppendixSVD}. One can then use the derived $\bar{\alpha}$ profile to compute the Parker-Yoshimura rule \citep{1955ApJ...122..293P,1975ApJ...201..740Y} and assess the consistency of a mean-field approach with our 3D turbulent model. We therefore compute 
\begin{equation}
    S_\theta = - \lambda \alpha \partial_r \left(\Omega/\Omega_0\right)\, ,
    \label{eq:PYrule}
\end{equation}
where $\lambda = r \sin\theta$. The time-latitude variations of $S_\theta$ are shown at the base of the convection zone of M09R3m in the top panel of Fig. \ref{fig:PYdiag}, with red/white denoting a southward migration rule and blue/black a northward migration rule. We overlay contours of $B_\phi$ as black contours (plain/dashed denoting positive/negative contours) in the top panel. We see that the derived Parker-Yoshimura dynamo wave rule does not agree with the observed latitudinal propagation, which strengthen our interpretation in terms of a cycle dominated by the Lorentz-force feedback on the differential rotation itself. 

We also find another type of cyclical behavior in our sample of models: short cycles, that seem to preferentially \ASmod{be sited} close the equator and in the upper part of the convection zone. Such types of cycles have already been reported in previous publications with numerical models \citep{2016A&A...589A..56K,2016ApJ...826..138B,2018ApJ...863...35S} and could be reminiscent of the possible quasi-biennial oscillations observed in the Sun \citep{2012MNRAS.420.1405B,2013ApJ...765..100S}. They oscillate on a yearly timescale as shown in the second column of Table \ref{tab:MagCyParams}. Short cycles are interestingly found in almost all of our models except the slowly-rotating cases. Two short cycles are illustrated in Fig. \ref{fig:DynamoTlTr} for case M07R5m and M09R3m. In the former fast rotating case, no long deeply-seated cycle is observed, and the short cycle clearly appears in both the latitude-time and radius-time diagrams. In the case of M09R3m, both types of cycle are found at the same time, and the short cycle appears clearly once the signal of the long cycle is removed (see zoomed panel). The short cycles are found to always show a poleward propagation branch, and to be concentrated close to the equator. We have performed the same SVD analysis and show the Parker-Yoshimura rule $S_\theta$ (Eq. \ref{eq:PYrule}) for model M07R5m, which is shown in the bottom panel of Figure \ref{fig:PYdiag}. In this case, the analysis is carried out in the upper part of the convective envelope, and contours of $B_r$ are overlaid above the propagation rule. The Parker-Yoshimura rule is found here to be in good qualitative agreement with the poleward branch, suggesting that an $\alpha-\Omega$ or an $\alpha^2-\Omega$ dynamo could be at the source of this type of cycles. The short cycles furthermore embed much less magnetic energy than the deeply-seated ones, and we do not find any clear DRKE beating associated with them. As a result, we find that the two types of cyclical behaviors likely originate from two different dynamo processes: the deep-seated cycle from the large-scale feedback loop between the magnetic field and the differential rotation through Maxwell torques, and the short cycles from the standard $\alpha-\Omega$ or $\alpha^2-\Omega$ dynamo loop. Finally, short cycles were also reported in the study of \citet{2018ApJ...863...35S}, with the same type of localization within the convective envelope. In this former study, the short cycles were only found at small Rossby number, \textit{i.e.} for the fastly rotating cases. Here we find short magnetic cycles much more ubiquitously in our sample models, as they only disappear at large Rossby numbers. It is possible that the coarse resolution used in \citet{2018ApJ...863...35S} with the EULAG code prevented models at intermediate Rossby number to develop such magnetic cycles. Additional modelling effort pushing the turbulence level of the simulations are required to properly assess this point, which is left for future work. 

Finally, some models in our sample do not present any cyclical behavior. Instead, they sustain a steady dynamo with stable magnetic wreaths within their convective envelope and tachocline. This is the case for instance of model M09R1m shown in the lower panels of Fig. \ref{fig:DynamoTlTr}. We obtain such solutions only in the high Rossby number regime, close and above the transition toward an anti-solar differential rotation.

\begin{figure}[!th]
\centering
\includegraphics[width=0.95\linewidth]{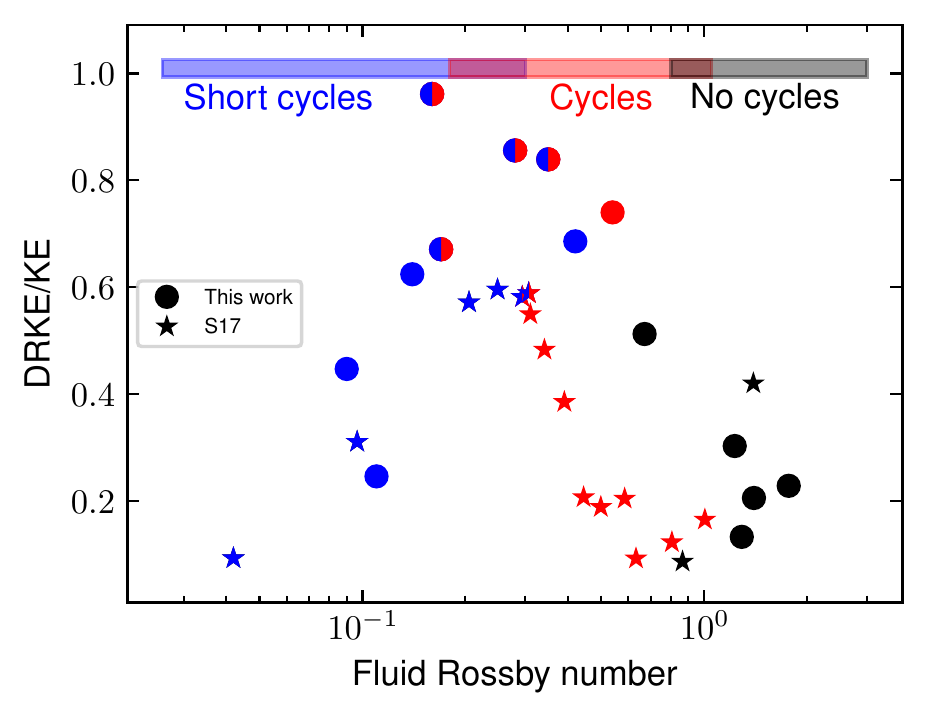}
\caption{Summary of the dynamo states found in our study (circles) and in the previous study of \citet{2017Sci...357..185S} (stars). In both studies, we find a clear trend in the type of cyclical behavior that models tend to produce as a function of the Rossby number. They are shown here by the ratio between the differential rotation and total kinetic energies. For small Rossby numbers, only short cycles are found. At intermediate Rossby numbers, decade long cycles resembling the solar cycle start to appear on a relatively narrow parameter space. At high Rossby numbers, magnetic cycles disappear and our models produce energetic stable wreaths of magnetic field in their convective envelopes. \label{fig:CycleSummary}}
\end{figure}

To summarize, we find that the different types of cyclical behaviors exist in specific Rossby number ranges in our sample. We illustrate this in Fig. \ref{fig:CycleSummary} where we follow \citet{1983ApJS...53..243G} and show DRKE/KE as a function of $Ro_{\rm f}$ in our set of models. Short cycles are found for \ASmod{$Ro_{\rm f} \lesssim 0.42$, deeply-seated solar-like cycle for $0.15 \lesssim Ro_{\rm f} \lesssim 0.65$, and steady magnetic fields for $Ro_{\rm f} \gtrsim 1.0$}. The exact boundaries between these cyclical behaviors regimes are not precisely defined and may depend on a number of factors. First, let us note that the same trend was found in \citet{2018ApJ...863...35S} with the EULAG code, as shown by the colored stars also plotted in Fig. \ref{fig:CycleSummary}. This is very important because it again demonstrates that the results discussed in this study are not code or setup dependent, but the results of genuine nonlinear convective dynamo action in a rotating spherical shell. It confirms that the Rossby number is one of the key parameters to characterize the various dynamo states found in the literature, and that cyclic convective dynamo solutions clearly exist in a parameter regime that our study helps to refine. The transitions between the different types of cycles were found at slightly different Rossby numbers, possibly due to different Reynolds, Prandtl and Rayleigh numbers regimes achieved in the two ensemble of simulations. Indeed, \citet{2013ApJ...762...73N} showed that fast-rotating models exhibiting stable wreaths of magnetism \citep{2010ApJ...711..424B} could produce reversals when the Reynolds and Rayleigh numbers are increased. Since the Rossby number of the more turbulent models nevertheless change significantly as well, it is therefore unclear whether this can be attributed to a fundamental change in the dynamo action or if it is the consequence of a change in Rossby number. Fundamental exploration aimed at predicting the Rossby number of turbulent numerical experiments such as \citet{2019ApJ...872..138A} are very promising in that respect, and need now to be extended to the full MHD regime. For the time being, we can conclude here that qualitatively the different regimes highlighted by our simulations are robust, yet simulations at much higher turbulent levels are required to assess the exact regime boundaries. Please note that case M11R5m is sometimes omitted in ensemble analysis in \S \ref{sec:MagneticProperties} and \S\ref{sec:EnergyContent} because it is not as well numerically converged as all the other cases and can sometimes be \ASmod{an outlier} in some analysis. This does not impact our conclusions in any of the results reported in the paper.

\begin{figure*}[!th]
\centering
\includegraphics[width=0.32\linewidth]{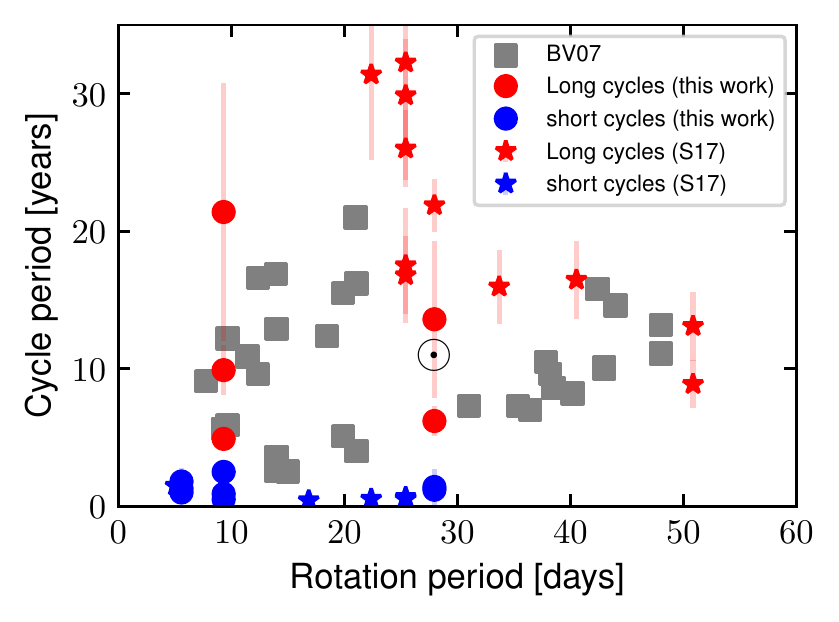}
\includegraphics[width=0.32\linewidth]{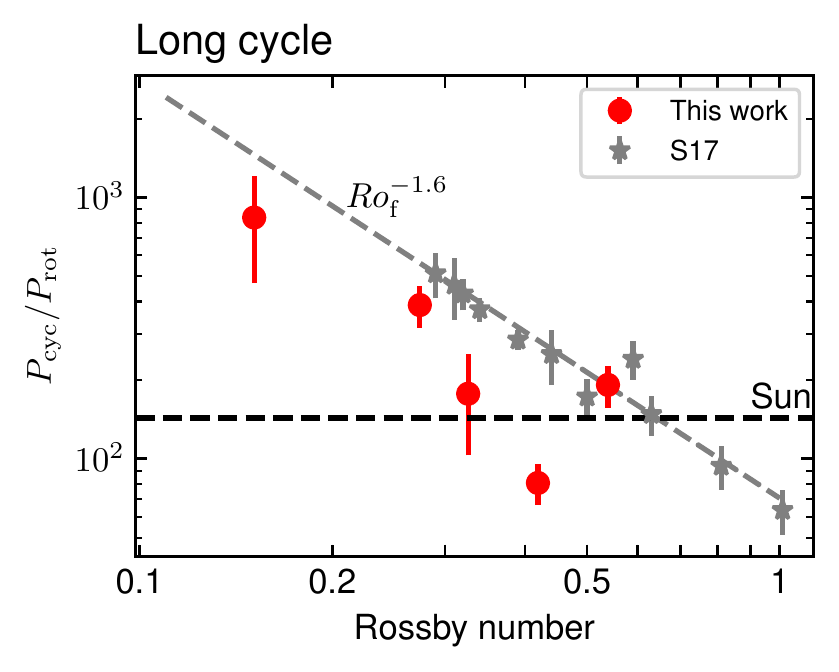}
\includegraphics[width=0.32\linewidth]{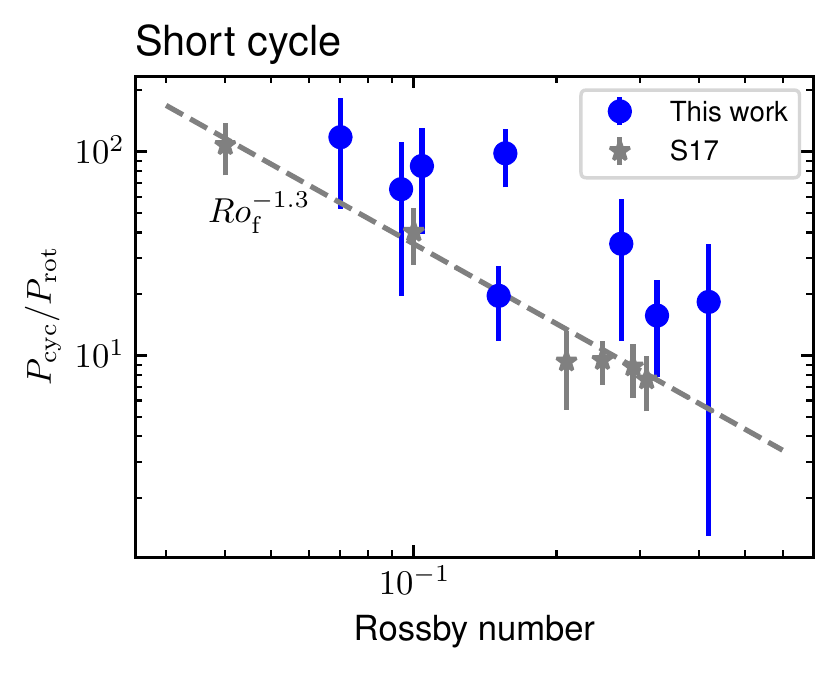}
\caption{Magnetic cycle periods trends. The left panel shows the classical cycle period-rotation period diagram with our models (circles), the models published in \citet{2017Sci...357..185S} (stars), and the stellar sample of \citet{2007ApJ...657..486B} (gray squares). The middle panel shows the trends of the long cycle period divided by the rotation period as a function of Rossby number. The right panel shows the same for the short cycles. In those two panels, the Rossby trend deduced in \citet{2018ApJ...863...35S} is indicated as gray dashed line. \label{fig:CycleTrend}}
\end{figure*}

\subsection{Dependencies of the cycle periods}
\label{sec:CycPeriodDep}

We have calculated the period of the short and long cycles and reported their values in the second and third columns of Table \ref{tab:MagCyParams}. We use the approach initially followed by \citet{2016A&A...589A..56K,2018ApJ...863...35S} and rely on an empirical-mode decomposition method \citep{Luukko2015} to identify quasi-periodic signals. Five of our models clearly exhibit a deeply-seated long cycle that can be identified by eye. We were nevertheless able to calculate accurately the associated period for four of them. The cycle period of the fourth model would require at least twice as long integration times to be identified. This would require an even more massive numerical effort and will be explored in future work.
Still we can deduce with some confidence what characterizes this long cycle nonlinear dynamo case. Conversely, the short cycles take place higher up in the convective envelope and their short periods allow us to determine the cycle periods for all the models exhibiting them. The error bars on the cycle periods are directly estimated with the empirical-mode decomposition method, as explained in \citet{2018ApJ...863...35S}.

The left panel of Fig. \ref{fig:CycleTrend} shows the cycle periods (in years) as a function of the rotation period (in days) of our models. We report both short and long cycles here, respectively in blue circles and red circles. We have also added the cycles found with the EULAG code and reported in \citet{2017Sci...357..185S,2018ApJ...863...35S} as red and blue stars. Finally, we have overlaid the detected cycles of distant stars reported by \citet{2007ApJ...657..486B} as gray squares, as well as the Sun right in the middle of the figure. Our three identified long cycles are achieved by models with different masses, which makes their direct comparison subject to caution in a $(P_{\rm cyc},P_{\rm rot})$ diagram. Overall, we do not recover the dichotomy between active and inactive branches as initially proposed by \citet{1999ApJ...524..295S} and \citet{2007ApJ...657..486B}. Rather, our sample of models combining ASH and EULAG simulation spans the whole diagram, including the hypothetical gap where the Sun stands. 

Using the EULAG sample of simulations only, we have previously shown that the cycle period is controlled by the effective Rossby number achieved by the simulated convection zone \citep{2017Sci...357..185S}. This is shown for the long and short cycles in the middle and right panels of Fig. \ref{fig:CycleTrend}. Here we find that our new ASH simulations are compatible with the trends obtains with the EULAG sample, which strengthens the similarities between the modelled dynamos in our two studies. This is moreover remarkable as the ASH simulations include a tachocline and a deeper radiative layer, whereas the EULAG sample considered only an isolated convective shell.  

The fact that the cycle period seems to decrease with the Rossby number has also been reported by other research groups using yet another code \citep[see \textit{e.g.}][]{2018A&A...616A..72W}. So far only one study relying on 3D turbulent simulations \citep{2019ApJ...880....6G} has shown some evidence for a cycle period increasing with rotation period. We believe this is due to how their differential rotation scales with rotation rate. Indeed, their simulations exhibit a differential rotation that strengthens as the rotation rate decreases (\textit{i.e.} the rotation period increases). This is at odds with all the aforementioned studies (including the present work), where we find it to increase with the rotation rate up to a point where magnetic feedback strongly back-reacts to suppress it. We suspect that the thermal treatment of the radiative-convective interface may produce this effect in the work of \citet{2019ApJ...880....6G}, albeit additional analyses are required to confirm this interpretation. 

\begin{figure*}[!th]
\centering
\includegraphics[width=\linewidth]{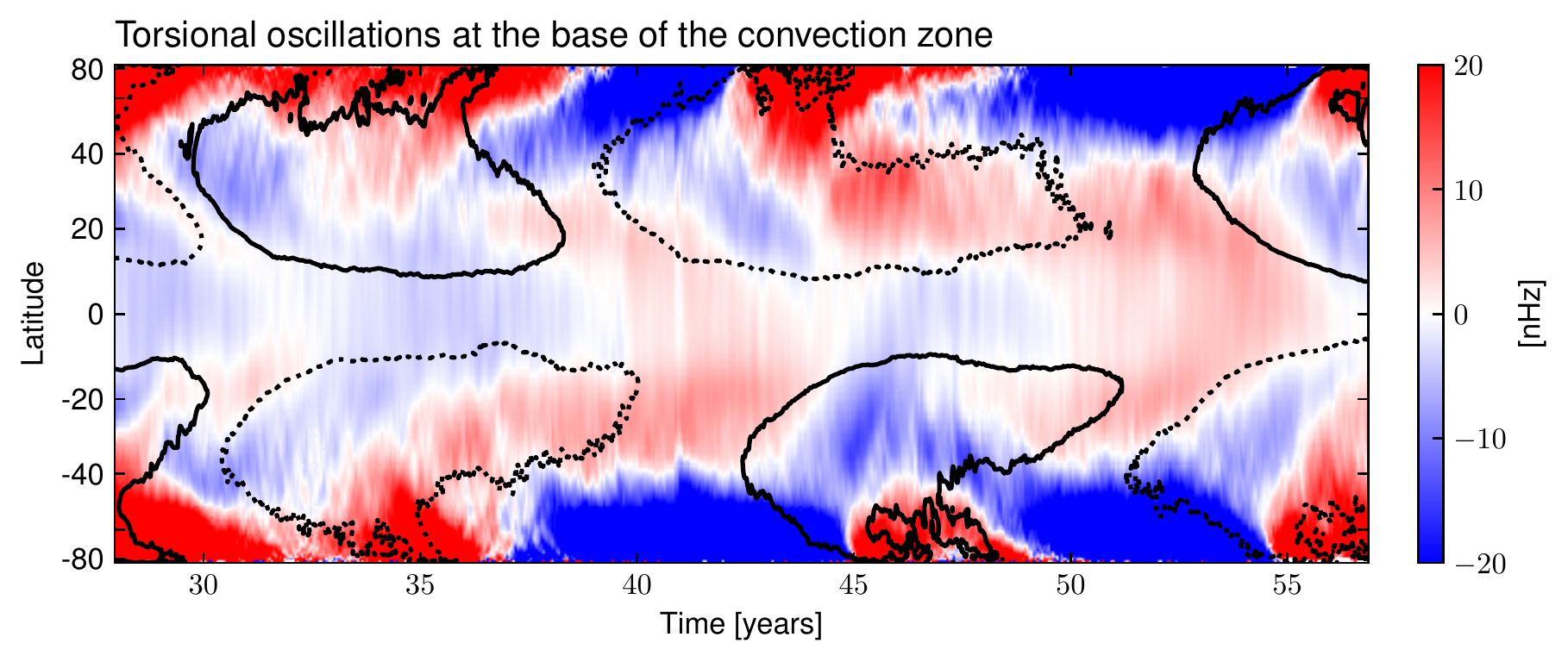}
\caption{Torsional oscillations (in nHz) in model M09R3m at the base of the convective envelope, as a function of latitude and time. The black lines label the contour of $\left\langle B_\phi\right \rangle_\phi$ at +4000 G (plain lines) and -4000 G (dashed lines). \label{fig:TOexample}}
\end{figure*}

Finally, it is worth noting that more complex dynamo states have also been reported in a similar Rossby number regime with the PENCIL code by \citet{2019ApJ...886...21V}. This warrants again caution in the interpretation of simulations results at moderate Reynolds number, and highlights the need of achieving more turbulent regimes in future work to confirm our trends.

\subsection{Torsional Oscillations in Cyclic Solutions}
\label{sec:TO}

We observe clear and strong torsional oscillations $\delta_t \Omega$ \citep{2019ApJ...883...93B} in all our models that exhibit a long, deeply-seated cycle. Torsional oscillations take the form of a modulation of the azimuthally-averaged rotation rate $\Omega(r,\theta,t)$ in both depth, latitude, and time. We illustrate the torsional oscillations at the base of the convection zone of model M09R3m in Fig. \ref{fig:TOexample}. The torsional positive/negative oscillations are shown in red/blue in nHz as a function of time and latitude. We have overlaid iso-contours of $B_\phi$ in black (plain lines correspond to 4000 G, dashed lines to -4000 G). The torsional oscillations are observed to {be in phase with long magnetic cycle}. At cycle minimum (in between black contours), the poles are rotating slower (blue) and the equator faster (red), meaning that the latitudinal differential rotation is strengthened as the magnetic field weakens and the associated magnetic torque stops inhibiting it. During cycle maximum, the opposite situations occur, and the differential rotation is found to \ASmod{decrease substantially}. We observe torsional oscillations very similar to what was found with EULAG simulations by \citet{2017Sci...357..185S,2018ApJ...863...35S} \ASmod{and previous ASH simulations by \citet{2013ApJ...762...73N,2015ApJ...809..149A}}. In all these simulations, the torsional oscillations are found to play a major role in producing the deeply-seated cycle. This is  reassuring because such nonlinear interplay between the flow and field seems independent of setup details such as BC's or numerical schemes. Moreover, torsional oscillations in our models are very energetic: they reach more the 20 nHz at the base of the convective envelope in model M09R3m, and their energy corresponds to the energy variations in the total magnetic energy (ME) seen in Fig. \ref{fig:KEME}. As a result, we find they play an active role in allowing deeply-seated cycles by reversing locally $\p \Omega/\p \theta$ and hence generating a toroidal field of opposite sign. 

We have also searched for torsional oscillations at the locations of short magnetic cycles, \textit{i.e.} at the surface and close to the equator of fast rotating models.  We find a temporal modulation of the local rotation rate at the surface in all our models. We have nevertheless not found any evidence for a correlation between these temporal variations and the short cycles themselves. This confirms that a different dynamo process sustains the short cycles, which is likely related to a more standard $\alpha-\Omega$ mechanism as we have seen in Sec. \ref{sec:DynamoSolutions}.

Finally, we have characterized the surface torsional oscillations in all our models and reported in Table \ref{tab:MagCyParams} the average values of $\delta_t \Omega$ within the activity band identified in Table \ref{tab:MagCyParams}. The surface torsional oscillations range from about 1 to 39 nHz in our sample of simulations, which corresponds to 0.4 to 6\% of the model rotation rates. Torsional oscillations associated with short cycles are found to be very weak, and the ones associated with the long cycle to be prominent deep inside the convective envelope. As a result, we do not observe any strong correlation between the amplitude of the surface torsional oscillations and the Rossby number of our models: a linear regression gives $\delta_t\Omega / \Omega_\star \propto Ro_{\rm f}^{1.1\pm 0.15} \simeq Ro_{\rm f}$.  
%\vspace{0.8cm}

 \begin{figure*}[!ht]
\centering
\includegraphics[width=0.32\linewidth]{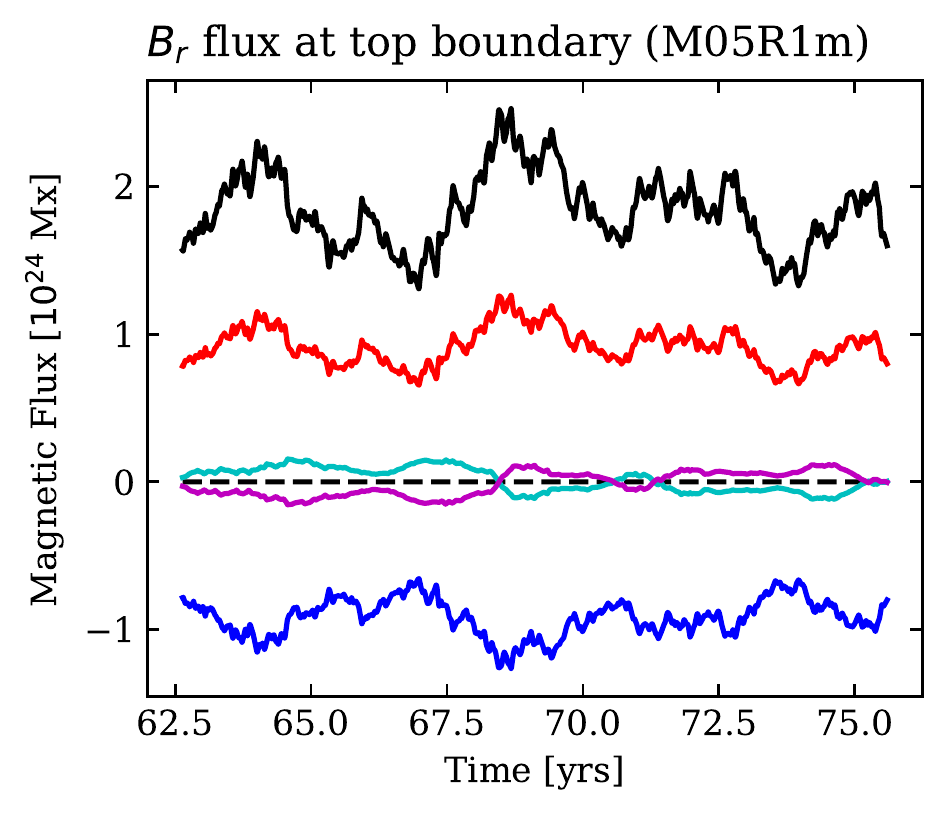}
\includegraphics[width=0.32\linewidth]{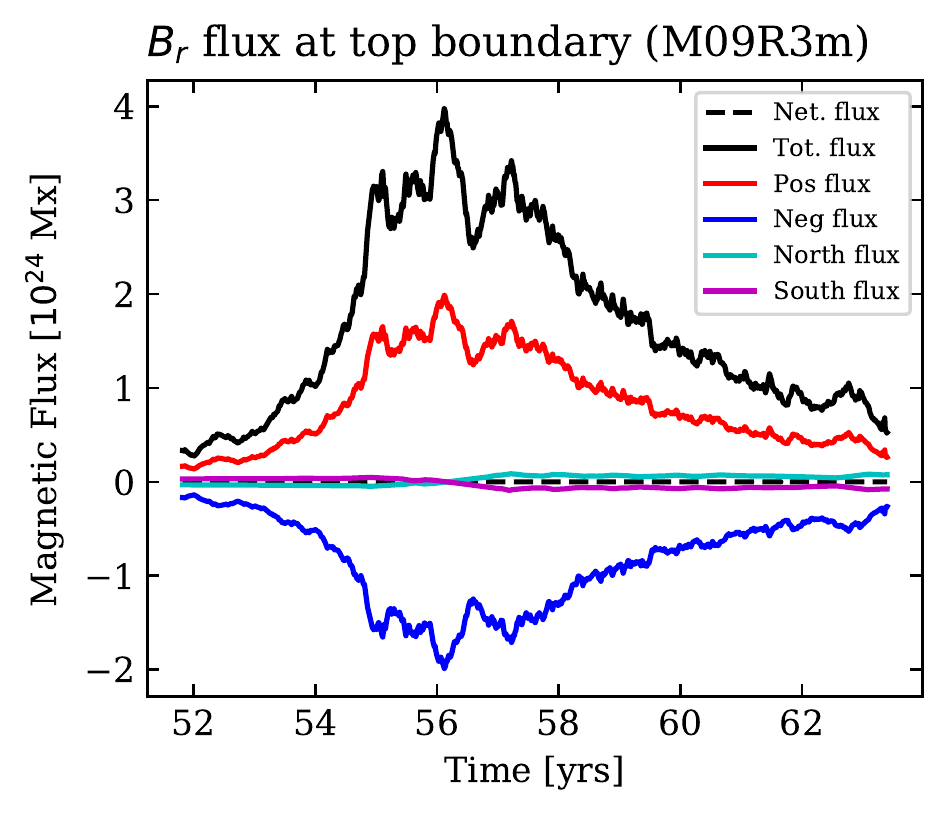}
\includegraphics[width=0.32\linewidth]{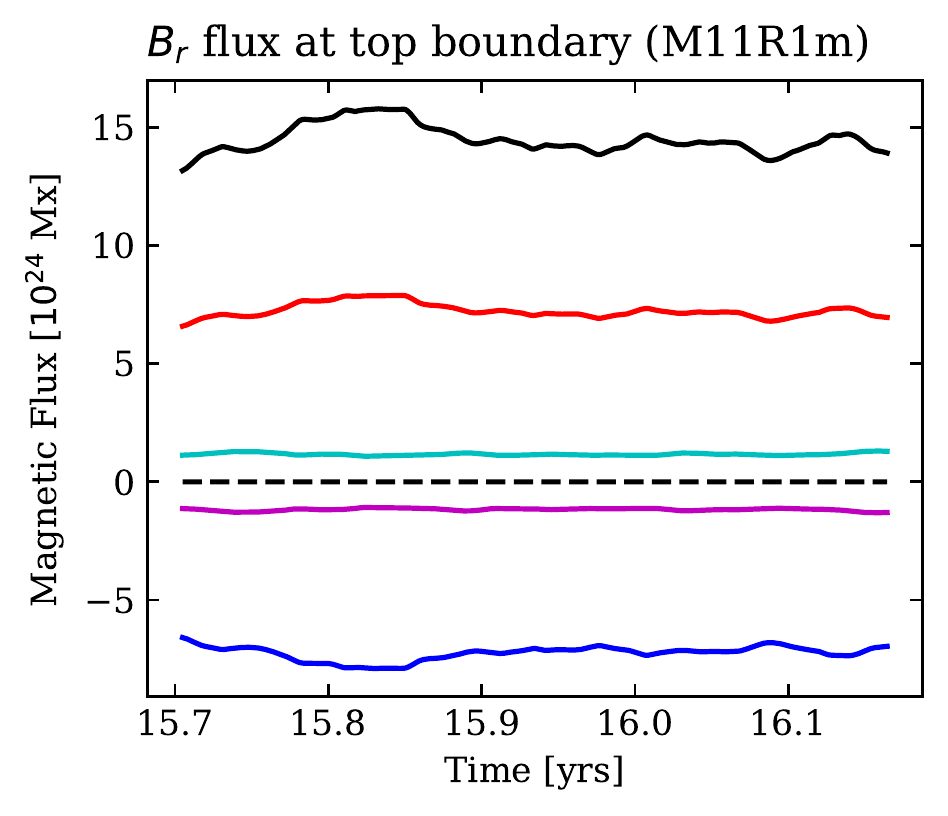}
\caption{Temporal evolution of various measures of the magnetic flux in representative dynamo cases studied. Plotted for models M05R1m, M09R3m and M11R1m are the positive (red) and negative (blue) fluxes $\Phi_+$, $\Phi_-$, total flux $\Phi_{tot}=|\Phi_+| + |\Phi_-|$ in black, the net flux $\Phi_{net}= \Phi_+ + \Phi_-$ as a dashed black line, and the southern (cyan) and northern (magenta) hemispheric fluxes. In M05R1m and M09R3m cases we clearly see the temporal modulation commensurable to their activity cycle.}
\end{figure*}\label{fig:magflux}

\subsection{Magnetic flux budget}
\label{sec:TotFlux}
% \ASB{Need to compute mag flux $\Phi_+$, $\Phi_-$ and total $\Phi_{tot}=|\Phi_+| + |\Phi_-|$}

 To further assess the magnetic properties of a dynamo solution, we display  
in Figure \ref{fig:magflux} for 3 representative cases (M05R1m, M09R3m, M11R1m), various measures of the magnetic flux available at the top boundary layer, that is $\Phi_+$, $\Phi_-$, the magnetic fluxes for $B_r|_{r=r_{top}} >0$ and $B_r|_{r=r_{top}} <0$ respectively, the total flux $\Phi_{tot}=|\Phi_+| + |\Phi_-|$, the net flux $\Phi_{net}= \Phi_+ + \Phi_-$ and the southern $\Phi_S$ and northern $\Phi_N$ hemispheric fluxes (\textit{i.e.} integrated only over the northern and southern hemisphere respectively). First, we see the very good conservation of divergenceless nature of the magnetic field, with $\Phi_{net}$ being systematically null (so implying that $\Phi_- = - \Phi_+$, as clearly evident). This is the direct consequence of solving the induction equation via a poloidal-toroidal field decomposition (see Eq. 6). Likewise, the two hemispherical measures of $\Phi$ have opposite signs, but a much smaller amplitude than $\Phi_+$, $\Phi_-$ by about a factor 10. This is likely due to a highly structured magnetic field, since for an axial dipole they are expected to be equal.
When adding up the absolute value of $\Phi_+$, $\Phi_-$, we can assess the total amount of magnetic flux generated by the dynamo. We find fluxes from $10^{24}$ to $10^{25}$ Mx, which are in good agreement with values observed in the Sun (see for instance Fig. 3 of \citealt{1994SoPh..150....1S}). We also note that in M05R1m and M09R3m cases, both of which possess a clear and long magnetic cycle, the temporal modulation of the magnetic fluxes is obvious. In M05R1m case, the modulation is about a factor of 2 from minimum to maximum of activity. In case M09R3m it reaches almost a factor of 8 (compared to 5 for the Sun). Here again the larger mass (luminosity) of M09R3m and its higher rotation rate leads to larger temporal modulation of the magnetic energy and hence the magnetic flux. Finally, for the steady dynamo case M11R1m, possessing an anti-solar differential rotation, a very small magnetic flux variability is observed. However, it is the model with the highest value of total magnetic flux, reaching about 10 times what is observed in the present Sun.

\ASmod{We furthermore see a tendency for $\Phi_{tot}$ to increase with both stellar mass and rotation rate, in good agreement with the level of magnetic energy found in the simulations. 
However, more robust tendencies appear on the rotation when one considers only model with $Ro_{\rm f}<1$. They are interestingly compatible with a simple linear dependency, with $\Phi_{tot} \simeq 2.3\, \Omega_\star^{0.84 \pm 0.42}$ for the rotation rate. When considering how the total magnetic flux scales with rotation rate $\Phi_{tot} \propto \Omega_*^n$, different values from n = 1.2 \citep{2001ASPC..223..292S} to n = 2.8 \citep{2003ApJ...590..493S} have been proposed \citep{2008JPhCS.118a2032R}. In our study we find a tentative scaling with the fluid Rossby number as 
\begin{equation}
    \label{eq:PhiVsRo}
    \ASmodmath{\Phi_{tot} \simeq 1.19\, Ro_{\rm f}^{-0.88 \pm 0.31}} 10^{24} {\rm Mx} \, ,
\end{equation}}
as shown in Fig. \ref{fig:PhiTotTrend}, where the time-averaged total flux of each model is considered (see also Table \ref{tab:tableDimParams}). Our models depart significantly from this trend when their Rossby number exceeds one, indicating a possible change for very slowly rotating stars. In this regime, our sample of models suggests that the total magnetic flux increases with Rossby number, as shown by the dash-dotted line. Additional models at large Rossby numbers are required to fully characterize this regime properly, which we leave for future work. To summarize, we find that the total magnetic flux follows a trend compatible with the one from \citet{2001ASPC..223..292S} for intermediate and small Rossby numbers, and that this trend reverses for slow rotators ($Ro_{\rm f} > 1$). \\

 \begin{figure}[!h]
\centering
\includegraphics[width=0.95\linewidth]{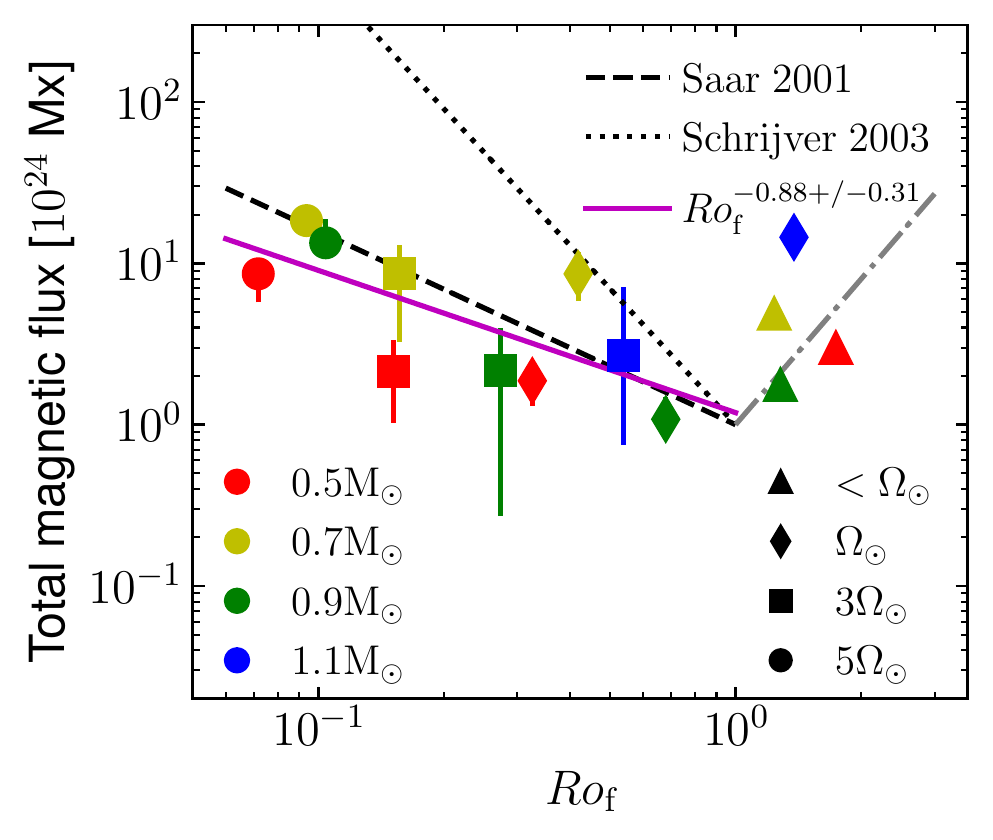}
\caption{Total magnetic flux as a function of Rossby number. The different models are labelled in the same way as in Fig. \ref{DR_fit}. The trends from \citet{2011IAUS..273...61S} and \citet{2003ApJ...590..493S} are shown respectively by the dashed and dotted black lines, assuming a constant stellar mass. An indicative trend proportional to $Ro_{\rm f}^3$ is indicated as dash-dotted line in the $Ro_{\rm f}>1$ regime.}
\end{figure}\label{fig:PhiTotTrend}

%\vspace{0.8cm}
\subsection{Dynamo families and $f_{dip}$}
\label{sec:fdipfquad}

We now turn to considering how the change of differential rotational state as a function of the Rossby number may influence the relative amplitude of the dynamo modes. We have seen in the previous sections that as we vary the Rossby number the type of dynamo solution changes, going from steady for large Rossby number to long period cyclic solutions for intermediate value of the Rossby number, to fast cyclic solutions for low Rossby numbers. 

\begin{figure}[!h]
\centering
\includegraphics[width=0.95\linewidth]{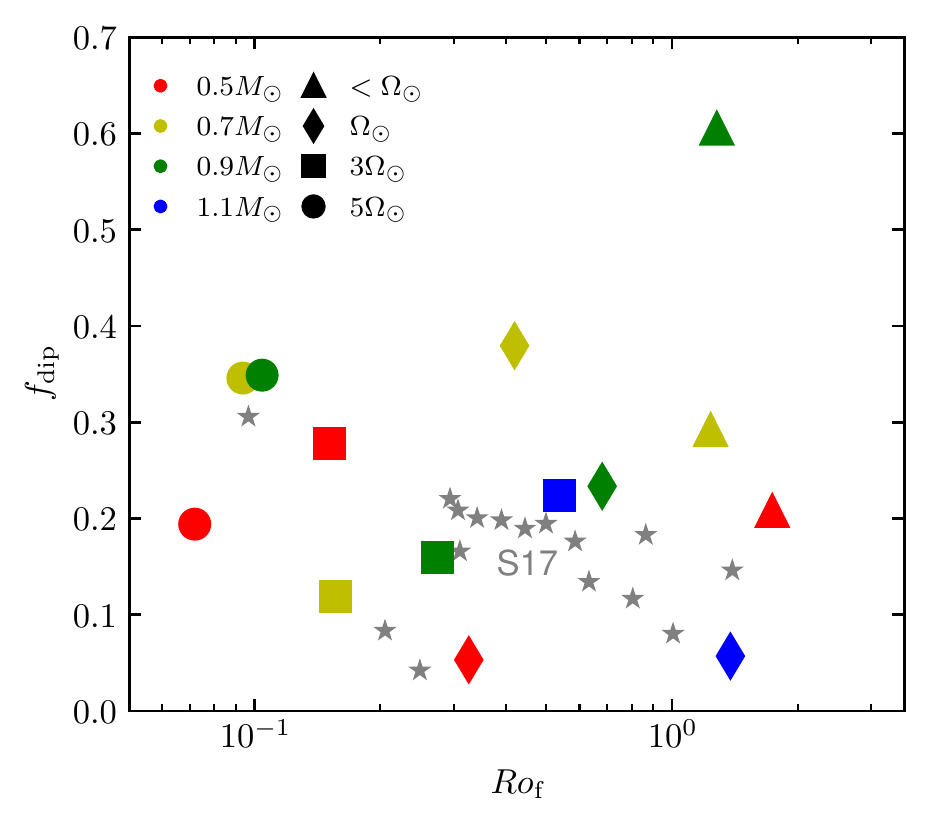}
\includegraphics[width=0.95\linewidth]{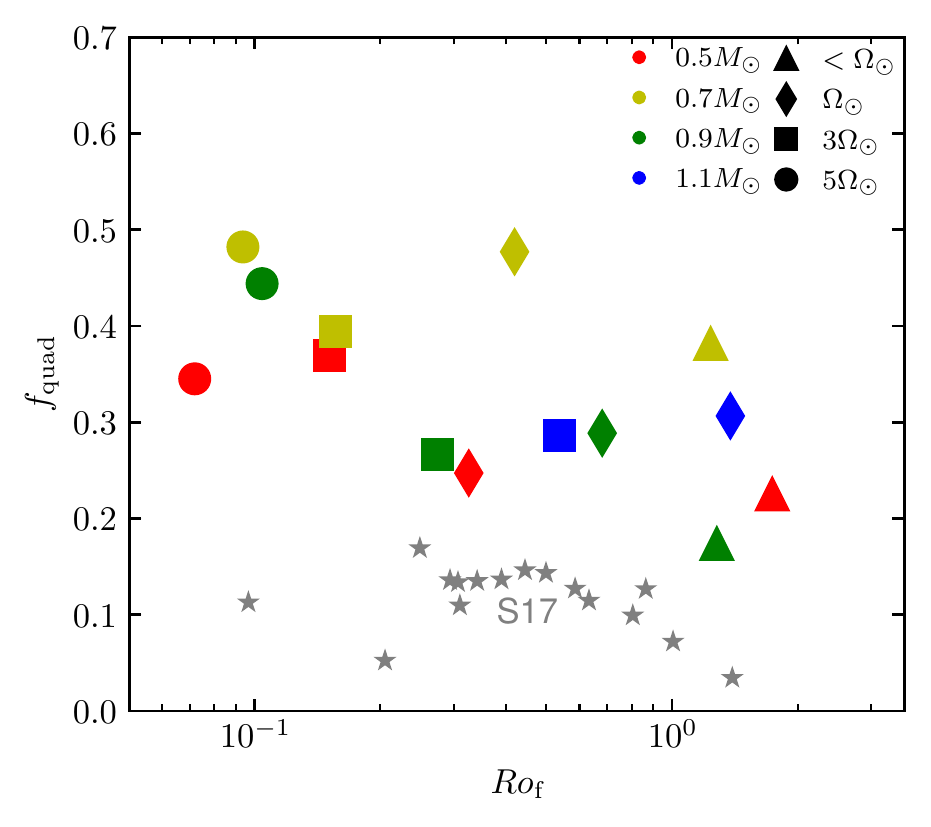}

\caption{$f_{\rm dip}$ and $f_{\rm quad}$ in all 15 models (color symbols) and those
published in 
\citet{2017Sci...357..185S,2018ApJ...863...35S} (gray stars). Models with Rossby number greater than 1 possess an anti-solar differential rotation. We see only a weak decreasing trend of $f_{\rm dip}$ and $f_{\rm quad}$ with Rossby number (for the parameter space explored). 
In addition, there does not seem to be a collapse of the large-scale magnetic field for slowly rotating stars. }
\end{figure}\label{fdip}

Such a variation of the temporal behavior of the dynamo solutions may or may not be associated to a change of dominant field geometry. This is of particular importance because it has been recently claimed by \citet{2016Natur.529..181V,2017SoPh..292..126M,2019ApJ...872..128V,2021NatAs.tmp...71H} that the Sun and solar-like stars older than the Sun may be undergoing a magnetic activity transition around a Rossby number of 1 (see \cite{2018A&A...619A..73L} for an alternative view). In particular, they argue that the wind braking efficiency may be collapsing around that rotational state transition. This would result in stars rotating more rapidly than what Skumanich law or gyrochronology would have predicted \citep{1972ApJ...171..565S,2003ApJ...586..464B}.
If for instance a collapse of the large-scale dynamo modes (mainly dipole and quadrupole) would occur after transiting to anti-solar differential rotation this would provide a very simple explanation, as it is well known that the most efficient wind braking for sun-like stars is found for the simplest magnetic field geometry \citep{1988ApJ...333..236K,2015ApJ...798..116R,2018ApJ...854...78F}.
In order to assess if such a change of magnetic geometry occurs at or near the $R_{of} \sim 1$ limit, we will use a measure called $f_{\rm dip}$, that was introduced by \cite{2006GeoJI.166...97C}, and that \ASmod{permits the assessment of} the energy content of the dipolar field with respect to the first 12 magnetic modes. We also introduce  $f_{\rm quad}$, using the same principle, as a quadrupolar field configuration is still quite efficient at spinning down a star via its associated wind braking. Both are defined as 
\begin{eqnarray}
    %\label{eq:BrYlm}{}
    %B_r(R_{\rm top},\theta,\varphi) &=& \sum_{l,m} a_{l,m} Y^m_l  \, ,\\
    \label{eq:fdip}
    f_{\rm dip} &=& \frac{\sum_m (a_{1,m})^2}{\sum_{l<12,m} (a_{l,m})^2 } \, ,  \\
    f_{\rm quad} &=& \frac{\sum_m (a_{2,m})^2}{\sum_{l<12,m} (a_{l,m})^2 } \, .
    \label{eq:fquad}
\end{eqnarray}
where $a_{l,m}$ are the spherical harmonics coefficient of the radial magnetic field \ASmod{at the upper boundary (surface) of our models}.

In Figure \ref{fdip} we show $f_{\rm dip}$ (top panel) and $f_{\rm quad}$ (bottom panel) from 32 dynamo cases: the 15 cases analyzed in details in this paper, to which we add the 17 published in \cite{2018ApJ...863...35S}, using the EULAG-MHD code \citep{2013JCoPh.236..608S}. This allows us to extend our database and to compare nonlinear dynamo solutions obtained with two different MHD codes using very different numerical techniques, hence giving us confidence that the trend found in our simulations is not due to a given code. We observe a relatively good agreement between the ASH and EULAG databases for $f_{\rm dip}$, and surprisingly find that the EULAG set of simulations produces systematically a weaker $f_{\rm quad}$ compare to the ASH database. In both series we find a weak trend for a decrease of $f_{\rm dip}$ and $f_{\rm quad}$ with Rossby number. Nevertheless, we do not find any hint of a collapse of $f_{\rm dip}$ or $f_{\rm quad}$ when the Rossby number exceeds 1 and the differential rotation realized in the simulations becomes anti-solar.
%It is clear that no clear trend vs Rossby number is found, and certainly no collapse of $f_{dip}$ amplitude is observed when crossing the $Ro=1$ limit. We may see a weak trend that low Rossby number solution have a slightly higher $f_{dip}$,
The weak decreasing trend is not significant enough to explain the stalling of stellar wind braking advocated by \cite{2016Natur.529..181V,2016ApJ...826L...2M}. Hence, it seems unlikely that field geometry is the source of the wind braking regime change for old solar-type stars. This is in agreement with the observational study of \cite{2016MNRAS.455L..52V}, whom have analyzed spectro-polarimetric inversion for a suite of sun-like stars, and they too did not find a collapse of the dipole strength as they crossed the $Ro_{\rm f}=1$ limit. So if such a stalling of stellar spin down occurs, it must come from another mechanism (see \S\ref{sec:AstrophysicalImplications}).

 In summary, we have shown in \S\ref{sec:MagneticProperties} that the dynamo solutions presented in this study possess very interesting magnetic properties that agree very well with observations and other theoretical studies. In particular, we have confirmed the key role of the Rossby number (and magnetic Reynolds number) in determining the type of dynamo realized. Now we wish to characterize better their energy content and how energies flows back and forth from kinetic to magnetic reservoirs.

%\vspace{1cm}
\section{Energy Content and Transfers in stellar convective dynamos}
\label{sec:EnergyContent}

 \ASmod{In the following section} we analyze the kinetic and magnetic energy contained in the models and how they are distributed between their various components.

\subsection{Global measure of kinetic and magnetic energies}
\label{sec:GlobEnerAnalysis}

%\ASB{Add Table with energies multiplied by volume for all models}

We now turn to discussing the global energy content in the convective envelope of the 15 dynamo cases presented in this study. In Table \ref{tab:tableEnergies} we list the kinetic (KE) and magnetic (ME) energy densities and their axisymmetric and non-axisymmetric components (see their definition \ASmod{in Appendix \ref{sec:AppendixTransfer} and}  \citealt{2004ApJ...614.1073B}). We first notice that as we increase the stellar mass, the KE is found to slightly decrease. This is due to the lower averaged mean density due to the shallower convective envelope in more massive stars. The averaged density over the simulated convective envelopes varies from 4 to 0.05 g/cm$^3$ when going from models M05m to M11m, so a drop by a factor of 80. This is in part compensated by the higher luminosity (convective velocity) of the more massive stars, leading to values of KE in the range of $10
^6$ to $10^7$ erg/cm$^3$. Note that the total kinetic energy (\textit{i.e.} the energy density multiplied by the volume) increases with stellar mass due to the much larger volume occupied by larger-mass stars. If we now decompose KE into its axisymmetric poloidal (MCKE) and toroidal (DRKE) and non-axisymmetric (CKE) components, we can further understand how the energy is being distributed in the various models.

\begin{table*}[!ht]
\begin{center}
\caption{Kinetic and magnetic energy densities. The explicit definitions of the different energy decomposition are given in Appendix \ref{sec:AppendixTransfer} and \citet{2004ApJ...614.1073B} \label{tab:tableEnergies}
}
\vspace{0.2cm}
\begin{tabular}{lrllllllll}
\toprule
{} &   KE &    DRKE (\%KE) &  MCKE (\%KE) &    CKE (\%KE) &      ME (\%KE) &    TME (\%ME) &    PME (\%ME) &    FME (\%ME) \\
{[erg/cm$^3$]} & $\times 10^{6}$ & $\times 10^{6}$  & $\times 10^{4}$  & $\times 10^{6}$ & $\times 10^{6}$  & $\times 10^{5}$  & $\times 10^{4}$ & $\times 10^{5}$  \\
\hline
M05Sm  &  9.3 &   2.1 (22.8\%) &  17.1 (1.8\%) &  7.0 (75.3\%) &   1.7 (17.7\%) &  6.7 (40.8\%) &     6.1 (3.7\%) &   9.2 (55.5\%) \\
M05R1m & 20.3 &  17.0 (83.9\%) &   2.3 (0.1\%) &  3.3 (16.0\%) &    0.8 (3.9\%) &  2.5 (31.9\%) &     0.6 (0.7\%) &   5.3 (67.4\%) \\
 M05R3m & 74.3 &  71.4 (96.1\%) &   0.9 (0.01\%) &   2.9 (3.9\%) &    1.1 (1.5\%) &  7.6 (68.2\%) &     0.1 (0.1\%) &   3.5 (31.6\%) \\
M05R5m &  2.7 &   1.2 (44.7\%) &   0.3 (0.1\%) &  1.5 (55.2\%) &   2.7 (98.2\%) &  8.8 (33.2\%) &    11.6 (4.4\%) &  16.7 (62.5\%) \\
M07Sm  &  4.0 &   1.2 (30.3\%) &   6.1 (1.5\%) &  2.7 (68.2\%) &    0.3 (7.6\%) &  2.0 (65.0\%) &     1.3 (4.3\%) &   0.9 (30.7\%) \\
M07R1m &  4.8 &   3.3 (68.5\%) &   1.0 (0.2\%) &  1.5 (31.3\%) &   0.9 (18.5\%) &  2.8 (31.5\%) &     1.4 (1.5\%) &   5.9 (67.0\%) \\
M07R3m &  3.5 &   2.3 (67.1\%) &   0.4 (0.1\%) &  1.1 (32.8\%) &   2.0 (57.8\%) &  9.4 (47.1\%) &     4.3 (2.2\%) &  10.2 (50.7\%) \\
M07R5m &  1.1 &   0.3 (24.6\%) &   0.2 (0.1\%) &  0.8 (75.3\%) &  1.3 (116.2\%) &   0.8 (6.4\%) &     5.1 (3.9\%) &  11.7 (89.8\%) \\
M09Sm  &  3.2 &   0.4 (13.3\%) &   3.0 (0.9\%) &  2.7 (85.8\%) &    0.04 (1.4\%) &  0.3 (68.2\%) &    0.7 (16.8\%) &   0.1 (15.0\%) \\
M09R1m &  4.0 &   2.1 (51.2\%) &   1.7 (0.4\%) &  1.9 (48.4\%) &    0.2 (5.7\%) &  1.9 (85.4\%) &     0.1 (0.6\%) &   0.3 (14.0\%) \\
M09R3m & 11.9 &  10.2 (85.5\%) &   0.8 (0.1\%) &  1.7 (14.4\%) &    0.5 (4.4\%) &  4.3 (83.5\%) &     0.8 (1.5\%) &   0.8 (15.0\%) \\
M09R5m &  2.6 &   1.6 (62.4\%) &   0.3 (0.1\%) &  1.0 (37.5\%) &   1.3 (49.3\%) &  4.1 (32.5\%) &     2.9 (2.2\%) &   8.3 (65.3\%) \\
M11R1m &  2.0 &   0.4 (20.5\%) &   2.0 (1.0\%) &  1.5 (78.4\%) &   0.2 (12.1\%) &  0.9 (37.7\%) &     1.9 (7.9\%) &   1.3 (54.4\%) \\
M11R3m &  3.9 &   2.9 (74.0\%) &   0.6 (0.2\%) &  1.0 (25.9\%) &    0.3 (8.8\%) &  2.7 (79.2\%) &     1.0 (2.8\%) &   0.6 (18.0\%) \\
M11R5m &  2.3 &   1.0 (43.3\%) &   8.9 (3.9\%) &  1.2 (52.7\%) &  4.4 (192.3\%) &   1.8 (4.1\%) &  384.3 (88.1\%) &    3.4 (7.8\%) \\

\hline
\end{tabular}

\end{center}
\end{table*}

First, as it is often the case, MCKE is found to play a minor role in all models independently of their mass or rotation rates. In most cases MCKE is of the order of $10^4$ erg/cm$^3$ so about 1\% or less of KE. This results in DRKE and CKE being the dominant components. Analyzing these two components, a clear trend is observed common to all masses. As the rotation rate is increased, going from Rossby number greater than 1 to value less than about 0.1, we note that DRKE first increases to constitute up to 96\% of KE. This means that most of the kinetic energy is in the differential rotation with both strong latitudinal and radial shear across the convective envelope and at its base (we refer the reader to \S\ref{sec:DR} where the angular velocity profiles of each model is discussed in details). Such a behavior is similar to what was observed in the purely hydrodynamic progenitors published in \citet{2017ApJ...836..192B}. Hence, up to a certain rotational influence, the presence of dynamo generated magnetic fields in the simulations does not modify significantly the trends observed before in the hydrodynamic cases. As a direct consequence, CKE is found to contribute less and less to the overall dynamics. CKE is found to be dominant for the slowly rotating cases, their convective motions having little azimuthal mean. As the Rossby number is decreased and the rotational influence on convective motions made stronger, we see that CKE drops to be less than a few percents of the total kinetic energy. However, this is not the case when the rotational influence increases even further. For all the fastest cases with the smallest Rossby numbers, we notice a sudden drop of DRKE both in percentage and absolute value, while CKE contributes relatively more to KE (but KE also undergoes a decrease of its amplitude). This is due to the strong feedback of the Lorentz force on the differential rotation, a phenomenon often called $\Omega$-quenching \citep{1985ApJ...291..300G,2004SoPh..220..333B,2005ApJ...629..461B,2015AandA...576A..26K} and seen only in global spherical rotating models by similitude to $\alpha$-quenching \citep{2001PhPl....8.2407B,2004PhRvL..93t5001S,2004ApJ...614.1073B} found in most local dynamo simulations (at the origin of the interface dynamo paradigm \citealt{1993ApJ...408..707P,2008MNRAS.391..467M}) and characterized in our simulations by the absolute concomitant drop of CKE.
 This significant drop of DRKE or ''$\Omega$-quenching'', accompanied by a smaller decrease of CKE or ''$\alpha$-quenching'', leads to a strong  decrease of KE. This confirms that dynamo simulations do not have \ASmod{the same rotational dependence as the purely} hydrodynamic cases. Since most solar-like stars are likely to have magnetic fields, such a finding indicates that scaling laws derived in this work will likely be more accurate when compared to observations. Since the influence of magnetic field becomes more and more dominant as we lower $Ro_{\rm f}$, it is also instructive to analyze how the magnetic energy content  evolves as well. 

\begin{figure*}[!ht]
\centering
\includegraphics[width=0.32\linewidth]{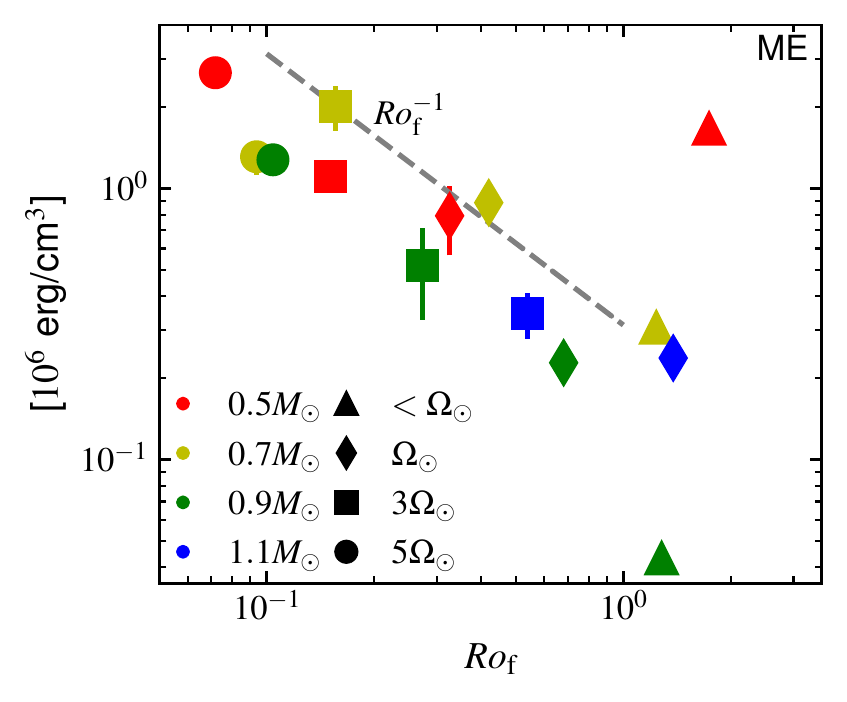}
\includegraphics[width=0.32\linewidth]{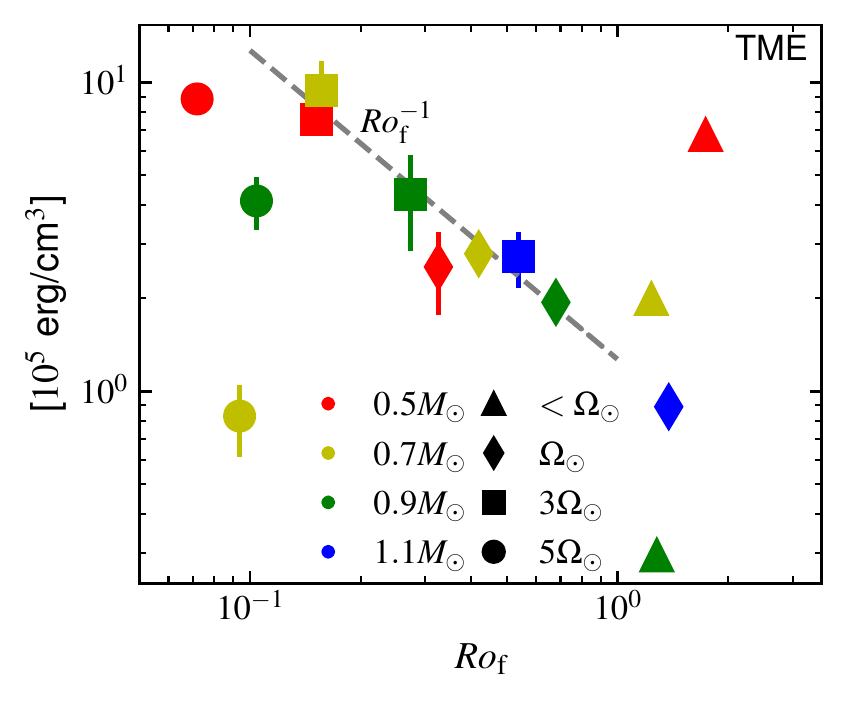}
\includegraphics[width=0.32\linewidth]{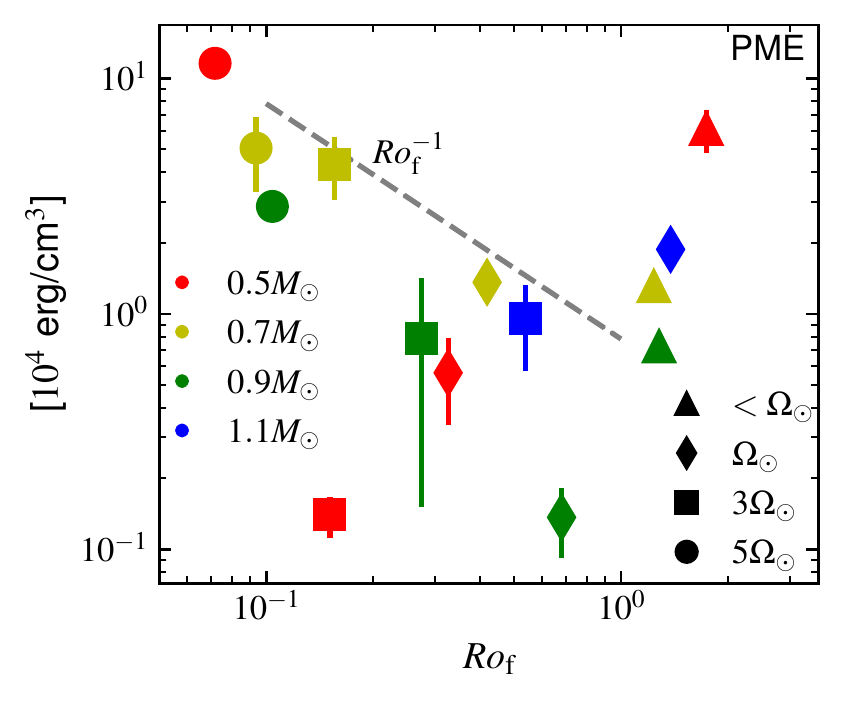}
\caption{Total energy density in the convective envelope of our models, shown in as a function of the fluid Rossby number. The magnetic energy (ME), toroidal magnetic energy (TME) and poloidal magnetic energy (PME) are displayed from left to right. The dashed gray lines indicate the $Ro_{\rm f}^{-1}$ trend.}
\end{figure*}\label{fig:MEdensity}

In Table \ref{tab:tableEnergies}, we also provide the value of the magnetic energy densities (total magnetic energy [ME], axisymmetric poloidal [PME] and toroidal [TME] components, and non axisymmetric components [FME]). Here there are some surprises given what we just discussed for their kinetic energy counterparts MCKE, DRKE and FKE. First, the axisymmetric poloidal component PME contributes more to total ME than MCKE contributes to KE. It often represents few \% of ME and in one case M11R5m it is even found to be dominant. Interestingly, PME is found to reach its lowest values for intermediate rotators close to the $Ro_{\rm f}=1$ regime. In the $Ro_{\rm f}>1$ we find that PME rises again, confirming the trend we observed on the total magnetic flux in \S\ref{sec:TotFlux}. TME somewhat follows DRKE, it first increases with rotation rate, more and more energy being pumped by the large-scale shear into toroidal magnetic energy via the dynamo $\Omega$-effect and also via complex convective motions. TME can reach values between 80 to 85\% of the total ME. \ASmod{However, the Lorentz force feedback is so strong past a certain point} that the large shear is quenched (the feedback destroying its generating source). In most of these highly rotationally constrained cases, the magnetic energy is found in the non-axisymmetric magnetic field. These trends are also illustrated in Fig. \ref{fig:MEdensity}. It is worth noting that the three magnetic energies show an overall similar trend: the total energy density ME decreases with \ASmod{an increasing Rossby number} until $Ro_{\rm f}\simeq 1$. The four models at  $Ro_{\rm f}> 1$ then exhibit a large scatter, and only PME shows an unambiguous increase with Rossby number in this regime. %This is the same trend as the one identified on the surface magnetic flux in \S\ref{sec:TotFlux}. 
We also see a hint of a saturation and possibly a slight decrease of TME at low Rossby numbers. Additional simulations at even lower Rossby numbers are required to confirm this trend, which is to be expected based on the observed saturation of magnetic activity for fast rotators (see \textit{e.g} \citealt{2011ApJ...743...48W,2014ApJ...794..144R}). In all panels, we have indicated the inverse Rossby number trend as a gray dashed line. We remark that the three magnetic energy densities are all compatible with $Ro_{\rm f}^{-1}$ trend at intermediate Rossby number, as expected from standard dynamo scaling laws in this regime (see \citealt{2019ApJ...876...83A}). This translates into a bulk magnetic field $B_{\rm bulk} \propto {\rm ME}^{1/2} \propto Ro_{\rm f}^{-0.5}$. We note that this scaling does not necessarily translate into the same scaling for the surface large-scale magnetic field, as will be made clear in \S\ref{sec:AstrophysicalImplications}. 

The relative energies (shown as percentages in Table \ref{tab:tableEnergies}) also present interesting trends. We note first that in the slowly rotating cases, ME is only a few percent of KE. As we lower $Ro_{\rm f}$, this value increases to reach equipartition by a subtle combination of both ME increasing while KE first increases and then decreases, as we have just seen. These variations inevitably lead to the fact that for the fastest rotating cases ME is even larger than KE and the simulations are in a so-called global super-equipartition state.
This is very interesting, because it means that the kinetic energy in the convective envelope is not the maximal value that the magnetic energy can reach. This is due to a change in the force balance in the Navier-Stokes equation between Lorentz, inertia, buoyancy and Coriolis forces. As the rotation rate is increased and the Coriolis force becomes stronger and stronger, the balance at first shifts from being between mostly inertia and Lorentz force to a magnetostrophic state that implies a balance between Lorentz and Coriolis forces. We refer the reader to these following studies for more detailed discussions of dynamo scaling laws \citep{2010SSRv..152..565C,2014GeoJI.198.1832D,2014GeoJI.198..828O,2015SSRv..tmp...30B,2019ApJ...876...83A}.

Overall we see that the dynamo states reached in our 15 cases do not show a strong difference as a function of mass, at least in the range studied here. However, both in terms of amplitude of the magnetic field and in the time variability of the magnetic field (cyclic, unsteady or steady solutions), we confirm that rotation plays a key role in determining the type of dynamo found in our simulations. We also note that the mean axisymmetric magnetic fields are not negligible in most of the models, often reaching values of 5 \% of the total energy content for the poloidal field and a large fraction for the toroidal magnetic field. For the latter, this has important consequences for the energy made available for the various magnetic phenomena occurring at the surface of solar-like stars (see \S 7).

Note that we did not look for hysteresis around the $Ro_{\rm f} = 0.1$ limit, by running various cases with different value of the seed magnetic field, as was done in some geophysical dynamo studies \citep{2014A&A...564A..78S}. We consider that stars acquired their magnetic field through a complex formation process, in which the seed magnetic field is likely very weak (interstellar medium magnetic field amplitude are on averaged about 10-100 microGauss) and that starting the dynamo process with a weak seed field is the most likely scenario \citep{2017ApJ...846....8E}. However, some studies have shown that weak and strong dynamo branches may exist under certain initial conditions (weak or strong seed magnetic field \citealt{2004IAUS..215..366C}) or parameters such as the magnetic Prandtl number \citep{2009EL.....8519001S,2018PEPI..277..113P}. Such weak or strong dynamo branches may explain some observed magnetic and rotational states seen in M dwarfs \citep{2011MNRAS.418L.133M}. Since this would depend on the local astrophysical context, we have decided to focus on the most common case of a weak seed magnetic field and refer the reader to these other complementary studies.

Having discussed how the kinetic and magnetic energies are distributed in our various models, we wish to go further in understanding exactly how these subtle balances come about. For this purpose we have computed the details of the energy transfers in our models, focusing on the mean axisymmetric components MCKE, DRKE, PME and TME, since large-scale fields and flows are of key astrophysical interest.

\subsection{Main transfer mechanisms between energy reservoirs}
\label{sec:Etransfers}

In this section we discuss the various energy transfers occurring in a rotating magnetized convective envelope. We refer the reader to  Appendix \ref{sec:AppendixTransfer} for the detailed derivation of the energy transfer equations, in which we have followed \citet{1966PApGe..64..145S,2006ApJ...647..662R}, generalizing their derivation to global 3D spherical geometry. We focus here on the energy budget for the mean (axisymmetric) fields in the convective envelope of our models. We decompose energies into toroidal (along the azimuth) and poloidal (in the meridional plane) components. The budgets can be summarized as
\vspace{-0.3cm}
\begin{eqnarray}
    \partial_t {\rm DRKE} &=& \underbrace{Q^{\rm DR}_{\rm RS}}_{\tiny \mbox{Reynolds stress}}  \overbrace{-Q_\Omega}^{\tiny \mbox{Omega effect}} \underbrace{- Q_C}_{\tiny\mbox{Coriolis force}} \overbrace{- Q_{\rm MS}^{\rm DR}}^{\tiny \mbox{Maxwell stress}} \nonumber \\ && \underbrace{- Q_\nu^{\rm DR}}_{\tiny \mbox{Viscosity}} \, \overbrace{+ C^{\rm DR}}^{\tiny \mbox{Curvature}} \underbrace{- S^{\rm DR}}_{\tiny \mbox{Boundaries}}\, ,
    \label{eq:DRKE_p}
\end{eqnarray}
\begin{eqnarray}
    \partial_t {\rm MCKE} &=& \underbrace{Q_{\rm RS}^{\rm MC}}_{\tiny \mbox{Reynolds stress}} \overbrace{+ Q_{\rm TM}^{\rm MC}}^{\tiny \mbox{Mixed advection}} \underbrace{+ Q_{\rm MS}^{\rm MC}}_{\tiny \mbox{Maxwell stress}} \overbrace{+ Q_{\nabla P}}^{\tiny \mbox{Pressure work}}
    \nonumber \\ && \underbrace{+ Q_C}_{\tiny \mbox{Coriolis force}} \overbrace{- Q_{\rm PM}^{\rm MC}}^{\tiny \mbox{Mixed stresses}} \underbrace{- Q_b}_{\tiny \mbox{Buoyancy}} \overbrace{- Q_\nu^{\rm MC}}^{\tiny \mbox{Viscosity}} 
    \nonumber \\ && \underbrace{+ C^{\rm MC}}_{\tiny \mbox{Curvature}} \, \overbrace{- S^{\rm MC}}^{\tiny \mbox{Boundaries}} \, , 
    \label{eq:MCKE_p}
\end{eqnarray}
\begin{eqnarray}
    \partial_t {\rm TME} &=& \underbrace{Q_\Omega}_{\tiny \mbox{Omega effect}} \overbrace{+ Q_{\rm emf}^{\rm TM}}^{\tiny \mbox{Elec. mot. force}} \underbrace{- Q_{\rm TM}^{\rm MC}}_{\tiny \mbox{Mixed advection}} \overbrace{- Q_\eta^{\rm TM}}^{\tiny \mbox{Ohmic diffusion}}
    \nonumber \\ && \underbrace{+ C^{\rm TM}}_{\tiny \mbox{curvature}} \, \, \overbrace{- S^{\rm TM}}^{\tiny \mbox{Boundaries}}\, , 
    \label{eq:TME_p}
\end{eqnarray}
\begin{eqnarray}
    \partial_t {\rm PME} &=& \underbrace{Q_{\rm PM}^{\rm MC}}_{\tiny \mbox{Mixed stresses}} \overbrace{+ Q_{\rm emf}^{\rm PM}}^{\tiny \mbox{Elec. mot. force}} \underbrace{- Q_\eta^{\rm PM}}_{\tiny \mbox{Ohmic diffusion}} \nonumber \\ && \overbrace{+ C^{\rm PM}}^{\tiny \mbox{Curvature}} \underbrace{- S^{\rm PM}}_{\tiny \mbox{Boundaries}}\, ,
    \label{eq:PME_p}     
\end{eqnarray}
where all the different terms are detailed in Appendix \ref{sec:AppendixTransfer}. We have computed individually each of the terms and show them normalized to the stellar luminosity in Fig. \ref{Ebudget}, as a function of the fluid Rossby number of the models. For each model, we have averaged the balances (\ref{eq:DRKE_p}-\ref{eq:PME_p}) over typically one hundred convective turnover time $\tau_c$ such that the sum of the terms is close to zero. Cyclic cases show large variations of the energy balance (we return to this point hereafter), in these cases we averaged on a shorter time span chosen at cycle maximum. In addition, we have tabulated the transfers for three representative cases in Table \ref{tab:tableEnergyDiff} in units of both \%$L_\star$ and \%$L_\odot$.

\begin{figure*}[!ht]
\centering
\includegraphics[width=0.48\linewidth]{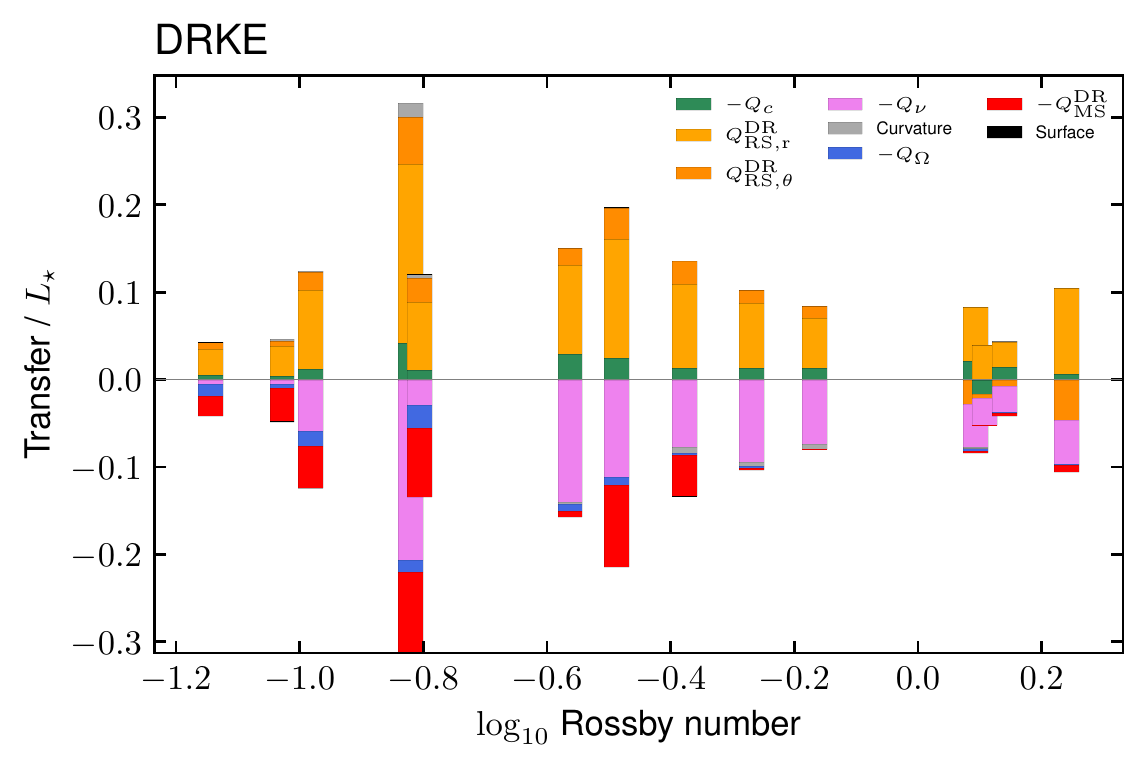}
\includegraphics[width=0.48\linewidth]{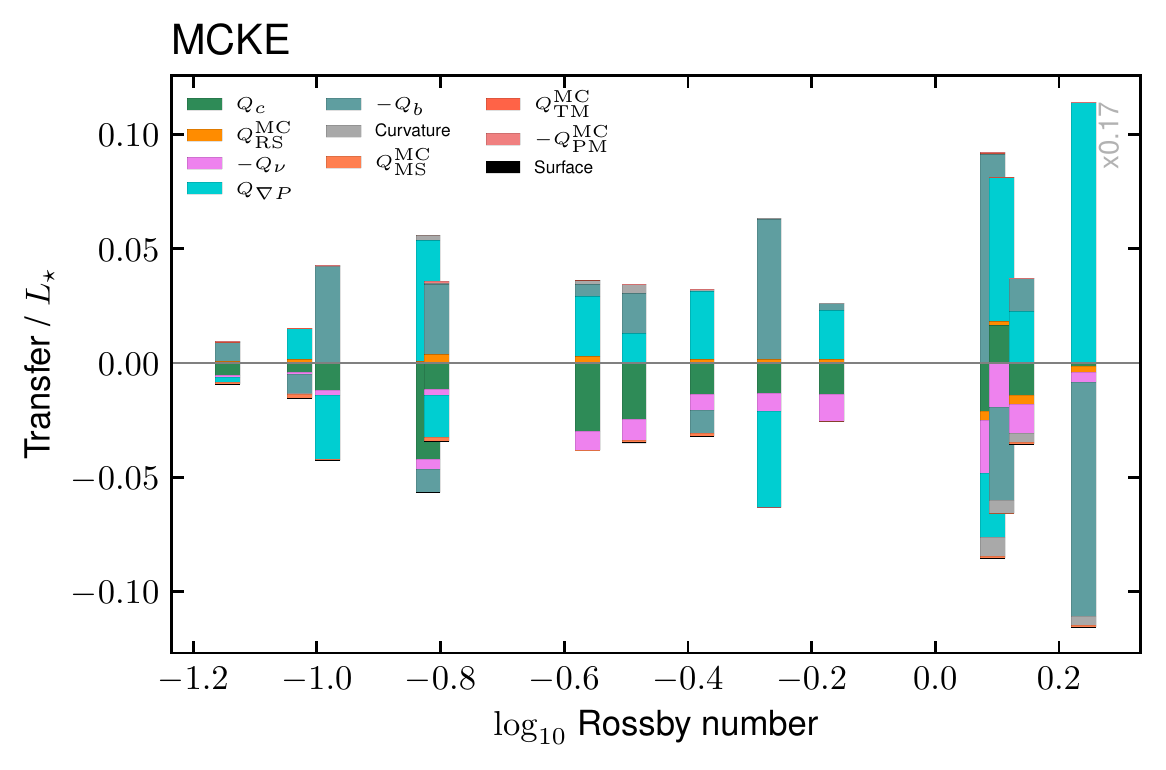}
\includegraphics[width=0.48\linewidth]{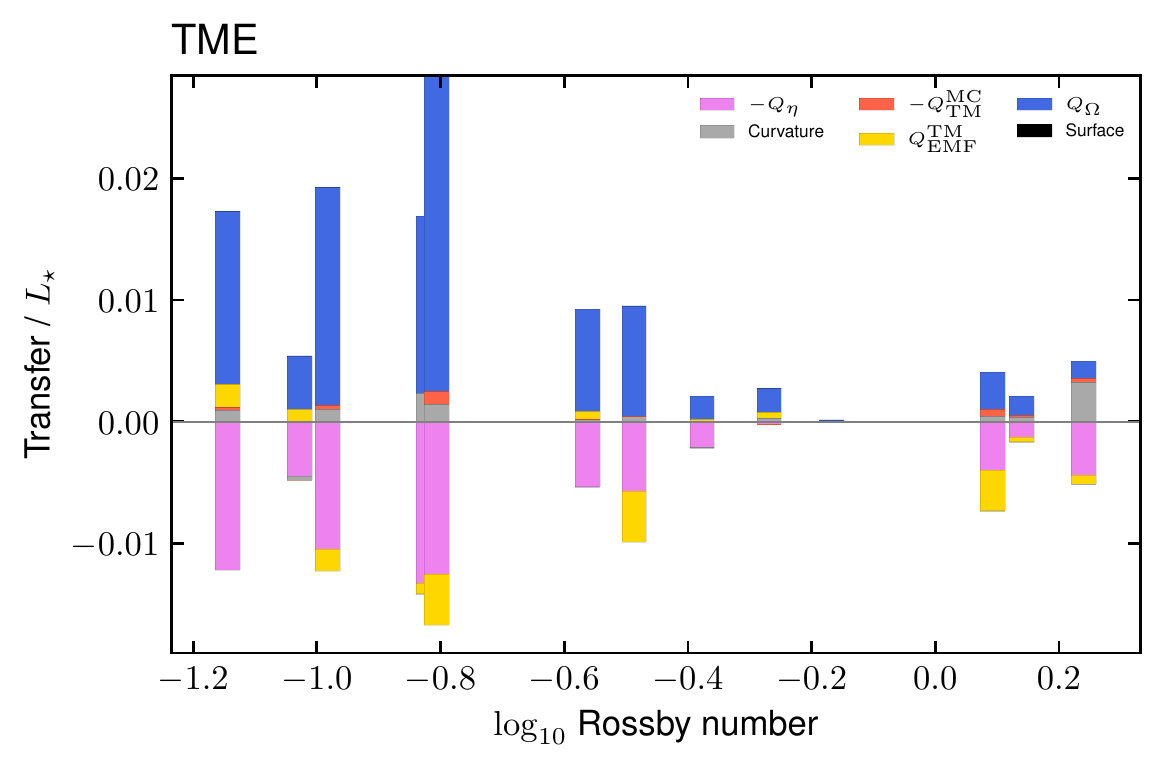}
\includegraphics[width=0.48\linewidth]{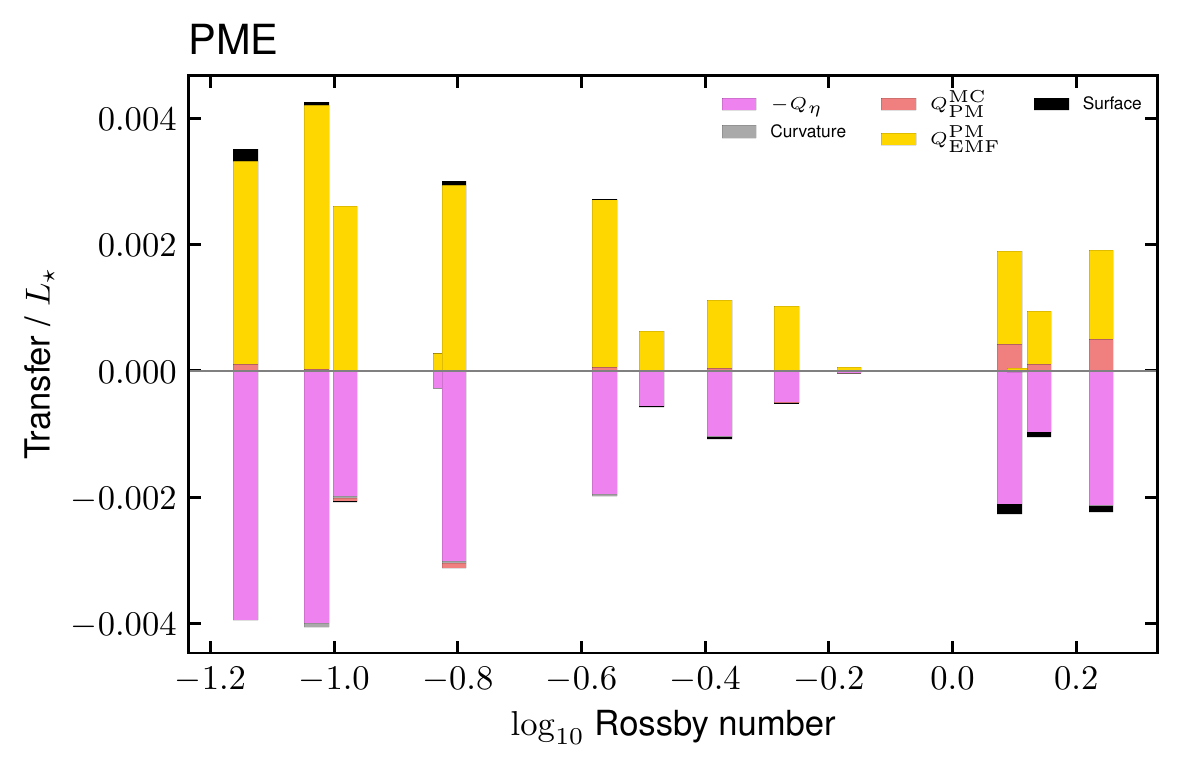}
\caption{Energy budgets as a function of fluid Rossby number. Transfers are normalized to the stellar luminosity and the logarithmic fluid Rossby number. From left to right and top to bottom, the energy budgets are shown for DRKE, MCKE, TME, and PME. The definitions of the various terms are given in Appendix \ref{sec:AppendixTransfer} \ASmod{and sketched in Eqs. \ref{eq:DRKE_p}-\ref{eq:TME_p}}.}
\end{figure*}\label{Ebudget}

\begin{table*}[!ht]
\begin{center}

\vspace{0.2cm}
\begin{tabular}{|r|rr|rr|rr|}
\hline
{} & \multicolumn{2}{c|}{M07R5m}  & \multicolumn{2}{c|}{M09R3m} & \multicolumn{2}{c|}{M09R1m}    \\
{} & $[\% L_\star]$ & $[\% L_\odot]$ & $[\% L_\star]$ & $[\% L_\odot]$ & $[\% L_\star]$ & $[\% L_\odot]$   \\
\hline

$-Q_C$                   &  0.38 &       0.06 &   {\bf 2.96} &       1.63 &  {\bf 1.35} &       0.75 \\
$Q_{\rm RS}^{\rm DR}$     &  {\bf 4.55} &       0.61 &  {\bf 12.06} &       6.63 &  {\bf 6.93} &       3.86 \\
$-Q_\nu^{\rm DR}$         & -0.53 &      -0.08 & {\bf -13.97} &     -7.69 & {\bf -7.43} &      -4.08 \\
$-Q_\Omega$  & -0.43 &      -0.06 &  -0.84 &      -0.46 & -0.004 &  -0.002 \\
$-Q^{\rm DR}_{\rm MS}$  & {\bf -3.84} &      -0.58 &  -0.70 &      -0.39 & -0.05 &      -0.03 \\
$C^{\rm DR}$     &  0.20 &       0.03 &  -0.26 &      -0.14 & -0.50 &      -0.28 \\
%DRKE\_+S\_dr     & -0.00 &      -0.00 &   0.00 &       0.00 &  0.00 &       0.00 \\
\hline
$Q_C$      & {\bf -0.38} &      -0.06 &  {\bf -2.96} &      -1.63 & {\bf -1.35} &      -0.75 \\
$Q_{\rm RS}^{\rm MC}$  &  0.15 &       0.02 &   0.29 &       0.16 &  0.15 &       0.08 \\
$-Q_\nu^{\rm MC}$     & -0.08 &      -0.01 &  {\bf -0.83} &      -0.46 & {\bf -1.17} &      -0.64 \\
$Q_{\nabla P}$     &  {\bf 1.36} &       0.20 &   {\bf 2.65} &       1.46 &  {\bf 2.17} &       1.20 \\
$-Q_b$      & {\bf -0.87} &      -0.13 &   0.50 &       0.28 &  0.29 &       0.16 \\
$Q_{\rm MS}^{\rm MC}$  & -0.17 &      -0.03 &  -0.03 &      -0.02 & -0.002 &      -0.001 \\
$C^{\rm MC}$     & -0.01 &      -0.001 &   0.19 &       0.10 &  0.01 &       0.01 \\
%MCKE\_+Q\_mc\_tm  &  0.00 &       0.00 &   0.01 &       0.00 & -0.00 &      -0.00 \\
%MCKE\_-Q\_mc\_pm  & -0.00 &      -0.00 &   0.00 &       0.00 & -0.00 &      -0.00 \\
%MCKE\_+S\_mc     & -0.00 &      -0.00 &   0.00 &       0.00 & -0.00 &      -0.00 \\
\hline
%{} & $[\% L_\star]$ & $[\% L_\odot]$ & $[\% L_\star]$ & $[\% L_\odot]$ & $[\% L_\star]$ & $[\% L_\odot]$ \hline
$-Q_\eta^{\rm TM}$     & {\bf -0.44} &      -0.07 &  {\bf -0.54} &      -0.29 &  {\bf -0.004} &   -0.002 \\
%TME\_+C\_tm      & -0.03 &      -0.01 &   0.02 &       0.01 &  0.00 &       0.00 \\
%TME\_-Q\_mc\_tm   & -0.00 &      -0.00 &   0.00 &       0.00 & -0.00 &      -0.00 \\
$Q_{\rm emf}^{\rm TM}$  &  {\bf 0.11} &       0.02 &   0.06 &       0.04 &  {\bf 0.008} &       0.005 \\
$Q_\Omega$   &  {\bf 0.43} &       0.06 &   {\bf 0.84} &       0.46 &  {\bf 0.004} &       0.002 \\
%TME\_+S\_tm      &  0.00 &       0.00 &  -0.00 &      -0.00 &  0.00 &       0.00 \\
\hline
$-Q_\eta^{\rm PM}$     & {\bf -0.40} &      -0.06 &  {\bf -0.20} &      -0.11 & {\bf -0.003} &      -0.002 \\
%PME\_+C\_pm      & -0.01 &      -0.00 &  -0.00 &      -0.00 & -0.00 &      -0.00 \\
%PME\_+Q\_mc\_pm   &  0.00 &       0.00 &   0.01 &       0.00 & -0.00 &      -0.00 \\
$Q_{\rm emf}^{\rm PM}$  &  {\bf 0.42} &       0.06 &   {\bf 0.27} &       0.15 &  {\bf 0.005} &       0.003 \\
%PME\_+S\_pm      &  0.01 &       0.00 &   0.00 &       0.00 & -0.00 &      -0.00 \\

\hline
\end{tabular}
\caption{Dominant energy transfer terms for three representative cases (M07R5m -- low Rossby number, M09R3m -- moderate Rossby number, and M09R1m -- high Rossby number). The strongest transfers for each case and each energy are identified in bold font. The four blocks of rows correspond in order to (i) the differential rotation kinetic energy balance (Eq. \ref{eq:DRKE_p}), (ii) the meridional circulation kinetic energy balance (Eq. \ref{eq:MCKE_p}), (iii) the toroidal magnetic energy balance (Eq. \ref{eq:TME_p}) and the poloidal magnetic energy balance (Eq. \ref{eq:PME_p}). Some transfer terms are tiny and have thus been omitted in the table.
 \label{tab:tableEnergyDiff}
}
\end{center}
\end{table*}

The differential rotation (upper left panel of Figure \ref{Ebudget}) is always sustained primarily by Reynolds stresses in the models (as discussed in \S4.2), with a dominant contribution of the radial component $\overline{v_r'v_\phi'}$ over the latitudinal component $\overline{v_\theta'v_\phi'}$. The cases exhibiting anti-solar differential rotation (Rossby number larger than 1) present a reversal of the latter term, showing that the latitudinal component of the Reynolds stress is detrimental to the differential rotation kinetic energy in these cases. The magnetic contributions $Q_\Omega$ (blue) and $Q_{\rm MS}^{\rm DR}$ (red) start playing a significant role for fast rotating cases (low Rossby numbers, see model M07R5m in Table \ref{tab:tableEnergyDiff}), sometimes even dominating completely viscous dissipation ($Q_\nu^{\rm DR}$, purple). In all cases the magnetic contributions tend to oppose differential rotation, as seen in \S\ref{sec:DR}. The power associated with the maintenance of differential rotation can reach about 30\% of the stellar luminosity, and drops at minimum to about 4\% in our sample of models. We remark that simulations with fluid Rossby numbers around $Ro_{\rm f} \sim 0.2$ achieve the most powerful maintenance of differential rotation that can reach values up to 17\% of the solar luminosity. At larger Rossby numbers, the star does not rotate fast enough and the differential rotation is weakly maintained. At lower Rossby numbers, the magnetic feedback from the dynamo field is so efficient that the power associated with the maintenance of differential rotation decreases significantly. 

The meridional circulation energy balance (upper right panel of Figure \ref{Ebudget}) is dominated by a balance between the work of pressure ($Q_{\nabla P}$, cyan), buoyancy ($Q_b$, blue-green) and Coriolis ($Q_c$, green) forces (see also Table \ref{tab:tableEnergyDiff} where the dominant transfer terms are highlighted in bold font). The latter almost always remains negative, indicating an energy transfer from the meridional flow to the differential rotation when models are in a steady-state. Viscous dissipation (purple) plays a much lesser role for MCKE compared to DRKE, and magnetic contributions can be considered as negligible, except maybe for small Rossby number cases possessing trans-equatorial meridional cells (see Fig. \ref{MCM07}). We find that the relative contribution of buoyancy and pressure gradients vary from model to model, and also vary in time for each model. We believe that is due to the anelastic approximation used in this study, and expect that a Lantz-Braginsky formulation \citep{2012ApJ...756..109B} would lead to more systematic relative contributions of these two important terms for MCKE. Finally, we note that the power associated with the meridional circulation maintenance increases with Rossby number, and does not go above 15\% of the stellar luminosity in our sample.      

Let us now turn to the power sustaining magnetism in our models. The toroidal (TME) and poloidal (PME) magnetic energy budgets are shown in the left and right lower panels of Figure \ref{Ebudget}. We immediately note that the power sustaining magnetism corresponds at maximum to 3\% of the stellar luminosity in our sample for TME. This corresponds to an absolute maximum of 6\% of the solar luminosity. A very large amount of power is therefore indeed channeled to sustain the large toroidal magnetic energy reservoir that the dynamo builds up in the simulations. Hence, it is expected that a significant proportion of this large magnetic energy reservoir will be accessible to trigger various surface magnetic activity events \citep{2013PASJ...65...49S}. The power associated with PME is  a bit weaker, but still reaches up to 0.4\% of the stellar luminosity. We find again that the most powerful transfers occur around $R_{of}\sim 0.2$. The power involved saturates for lower Rossby numbers, which is reminiscent of the saturation of magnetic activity observed in the X-ray luminosity of fast-rotating stars (e.g. \citealt{2011ApJ...743...48W}). It slowly drops for large Rossby numbers, but the power maintains a value of at least 0.01\% of the stellar luminosity even in our most slowly rotating models. These figures are in good qualitative agreement with the value of 0.1\% found for the Sun by \cite{2006ApJ...647..662R} using 2.5D mean field dynamo models. Let us stress again that with values ranging in our sample between 0.01\% and 3\% of the star's luminosity, this is a massive reservoir of magnetic energy extracted by dynamo action.

\begin{figure*}[!th]
\centering
\includegraphics[width=0.49\linewidth]{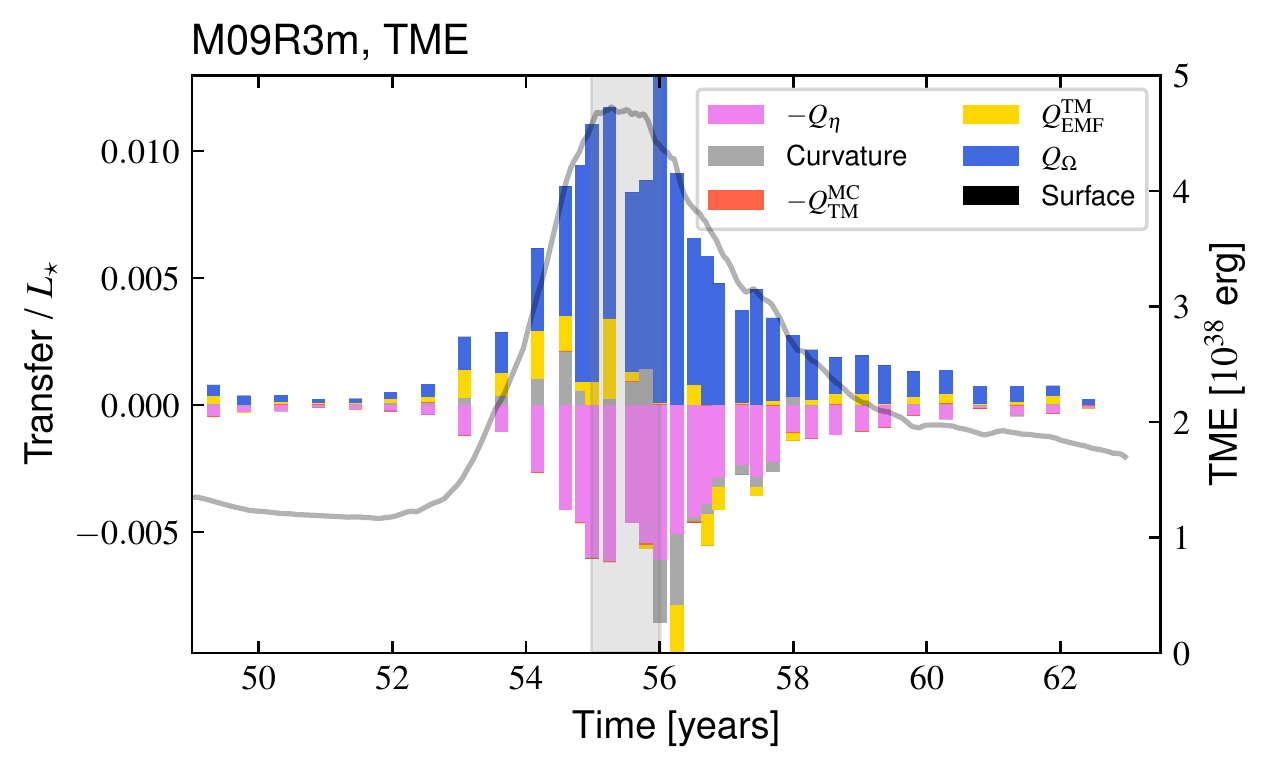}
\includegraphics[width=0.49\linewidth]{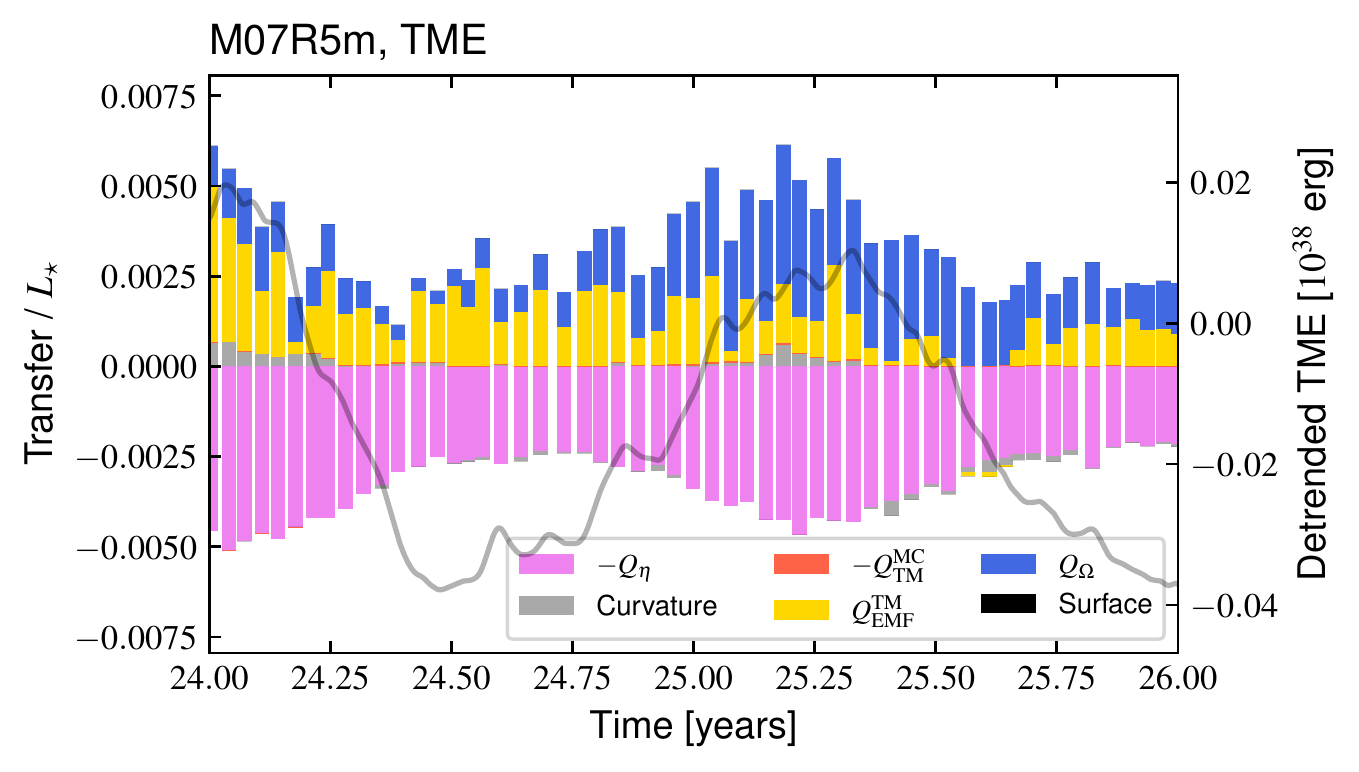}
\caption{Energy budget as a function of time for model M09R3m. The labels are the same as the lower left panel in Fig. \ref{Ebudget}. The toroidal magnetic energy (TME) is shown by the black line. The gray area in the left panel corresponds to the time-average interval used for M09R3m in Fig. \ref{Ebudget}.}
\end{figure*}\label{Ebudget_vs_time}

The poloidal magnetic energy balance is relatively straightforward: it is sustained primarily by the turbulent electromotive force originating from the convective motions ($Q_{\rm EMF}^{\rm PM}$, yellow) and opposed by Ohmic dissipation (purple). Mixed stresses involving the mean meridional flow ($Q_{\rm PM}^{\rm MC}$, salmon) are not observed to play any major role here. The toroidal magnetic energy balance is slightly more complex. In most of our models, it is primarily sustained by the Omega-effect ($Q_\Omega$, blue), and saturated by Ohmic dissipation (purple). Interestingly, we find that the role of the turbulent electromotive force can change from one model to the other (see Table \ref{tab:MagCyParams}), and it can even change sign with time in our cyclic solutions. 

This is highlighted in Fig. \ref{Ebudget_vs_time} where we observe how the various transfer terms for TME vary during one long cycle for model M09R3m in the left panel (TME is over-plotted in black), and one short cycle for model M07R5m in the right panel. First, we observe that the amplitude of the transfers vary by an order of magnitude along the long cycle (left panel), being maximum when the magnetic energy is maximum as one may expect. We also see that electromotive force (yellow) plays a dominant role when TME increases right after cycle minimum, and then switches sign and draws energy from TME when TME decreases. This striking behavior is at odds with the classical picture of constant-in-time parametrization of mean-field coefficients. It furthermore supports our interpretation that the dynamo processes behind the decadal magnetic cycles observed in some models involve a complex interplay between sources and sinks of magnetic energy that vary at different stages of the cycle. This is important because it reinforces the conclusions drawn in \S\ref{sec:MagneticProperties} about the special nature of the long cycle period dynamo simulations presented in this study. We also see that the short cycle (right panel) behaves differently than the long cycle on the left. In model M07R5m, the electromotive force sometimes equates or even dominates the $\Omega-$effect \ASmod{while still being balanced by} Ohmic dissipation. In this case, the amplitude of the transfer terms vary much less with time, and we recover a behavior expected for $\alpha^2-\Omega$ dynamos. These simulations could therefore be categorized either as $\alpha-\Omega$ or $\alpha^2-\Omega$ dynamos depending on the phases of evolution. We observe that the SVD analysis discussed in  \S\ref{sec:DynamoSolutions} and Appendix \ref{sec:AppendixSVD} shows coherent results when we take into account these temporal variations of the production terms, as shown in Fig. \ref{Ebudget_vs_time}. Given the highly time-dependent nature of these nonlinear convective dynamo simulations, the analysis presented in this section about their dynamical properties is more robust than the SVD decomposition we performed in Appendix \ref{sec:AppendixSVD} as a companion analysis, since it does not assume any scale-separation approximation.

\vspace{0.8cm}
\section{Astrophysical implications and Conclusion}
\label{sec:AstrophysicalImplications}

We have shown in the previous sections how various magnetic properties of solar-type stellar dynamo simulations change as a function of stellar mass and rotation. Often such variations can be understood using the Rossby number as a key control parameter. We here wish to reflect upon these findings and what are their astrophysical implications. 
There are several properties of solar-like stars such as their convective power and spectra, rotation profile, level of activity and presence of a magnetic cycle to cite only a few, that are of keen interest to be characterized. Our set of simulations can help us discuss some of these properties and provide clues to understand the physical mechanisms acting behind them.

Take for instance their interior rotation profile, we have seen in \S \ref{sec:LargeScaleFlows} that various states can be achieved in our set of simulations. We have further confirmed that such states depend on the Rossby numbers of the simulations. In \cite{2017ApJ...836..192B} it was advocated, based on the hydrodynamic counterpart of the dynamo cases studied here, that three states of internal rotation could be found: solar-like (fast equator-slow poles), Jupiter-like (cylindrical profile with alternations of prograde and retrograde zonal jets) and anti-solar like (slow equator, fast poles). How is the presence of a self-sustained dynamo field changing this statement? We find that two states are retained: solar-like and anti-solar, and that the third one found for small Rossby numbers has been replaced by a new state.
Indeed, we find that as the Rossby number decreases the feedback of the Lorentz force on the convective motion (via Maxwell stresses opposing Reynolds stresses in the angular momentum transport balance) yields smaller angular velocity contrast. \ASmod{This comes about because the rotation state tends} towards uniform rotation (see \S \ref{sec:LargeScaleFlows}). So for very small Rossby numbers, cases such as M11R5m or M09R5m are mostly showing a solid body rotation in their convective envelope, in sharp contrast with the banded profile of their hydrodynamics counterpart.
However, the disappearance of cylindrical banded differential rotation state may be due to the range of Reynolds and magnetic Reynolds numbers considered in our study. The strong Lorentz force feedback may be due to our moderate state of turbulent convection. It is possible that at higher Reynolds numbers a cylindrical state would be retained even for a state near super-equipartition between kinetic and magnetic energy. This is a point that needs to be investigated further with a dedicated low Rossby/high Reynolds numbers  study. Said differently: is there a level at which the magnetic energy contained in the convective envelope is so high that quasi-uniform internal rotation is inevitable? We believe this is a reasonable assumption given the tendency of magnetic field to quench differential rotation as identified by many authors \citep[][and references therein]{1982ApJ...256..316G,2004IAUS..215..366C,2005ApJ...629..461B,2015AandA...576A..26K,2020A&A...642A..66W}. So in summary, we find that the likely rotation states of solar-type stars depend on their increasing Rossby number: quasi-uniform, banded/cylindrical, solar-like and anti-solar.
Such variations of the differential rotation states translated into an overall variation of surface angular velocity contrast being less sensitive to the bulk rotation rate, with $\Delta \Omega \propto \Omega^{0.46}$, down from $\Omega^{0.66}$ as in \cite{2017ApJ...836..192B}. 
We also find another potential interesting property for the differential rotation of solar-like stars: 
a scaling law may not be the best fit to our simulations database. As in \cite{2011IAUS..273...61S}, we find that there is a clear change of trend for small Rossby number (see Fig. \ref{DR_fit2}). This is interestingly the change of rotation state from solar to almost uniform rotation.
Determining for these various rotation states the exact transition in Rossby number will require more numerical study at higher levels of turbulence and continued dedicated observations. We intend to contribute to this effort with dedicated new simulations but also in preparing the scientific exploitation of PLATO \citep{2014ExA....38..249R}.

%\noindent - Non-collapse of low order magnetic field for anti-solar cases \\
These various transitions of rotation profiles must impact the resulting dynamo and field properties. We have shown in the paper (sections \ref{sec:MagneticProperties} and \ref{sec:EnergyContent}) that this is indeed the case. Going from low to high Rossby number we find that dynamo action yields short cycle, long cycle and statistically steady (yet irregular) magnetic field evolution. This is very interesting because we can guide observations to search for these transitions in rotation state or temporal variability of the magnetic field. This will also help us discriminate between various dynamo scenarios.

%\noindent - Energy in emerging toroidal ribbons - Magnetic flux \\
Our set of dynamo solutions can help us characterize the mechanisms at work to generate and maintain magnetic fields for different sets of global stellar parameters. The rich range of magnetic phenomena occurring in stars rely on the free energy available in magnetic structures created by dynamo mechanism. In this study we have focused our analysis on a key aspect of the convective dynamo: energy transfers. We have done an extensive study on how the energy flows to and from the kinetic and magnetic energy reservoirs, separating them into their toroidal and poloidal components.  
The first key result is that a significant amount of the star's luminosity is being transferred into kinetic and magnetic energies. 
In Table \ref{tab:tableEnergyDiff} we listed as a function of the star luminosity (also with respect to the solar one) the amount of accessible energy. We demonstrated that typical numbers for the kinetic energy contained in the differential rotation are of order 10\%, for the meridional circulation 1\% of the star's luminosity. We also showed that for the toroidal magnetic energy, the energy available is also around 1\% (with a maximum of 3\%) and of the order of 0.1\% for the poloidal magnetic energy. Having access to 1\% of the star luminosity to power stellar magnetism via collective emergence of toroidal structures is significant. \ASmod{This means that there is large reservoir of magnetic energy accessible for the  manifestation of various magnetic phenomena at the star's surface}. We find for instance that our modelled stars can power dynamos such that they reach a global magnetic energy content from $10^{37}$ to $10^{39}$ erg. Part of this energy is found to be stored in the mean toroidal magnetic field (up to $6\times 10^{38}$ erg), and the mean poloidal magnetic field is generally found to be much less energetic (reaching at most $4\times 10^{37}$ erg). 
The corresponding total (unsigned) magnetic flux $\Phi_{tot}$ is found to vary between $10^{24}$ to $10^{25}$ Mx over the range of mass and rotation covered by our study, thus very similar to observations of the Sun and other solar-type stars. In dynamo cases with long cycles such as case M09R3m, $\Phi_{tot}$ is found to vary by a factor between 7 and 8 (see Figure  \ref{fig:magflux}) which is slightly more than what is found for the Sun (a factor of about 5 has been found for cycle 21 \citealt{1994SoPh..150....1S}). 

We also found that $\Phi_{tot}$ follows a scaling law with the Rossby number $\Phi_{tot} \sim R_{of}^{-0.88}$ in qualitative agreement with observations (see Figure \ref{fig:PhiTotTrend}). 

Another interesting finding of our study, which confirms results published in \cite{2015ApJ...809..149A} with the same ASH code and in \citet{2017Sci...357..185S,2018ApJ...863...35S} with the Eulag-MHD code, is the existence of a so-called nonlinear cyclic dynamo. Of course, convective dynamos are nonlinear in essence but \ASmod{what is meant here is that through the feedback of the Lorentz force on the flow,} a cyclic behavior of the dynamo arises. Standard kinematic $\alpha-\Omega$ mean field dynamos follow the Parker-Yoshimura (P-Y) rule \citep{1955ApJ...122..293P,1975ApJ...201..740Y} and do not take into account nonlinear retroaction or do so in a limited way via the so-called Malkus-Proctor approach \citep[][and references therein]{2005AandA...429..657C,2006MNRAS.371..772B,2014SSRv..186..535L}. By contrast, more and more 3D MHD convective dynamo simulations find that in a limited range of the parameter space, the P-Y rule does not apply anymore. This is the case in this study, where we find that for intermediate values of the Rossby number, typically $0.15 \lesssim Ro_{\rm f} \lesssim 0.65 $, the long cycle periods are due to a subtle interplay between the large-scale flow and the field. As the rotation rate is increased and the toroidal component of the dynamo generated magnetic field becomes more and more dominant via an efficient $\Omega$-effect acting on the large scale poloidal field, the associated Lorentz force starts to quench the differential rotation via the action of Maxwell stresses opposing Reynolds stresses. This quenching of the differential rotation in turn implies that the $\Omega$-effect is modified to the point that locally its latitudinal variation $\p\Omega/\p \theta$ reverses sign, leading to the generation of a toroidal field of opposite polarity, \ASmod{and through the action of turbulent convection}, a reversed poloidal field.
This nonlinear cyclic dynamo behavior is in sharp contrast with P-Y mechanism. Note that this is a delicate dynamo state to achieve, as the magnetic energy needs to be neither too weak nor too strong as discussed in \cite{1983ApJS...53..243G} (see for instance their Figure 31 or in  \cite{2005ApJ...629..461B} where such a modulated dynamo state was also found in stellar core dynamos). To demonstrate that further, we have computed in Figure \ref{fig:PYdiag} the P-Y rule for one typical long cycle period dynamo case of our study and confirm that it is unable to explain the dynamo wave and cyclic behavior of this subset of dynamo cases (M09R3m, M11R3m for instance). However, we do find that for low Rossby number ($Ro_{\rm f} < 0.42$), the P-Y rule still works, and for instance in a case such as M07d5m also shown in Figure \ref{fig:PYdiag}, we clearly have poleward dynamo waves compatible with the radial shear and the $\alpha$-effect. 
Hence, we may have been able in this study to identify when P-Y vs nonlinear cyclic dynamos (in the sense defined in this study, e.g. feedback of the magnetic field on the local shear) take place. This is very important as it tells us how to reconcile various recent publications in the community that sometimes were finding that global convective dynamo could be interpreted as classical $\alpha$-$\Omega$ dynamos \citep[][and references therein]{2018A&A...616A..72W,2018A&A...616A.160V,2019ApJ...886...21V}, whereas others did not \citep{2015ApJ...809..149A,2017Sci...357..185S,2018ApJ...863...35S}.
We propose that it is linked to different effective values of the Rossby number used in these various dynamo simulations.

As we have seen above, it is instructive to make the link between full 3D MHD convective dynamo simulation and mean field dynamo concepts. Mean field dynamo theory usually uses the $\alpha$-effect to parameterize turbulent magnetic field generation. In this study, we have estimated it through both the kinetic helicity (see \S\ref{sec:KH} and \citealt{1976JFM....77..321P}) and an SVD decomposition (see \S\ref{sec:DynamoSolutions} and \citealt{0004-637X-735-1-46,2013ApJ...775...69D,2015ApJ...809..149A,2017ApJ...846....8E}). In the former case, we do not find a significant change of sign nor amplitude in the kinetic helicity of models possessing an anti-solar differential rotation. In the range of parameters considered in this study, this means that anti-solar-like stars need to be modeled with an $\alpha$-effect similar to solar-like stars at least in their radial dependency, if not in amplitude. In the mean field $\alpha-\Omega$ dynamo paradigm this implies that anti-solar-like stars will have a dynamo wave with a propagation reversed to that of the Sun, e.g. poleward from the equator to mid-latitudes as imposed by the P-Y rule. In our 3D simulations, we do not find such cyclic poleward dynamos for slowly rotating simulations, instead we find that they are statistically steady (but highly time dependent on short time scales). This is likely due to a less favorable phasing between poloidal and toroidal magnetic field generation in the convective envelope of these slowly rotating case that develops via complex nonlinear interactions between the fields and flows, which are not fine-tuned but instead evolves depending on the global parameters considered. 

\begin{figure*}[!ht]
\centering
\includegraphics[width=0.32\linewidth]{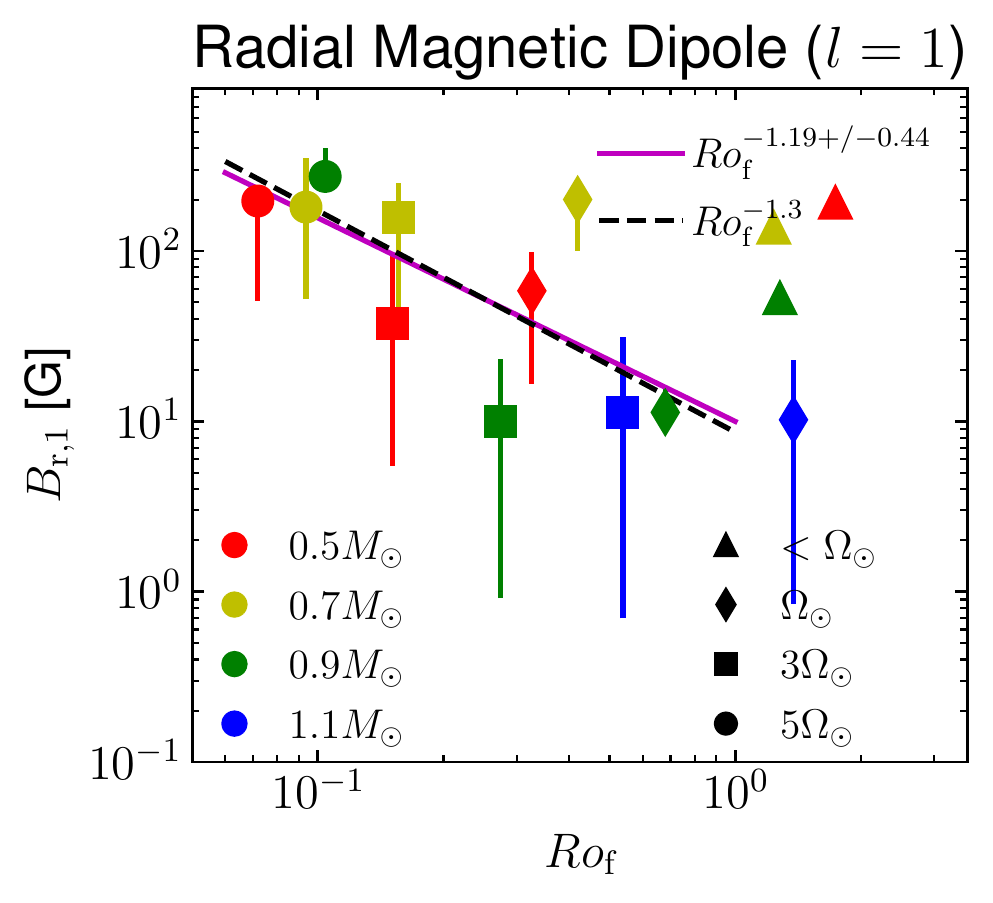}
\includegraphics[width=0.32\linewidth]{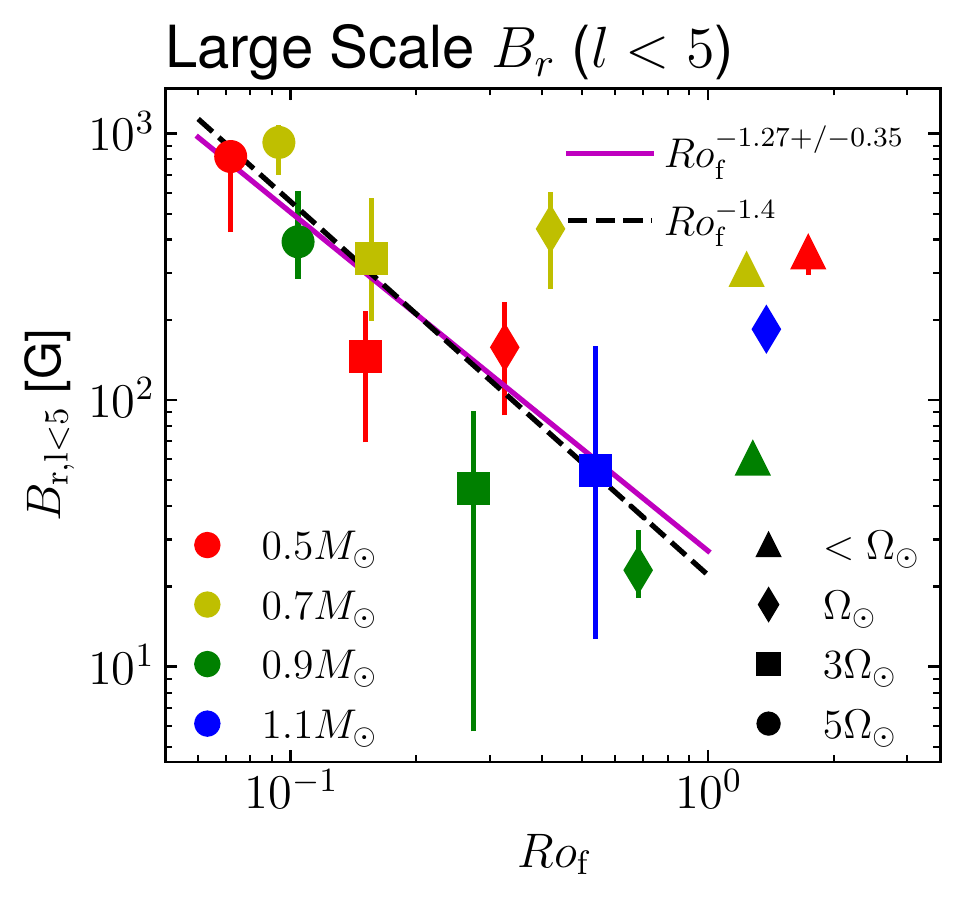}
\includegraphics[width=0.32\linewidth]{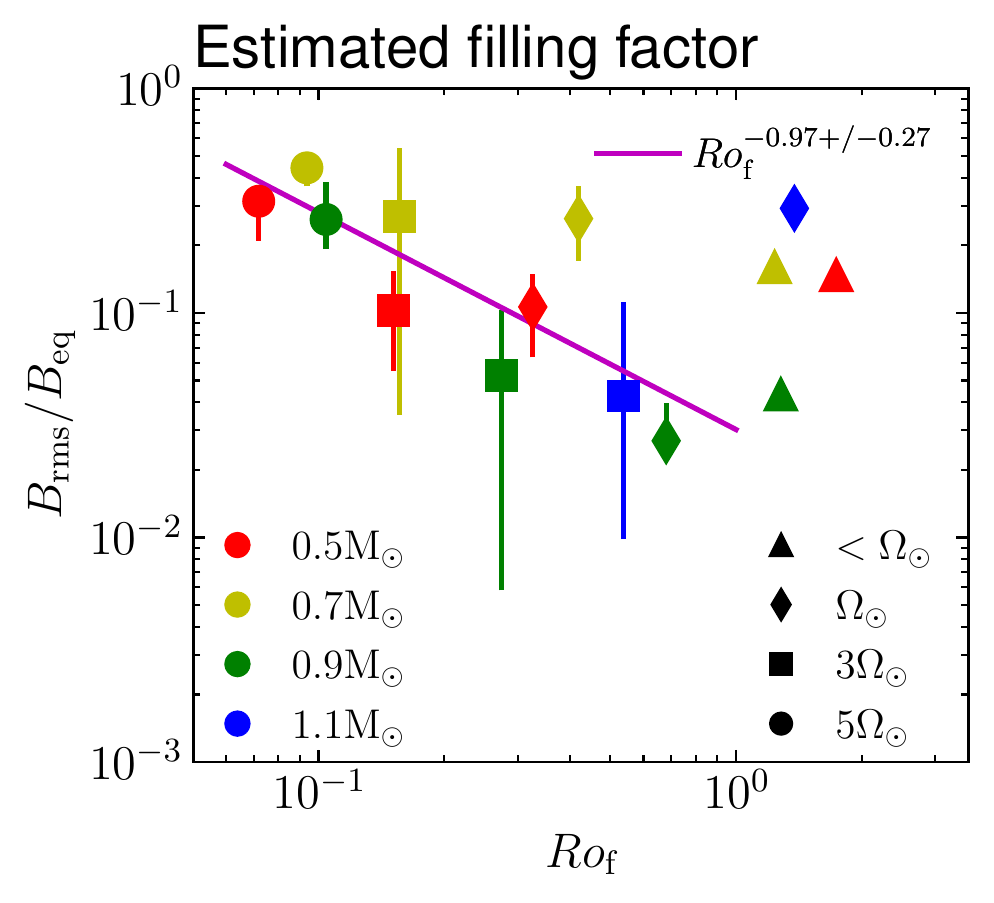}
\caption{Large-scale field at the surface of our modelled convective envelopes as a function of the Rossby number. The first panel shows the total dipole field, the second panel the large-scale fields (spherical harmonics $\ell<5$, including the non-axisymmetric components ($m \neq 0$)). The third panel shows the ratio between the total root-mean-square (rms) field at the surface, and the equilibrium field based on the gas pressure at the photosphere. It can be considered as a measure of the filling factor $f$ (see \citealt{2011ApJ...741...54C,2019ApJ...876..118S}). The symbols used in the panels are the same as in Fig. \ref{DR_fit}}
\end{figure*}\label{fig:BdipBall}

Another interesting aspect is to assess how the dynamo-generated magnetic field is distributed over spatial scales. It is well known that there is a nonlinear feedback loop between rotation, dynamo, stellar wind and magnetic braking over secular time scales  \citep{1972ApJ...171..565S,2014ApJ...789..101B,2015ApJ...799L..23M,2017LRSP...14....4B,2017SoPh..292..126M,2020ASSP...57...75B,2021arXiv210315748V}. It has been demonstrated that the magnetic torque provided by stellar winds are mostly controlled by the dipolar and quadrupolar modes \citep{2015ApJ...798..116R,2015ApJ...807L...6G,2018ApJ...854...78F}. Hence, one key question is to assess what happens with dipolar and quadrupolar modes when the dynamo changes its properties. To this end we showed in Figure \ref{fdip} how magnetic geometry changes by computing quantities known as $f_{dip}$ and $f_{quad}$. This allows us to assess the overall contribution of these two  dynamo modes to the overall magnetic energy spectra. We found that they are key contributors to the overall magnetic energy with values ranging from 0.05 to 0.6, with most of the cases studied possessing $f_{dip}$ and $f_{quad}$ around 0.2 - 0.3. We do not see any clear trend with Rossby number. Fast rotators and slows rotators both possess large dipolar and quadrupolar components. So from a stellar dynamo point of view it is difficult to invoke a drop in the large-scale magnetic field to explain a possible break of stellar spin down for slow rotators as proposed by \cite{2017SoPh..292..126M}. Similar findings are obtained from observations of magnetic fields in cool stars as shown in \cite{2016MNRAS.455L..52V}.
The advocated Rossby number transition in magnetic field geometry to explain a collapse of magnetic breaking is thus unlikely. This study suggests that we must find a different explanation, maybe a less efficient heating mechanism inducing a sudden drop of coronal temperature and wind mass loss \citep{2018MNRAS.476.2465O} which directly impacts angular momentum loss. Self-consistent rotating wind models with detailed treatment of the coronal heating mechanism are needed \citep[see for instance][]{2020ApJ...896..123S,2021ApJ...910...90H} in order to confirm the existence or not of such a transition in mass loss at slow rotation rates.

We have focused our analysis on the global energetics of the dynamo, and showed that the global dynamo field followed roughly a $B_{\rm bulk} \simeq Ro_{\rm f}^{-0.5}$ trend (see \S\ref{sec:GlobEnerAnalysis}) in agreement with previously published dynamo scaling laws \citep{2019ApJ...876...83A}. It is also useful to interpret our simulations only considering the top of the dynamo domain, making a more direct link with stellar observations of surface magnetism. In this context, we show in Fig. \ref{fig:BdipBall} the trend in Rossby number for the surface dipole field (first panel), surface large-scale field (second panel, see Table 5), and the ratio of the root-mean-square (rms) surface field to the equipartition field (third panel). The error-bars were deduced from the temporal variability of the fields, and the values are reported in Table \ref{tab:MagCyParams}. The first striking observation is that the scaling law of the surface large-scale field differs from the global volume-averaged dynamo field e.g. including all scales with a steeper slope. Indeed, we find for low and intermediate Rossby numbers that
\begin{eqnarray}
    \ASmodmath{B_{r, \rm dip} \simeq 10 \, Ro_{\rm f}^{-1.16\pm 0.47}} G\, ,\\
    \ASmodmath{B_{L, {\rm surf}} \simeq 28\, Ro_{\rm f}^{-1.27\pm 0.35}} G\, .
\end{eqnarray}
Both trends are compatible with the trends deduced from Zeeman-Doppler Imaging surveys, that generally find
the large-scale surface magnetic field to follow a $Ro_{\rm f}^{-1.3}$ trend at intermediate Rossby numbers \ASmod{\citep[][]{2019ApJ...886..120S}}. Finally, it is also instructive to assess the level of equipartition at the surface through the ratio between the surface rms field $B_{\rm rms}$ and the equipartition field $B_{\rm eq}$ \citep[as defined in][]{2000ASPC..198..371J} deduced here from the gas pressure at the surface of the stellar models we considered. Indeed, \citet{2011ApJ...741...54C} have proposed that this ratio measures the filling-factor $f$ of the large-scale field that shapes the lower stellar corona and ultimately determines the angular momentum loss rate of stars. \citet{2019ApJ...876..118S} have found observationally that this ratio decreases with Rossby number. We find a similar trend here as seen on the third panel of Fig. \ref{fig:BdipBall}, with $\ASmodmath{f \simeq B_{\rm rms}/B_{\rm eq} \simeq 0.03\, Ro_{\rm f}^{-0.97\pm 0.27}}$. Finally, we note that the three magnetic field measures shown in Fig. \ref{fig:BdipBall} all exhibit an increase in amplitude at high Rossby number. This again strengthens the case that dynamo action within cool-stars does not exhibit any significant decrease of the large-scale magnetic field for slow rotators. 

How are these results informing us about our star, the Sun? First, we note that the study of \citet{2017Sci...357..185S,2018ApJ...863...35S} is about 1 solar mass stars and is taken into account in the analysis presented in this study. Given the good agreement seen in many of the plots discussed in \S \ref{sec:MagneticProperties} between the study done with the Eulag-MHD code and the one presented here with the ASH code (independently of models details), we are confident that the dynamo solutions discussed in this study are useful to understand the physical nature of the cyclic activity of a 1 solar mass star such as the Sun. Second, in this parametric stellar dynamo study we are proposing that in order to get both a solar-like conical differential rotation and a deep slow decadal-long magnetic cycles, the Rossby number must be between 0.15 and 0.65. Hence, we here acknowledge that cases M09R1m and M11R1m rotating at the solar rate do not show behaviors that are sun-like with respect to their magnetic activity (no cycles present) because their Rossby number is not falling in the 0.15-0.65 range. Instead, we believe that M09R3m or M11R3m are better, closer representations of the Sun even though their rotation rate is faster than the Sun, because their Rossby number is in the correct range of values. This means that while the overall trends found in our study are robust, the specific location of any given star must be thought with extreme care due to the so-called {\it convective-conundrum}, i.e a mismatch between global convection simulations and solar helioseismic inversion regarding the amplitude of giant convection cells \citep{2016AnRFM..48..191H,2021NatAs.tmp..165H}. This is likely due to the fact that for any given rotation rate because of the {\it convective-conundrum}, the Rossby number achieved in the rotating convection simulation is slightly too large. So in order to be likely closer to the solar state and to aim for the correct value of the solar Rossby number, models rotating faster such as M09R3m or M11R3m cases are somewhat a better match to model the Sun than M09R1m or M11R1m. Thanks to this knowledge, we will next build a new global convective dynamo model of the Sun with an improved set of parameters by keeping the rotation rate to the solar one while controlling the effective Rossby number achieve in the simulation to be in the right range of values. We will report our finding in a future work.

To conclude, our study has confirmed the richness of dynamo solutions in parameter regimes that are likely to be found in solar-like stars and the large amount of magnetic energy and flux made available to the star and its surface activity by dynamo action. We have also identified the Rossby number regimes for different realization of differential rotation profiles and magnetic temporal modulations (cyclic or not), generalizing in an MHD context what we published in \cite{2017ApJ...836..192B}. Two key transitions in parameter space seem to be present, one at low Rossby number ($Ro_{\rm f} < 0.1$), another at high Rossby number ($Ro_{\rm f} > 1$). We need to study them with even more detail and at higher resolution and turbulence level to confirm the trends and scaling laws we have reported here. We intend to do so in the near future \ASmod{as well as study in more details the influence of a realistic atmosphere and of a wind \citep{2021ApJ...910...50P} on the dynamo properties}.

\begin{acknowledgements}
We acknowledge partial financial supports by: DIM-ACAV+ ANAIS2 project, ERC Whole Sun Synergy grant \#810218 and STARS2 starting grant \#207430, ANR Toupies, INSU/PNST, Solar Orbiter and PLATO CNES funds. We thank GENCI via project 1623 for having provided the massive computing resources needed to perform this extensive study.
J.V. acknowledge Project 2019-T1/AMB-13648  and  multiannual agreement with UC3M  (``Excelencia para el Profesorado Universitario'' - EPUC3M14) - Fifth regional research plan 2016-2020 founded by the Comunidad de Madrid. This work benefited from discussions within the international team "The Solar and Stellar Wind Connection: Heating processes and angular momentum loss", supported by the International Space Science Institute (ISSI). \ASmod{A.S.B is thankful to B.P. Brown, M. Derosa, N. Featherstone, B. Hindman, K. Shibata, S.-i Takehiro and M. Yamada for useful discussions}.
\end{acknowledgements}

\newpage
\appendix

\section{Equation for kinetic and magnetic energy transfers in MHD anelastic systems}
\label{sec:AppendixTransfer}

In this appendix we list the set of equations describing the energy transfer occurring in a star, focusing on mean energy quantities such as the poloidal and toroidal mean axisymmetric kinetic and magnetic energies.
Following \citet{1966PApGe..64..145S, 1996JFM...306..325B, 2002ApJ...581.1356D,2006ApJ...647..662R}, we derive the set of equations of full energy transfers in spherical MHD configurations.

Let us denote the azimuthal average by a bar, and the derivation from it by a prime. For example, the radial velocity component will be written as $v_r = \vrb + \vrp$. In order to characterize the axisymmetric magnetic ($\overline{E_m}$) and kinetic ($\overline{E_k}$) energy transfers between the various reservoirs of energy (thermal, potential, kinetic and magnetic) we will split $\overline{E_m}$ and $\overline{E_k}$ into three components:
\begin{eqnarray}
\overline{E_m} &=& \frac{1}{4\pi} (\br2 + \bt2 + \bp2) \\ \nonumber
&=&  \underbrace{\frac{1}{4\pi}(\bpb^2)} + \underbrace{\frac{1}{4\pi} (\brb^2 + \btb^2)} + \underbrace{\frac{1}{4\pi} (\overline{\brp^2} + \overline{\btp^2} + \overline{\bpp^2})} \\ \nonumber
& & \;\; {\rm TME} \;\;\;\;\;\;\;\;\;\;\;\;\;\; {\rm PME} \;\;\;\;\;\;\;\;\;\;\;\;\;\;\;\;\;\;\;\;\;\;\; {\rm FME}  \, ,\\ \nonumber
\\
\overline{E_k} &=& \frac{\hat{\rho}}{2} (\vr2 + \vt2 + \vp2) \\ \nonumber
&=& \underbrace{\frac{\hat{\rho}}{2}(\vpb^2)} + \underbrace{\frac{\hat{\rho}}{2} (\vrb^2 + \vtb^2)} + \underbrace{\frac{\hat{\rho}}{2} (\overline{v_r^{'2}} + \overline{v_{\theta}^{'2}} + \overline{v_{\phi}^{'2}})} \\ \nonumber
& & {\rm DRKE} \;\;\;\,\;\;\; {\rm MCKE} \;\;\;\;\;\;\;\;\;\;\;\;\;\;\;\;\; {\rm FKE} \, , \nonumber
\end{eqnarray}
with DRKE and TME the mean axisymmetric toroidal energies, MCKE and PME the mean axisymmetric poloidal energies and FKE and FME the non-axisymmetric energies. To find the energy transfer equation for these various components we project the Navier-Stokes or induction equation onto the direction we wish to write the energy equation for, e.g. $\phi$ for TME for instance and inject the decomposition between mean and prime quantities. Then we perform an azimuthal average, thereby eliminating all terms that are linear in prime quantities. For each energy equations, we then multiply by a bar quantity (for instance $\bpb$ for TME) and rearrange the terms. For MCKE and PME, we combine the radial and latitudinal equations. Doing so systematically leads to the following set of equations\footnote{Since we focus our study on the mean flows and magnetic fields, we will not show the equations for FKE and FME}.

%\section{Energy budgets}

\subsection{Overall Energy budgets}

We follow the approach of \citet{1966PApGe..64..145S} and write 
the energy budgets in the following way (see the schematic in Fig. \ref{fig:Ebudgets}):

\begin{eqnarray}
    \partial_t {\rm DRKE} &=& Q^{\rm DR}_{\rm RS} - Q_\Omega - Q_C - Q_{\rm MS}^{\rm DR} - Q_\nu^{\rm DR} + C^{\rm DR} - S^{\rm DR}\, ,
    \label{eq:DRKE}
    \\
    \partial_t {\rm MCKE} &=& Q_{\rm RS}^{\rm MC} + Q_{\rm TM}^{\rm MC} + Q_{\rm MS}^{\rm MC} + Q_{\nabla P} + Q_C - Q_{\rm PM}^{\rm MC} - Q_b - Q_\nu^{\rm MC} + C^{\rm MC} - S^{\rm MC} \, , 
    \label{eq:MCKE}
    \\
    \partial_t {\rm TME} &=& Q_\Omega + Q_{\rm emf}^{\rm TM} - Q_{\rm TM}^{\rm MC} - Q_\eta^{\rm TM} + C^{\rm TM} - S^{\rm TM}\, , 
    \label{eq:TME}
    \\
    \partial_t {\rm PME} &=& Q_{\rm PM}^{\rm MC} + Q_{\rm emf}^{\rm PM} - Q_\eta^{\rm PM} + C^{\rm PM} - S^{\rm PM}\, .
    \label{eq:PME}     
\end{eqnarray}

\begin{figure}[!h]
\centering
\includegraphics[width=0.95\textwidth]{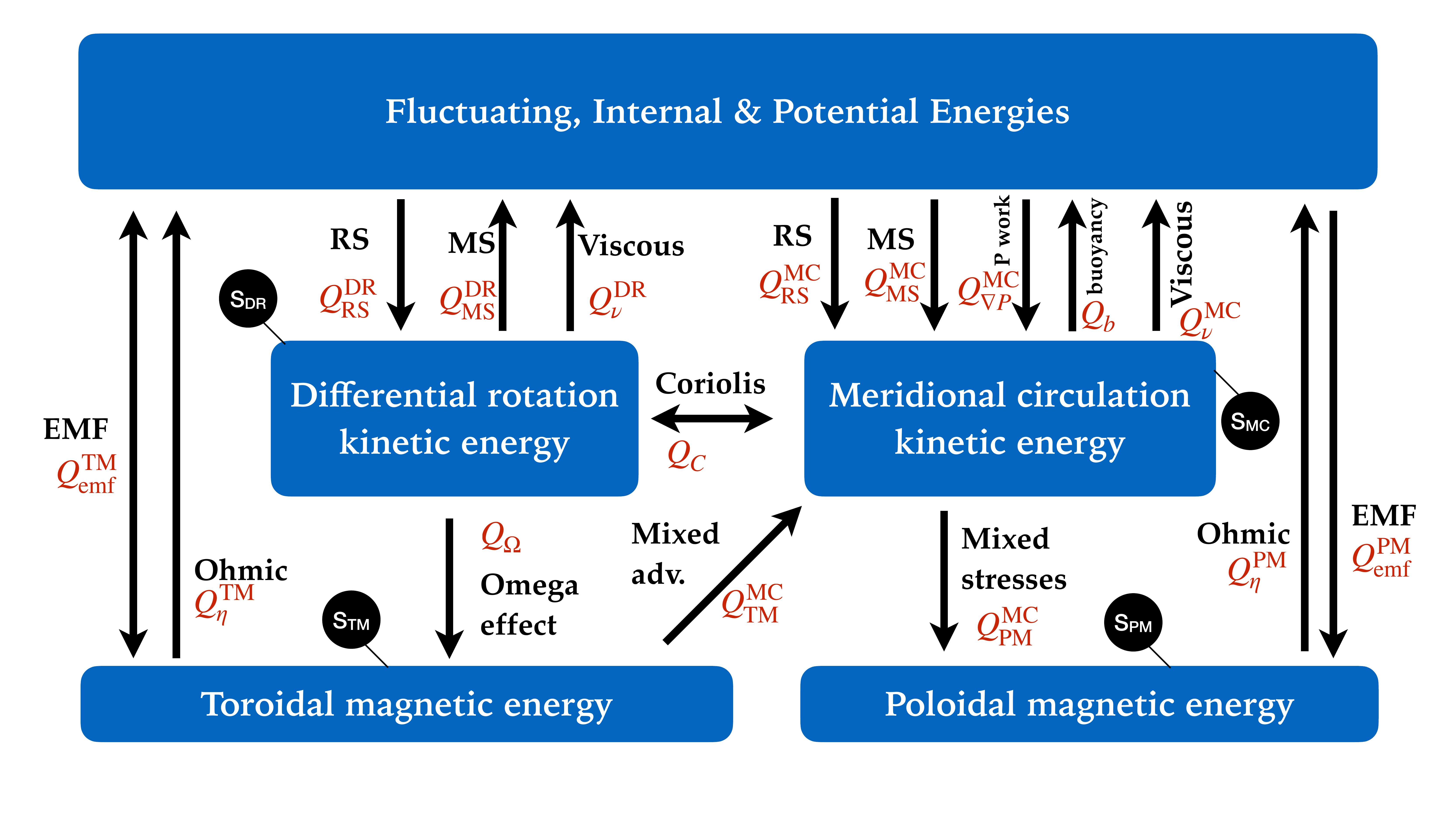}
\caption{Global energy budget schematic. In red, we list all the key energy transport terms (see appendix \ref{sec:AppendixSVD}). Black arrows correspond to the direction of transport between the various energy reservoirs. Surface terms are indicated as black disks. We omit curvature terms to avoid crowding the figure.}
\end{figure}\label{fig:Ebudgets}

In all that follows, quantities are separated into mean and fluctuating components through
\begin{equation}
    A = \bar{A} + A'\, ,
\end{equation}
and the corresponding terms in the original derivation of \citet{1966PApGe..64..145S} are given by the blue \ASmod{'SG66: [XX]' labels at the end of each equation, where XX is the term or equation number in \citet{1966PApGe..64..145S}}. Note that we have extra curvature terms $C^{x}$ due to our choice of spherical coordinates.

\subsection{Axisymmetric differential rotation kinetic energy equation (DRKE)}

The various terms of Eq. \ref{eq:DRKE} are

\begin{eqnarray}
    Q^{\rm DR}_{\rm RS} &=& \int \hat{\rho} \left[ \vrvpp \partial_r \vpb + \vtvpp \frac{1}{r}\partial_\theta \vpb \right] {\rm d}V \, , \, \mbox{{\color{blue} SG66: [6c+6d]}} \\
    Q_\Omega &=& \int \frac{1}{4\pi}\bpb \left[ \brb\partial_r\vpb + \frac{\btb}{r}\partial_\theta \vpb \right] {\rm d}V \, , \label{eq:Qomv2} 
    \, \mbox{{\color{blue} SG66: [5a+5b]}} \\
    Q_C &=& \int 2\Omega\hat{\rho} \vpb \left[ \cos\theta \vtb + \sin\theta \vrb \right] {\rm d}V \, , \label{eq:Qcv2}
    \, \mbox{{\color{blue} SG66: [6a+6b]}} \\
    Q_{\rm MS}^{\rm DR} &=& \int \frac{1}{4\pi}\brbpp \partial_r \vpb + \frac{1}{4\pi} \frac{\btbpp}{r}\partial_\theta \vpb {\rm d}V \, , \, \mbox{{\color{blue} SG66: [7a+7b]}} \\
    Q_\nu^{\rm DR} &=& \int \nu\hat{\rho} \left\{ \left[r\partial_r\left(\frac{\vpb}{r}\right)\right]^2 + \left[\frac{\sin\theta}{r}\partial_\theta\left(\frac{\vpb}{\sin\theta}\right)  \right]^2 \right\} {\rm d}V \, , \, \mbox{{\color{blue} SG66: [Fz]}}\\
    C^{\rm DR}&=& \int - \frac{\hat{\rho}\vpb}{r}\left[\vrvp + \cot\theta\vtvp \right] + \frac{1}{4\pi}\frac{\vpb}{r}\left[\brbp + \cot\theta\btbp \right] {\rm d}V \, , \\
    S^{\rm DR} &=& -\frac{1}{4\pi}\int_{r=R_{\rm top}} \brbp\, \vpb {\rm d}S\, .  \, \mbox{{\color{blue} SG66: [Eq. 6]}}
\end{eqnarray}

\subsection{Axisymmetric meridional circulation kinetic energy equation (MCKE)}

$Q_C$ was defined previously in Eq. \ref{eq:Qcv2}. The remaining terms in Eq. \ref{eq:MCKE} are

\begin{eqnarray}
    Q_{\rm RS}^{\rm MC} &=& \int \hat{\rho} \vrvtp\left[\partial_r \vtb + \frac{1}{r}\partial_\theta \vrb \right] + \hat{\rho}\vrvrp\partial_r\vrb + \hat{\rho}\vtvtp\frac{1}{r}\partial_\theta \vtb {\rm d}V \, , 
    \, \mbox{{\color{blue} SG66: [3a+3b+3c]}} \\
    Q_{\rm TM}^{\rm MC} &=& \int \frac{1}{8\pi} \bpb^2\left[\frac{1}{r^2}\partial_r \left( r^2 \vrb \right) + \frac{1}{r\sin\theta}\partial_\theta\left(\sin\theta\vtb\right) \right]{\rm d}V \, ,
    \label{eq:Qtmmcv2}
    \, \mbox{{\color{blue} SG66: [4a]}} \\
    Q_{\rm MS}^{\rm MC} &=& \int \frac{1}{8\pi}\left[-\brbrp + \btbtp + \bpbpp\right]\partial_r\vrb +\frac{1}{8\pi}\left[\brbrp - \btbtp + \bpbpp\right]\frac{1}{r}\partial_\theta\vtb \nonumber \\ &-& \frac{1}{4\pi}\brbtp\left[\partial_r\vtb + \frac{1}{r}\partial_\theta\vrb \right]{\rm d}V \, , 
    \, \mbox{{\color{blue} SG66: [8a]}} \\
    Q_{\nabla P} &=& \int -\left[\vrb\partial_r\overline{p} + \vtb\frac{1}{r}\partial_\theta\overline{p} \right]{\rm d}V \, , 
    \, \mbox{{\color{blue} SG66: [1a]}} \\
    Q_{\rm PM}^{\rm MC} &=& \int \frac{1}{4\pi}\left[\brb\vtb-\vrb\btb\right]\cdot\left[-\partial_r\btb + \frac{1}{r}\partial_\theta\brb\right] {\rm d}V \, ,
    \label{eq:Qpmmcv2}
    \, \mbox{{\color{blue} SG66: [9a]}} \\ 
    Q_b &=& \int \bar{\rho}g\vrb {\rm d}V\, ,
    \, \mbox{{\color{blue} SG66: [2a]}} \\
    Q_\nu^{\rm MC} &=& \int 2\, \nu \rb \left\{ \left[\partial_r \vrb - \frac{1}{3} \overline{(\nab \cdot \vv)} \right]^2 +  \left[\frac{1}{r}\partial_\theta \vtb + \frac{\vrb}{r} - \frac{1}{3} \overline{(\nab \cdot \vv)} \right]^2 \right. \nonumber \\ 
&+&  \left. \left[\frac{\vrb}{r} + \frac{\vtb \cos\theta}{r \st} - \frac{1}{3} \overline{(\nab \cdot \vv)} \right]^2 + \frac{1}{2}\left[\frac{1}{r}\partial_\theta \vrb + r\partial_r \left(\frac{\vtb}{r}\right) \right]^2 \right\} {\rm d}V \, , 
    \, \mbox{{\color{blue} SG66: [Fm]}} \\
    C^{\rm MC} &=& \int \frac{\hat{\rho}}{r}\left[ \vrb\left(\vtvt + \vpvp \right) - \vtb\left(\vrvt - \cot\theta \vpvp \right) \right] \nonumber \\ &-& \frac{\vrb}{4\pi r}\left(\btbt + \bpbp\right) + \frac{\vtb}{4\pi r}\left(\brbt - \cot\theta \bpbp\right) \nonumber \\ &+& \frac{1}{8\pi r}\overline{B'^2}\left(2\vrb + \cot\theta \vtb \right) {\rm d}V \, , \\
    S^{\rm MC} &=& -\frac{1}{4\pi}\int_{r=R_{\rm top}} \brbtp\, \vtb {\rm d}S \,. \, \mbox{{\color{blue} SG66: [Eq. 8]}}
\end{eqnarray}

\subsection{Axisymmetric toroidal magnetic equation (TME)}

$Q_\Omega$ and $Q_{\rm TM}^{\rm MC}$ were defined previously in \ref{eq:Qomv2} and \ref{eq:Qtmmcv2}. The remaining terms in Eq. \ref{eq:TME} are
\begin{eqnarray}
    Q_{\rm emf}^{\rm TM} &=& \int \frac{1}{4\pi}\left[\bpvrp-\brvpp\right]\partial_r\bpb + \frac{1}{4\pi}\left[\bpvtp-\btvpp \right]\frac{1}{r}\partial_\theta \bpb {\rm d}V \, , 
    \, \mbox{{\color{blue} SG66: [4b + 4c]}} \\
    Q_\eta^{\rm TM} &=& \int -\frac{\bpb}{4\pi r}\left[ \partial_r\left\{\eta\partial_r\left(r\bpb\right) \right\} + \frac{\eta}{r}\partial_\theta\left\{\frac{1}{\sin\theta}\partial_\theta\left(\sin\theta \bpb\right) \right\} \right] {\rm d}V \, , \, \mbox{{\color{blue} SG66: [Jz]}} \\
    C^{\rm TM} &=& \int -\frac{\bpb}{4\pi r}\left[\brvp + \cot\theta \btvp - (\bpvr + \cot\theta \bpvt )  \right] {\rm d}V \, , \\ 
    S^{\rm TM} &=& -\frac{1}{4\pi}\int_{r=R_{\rm top}} \brvpp\, \bpb {\rm d}S  \,. \, \mbox{{\color{blue} SG66: [Eq. 7]}}
\end{eqnarray}

\subsection{Axisymmetric poloidal magnetic equation (PME)}

$Q^{\rm PM}_{\rm MC}$ have already been defined in Eq. \ref{eq:Qpmmcv2}. The remaining terms in Eq. \ref{eq:PME} are

\begin{eqnarray}
    Q_{\rm emf}^{\rm PM} &=& \int \frac{1}{4\pi}\left[\brvtp-\btvrp\right]\cdot\left[-\partial_r\btb + \frac{1}{r}\partial_\theta\brb\right] {\rm d}V \, ,
    \, \mbox{{\color{blue} SG66: [9b]}} \\
    Q_\eta^{\rm PM} &=& \int \frac{\brb}{4\pi r^2\sin\theta}\eta\partial_\theta\left\{\sin\theta\left[ \partial_r\left(r\btb\right)-\partial_\theta\brb \right]  \right\} - \frac{\btb}{4\pi r}\partial_r\left\{ \eta\left[\partial_r\left(r\btb\right)-\partial_\theta\brb\right]\right\}  {\rm d}V \, , \, \mbox{{\color{blue} SG66: [Jm]}}\\
    C^{\rm PM}&=& \int \frac{\btb}{4\pi r}\left[ \btvr - \brvt \right] {\rm d}V \, , \\
    S^{\rm PM} &=& -\frac{1}{4\pi}\int_{r=R_{\rm top}} \brvt\, \btb {\rm d}S \,. \, \mbox{{\color{blue} SG66: [Eq. 9]}}
\end{eqnarray}\\

\section{Mean Field SVD Decomposition of Dynamo Solution}
\label{sec:AppendixSVD}

It is instructive to compare our 3D simulation results with the concepts used in mean field dynamo theory (see \S\ref{sec:MagneticProperties}).
For instance, the generation of poloidal magnetic field in the simulation is dominated by the action of the fluctuating EMF: $E_{\rm{FI}} = \mathcal{E}' = \langle \mathbf{v}' \times \BB'\rangle$.
This process can also be interpreted through the $\alpha$-effect approximation, which is a first order expansion of $\mathcal{E}'$ around the mean magnetic field and its gradient:
\begin{equation}
\langle \mathcal{E}' \rangle_i = \alpha_{ij} \langle \BB \rangle_j + \beta_{ijk} \partial_j \langle \BB\rangle_k + \mathcal{O} \left( \partial \langle \BB \rangle /\partial t, \nabla^2 \langle \BB \rangle \right)
\end{equation}
with $\alpha_{ij}$ a rank-two pseudo-vector and $\beta_{ijk}$ a rank-three tensor. In the following, we will neglect the $\beta$ term. 
However, this will increase the systematic error when estimating the $\alpha$ term. 
Thus, a single-value decomposition (SVD) including the $\beta$-effect has been calculated in order to provide a lower-bound on the systematic error as discussed in \cite{2015ApJ...809..149A}. 
In the following analysis, $\alpha$ has been decomposed into its symmetric and antisymmetric components 
\begin{equation}
\label{eq:alphagamma}
\alpha \langle \BB\rangle = \alpha_S \langle \BB \rangle + \gamma \times \langle \BB \rangle 
\end{equation}
with
\begin{equation}
\alpha_S=
\left[
\begin{matrix}
\alpha_{(rr)} & \alpha_{(r\theta)} & \alpha_{(r\varphi)} \\
\alpha_{(r\theta)} & \alpha_{(\theta\theta)} & \alpha_{(\theta\varphi)} \\
\alpha_{(r\varphi)} & \alpha_{(\theta\varphi)} & \alpha_{(\varphi\varphi)} \\
\end{matrix}
\right]
\quad\hbox{ and }\quad \gamma=
\left[
\begin{matrix}
\gamma_{r} \\
\gamma_{\theta} \\
\gamma_{\varphi} \\
\end{matrix}
\right].
\end{equation}

Thanks to the SVD decomposition we can quantify the relative efficiency of the $\alpha$-effect in generating the mean magnetic field and characterize the type of dynamo through the relative influence of its regenerating terms. 
We can start by evaluating how the convective flows regenerate mean magnetic fields. This can be determined by finding the amplitude of an estimated $\alpha$-effect relative to the rms value of the non-axisymmetric velocity field 
\begin{equation} \label{eq:alphaefficiency}
E \simeq \left\langle \frac{\alpha}{\text{v}_{\rm{rms}}} \right\rangle = \frac{3}{2(r_{top}^3-r_{bcz}^3)} \times \sum_{i,j} \iint dr d\theta r^2 \sin \theta \sqrt{\frac{\alpha_{ij} \alpha^{ij}}{\lbrace \mathbf{v}'\cdot \mathbf{v}'\rbrace}}
\end{equation}
where $\lbrace \mathbf{v}'\cdot \mathbf{v}'\rbrace$ is the sum of the diagonal elements of the Reynolds stress tensor averaged over time and over all longitudes. 
If we want to refine the analysis, we can use the equation \ref{eq:alphaefficiency} to provide a measure of the importance of each component of $\alpha$ as
\begin{equation}
\label{eq:alphacomponentefficiency}
\begin{matrix}
\displaystyle{\varepsilon_{ij}} & = & \displaystyle{\frac{E_{ij}}{E}} & & \\
&\simeq & \displaystyle{\frac{1}{E} \left\langle\frac{\alpha_{ij}}{\text{v}_{\rm{rms}}} \right\rangle} & = & \displaystyle{\frac{3}{2E(r_{top}^3-r_{bcz}^3)} \iint dr d\theta r^2 \sin \theta \sqrt{ \frac{\alpha_{ij} \alpha^{ij}}{\lbrace \mathbf{v}' \cdot \mathbf{v}'\rbrace}}} \\
& & \\
& & & = & \left[\begin{matrix} \varepsilon_{(rr)} & \varepsilon_{(r\theta)} & \varepsilon_{(r\varphi)} \\\varepsilon_{\gamma_{\varphi}} & \varepsilon_{(\theta\theta)} & \varepsilon_{(\theta\varphi)} \\ \varepsilon_{\gamma_{\theta}} & \varepsilon_{\gamma_r} & \varepsilon_{(\varphi\varphi)} \end{matrix} \right]
\end{matrix}
\end{equation}
with $\varepsilon_{(xx)} = \displaystyle{\frac{\alpha_{(xx)}}{E}}$ and $E_{\gamma x} = \displaystyle{\frac{\gamma_x}{E}}$. 
By calculating this matrix, see Table \ref{alphaomega}, we notice that for the antisymmetric part $\gamma$, the predominant term is $\gamma_{\varphi}$ that impacts the poloidal component of the magnetic field. Only for M07R5m are the three components of the same order of magnitude. In the three other cases shown, $\gamma_r$ and $\gamma_{\theta}$ have roughly the same order of magnitude and are smaller by a factor 2 to 3 compared to $\gamma_{\varphi}$. 
By looking at the symmetric part $\alpha_{\rm{S}}$, we see the same trend. 
The predominant term is $\alpha_{(rr)}$ with $\alpha_{(r\theta)}$ and $\alpha_{(\theta\theta)}$ close second. They all act on the poloidal component of the magnetic field. The smallest term is in most cases $\alpha_{(\varphi\varphi)}$ which is at least 5 times smaller than the predominant term except once more in case M07R5m where it is of the same order of magnitude. 
The sum of all $\alpha$ terms varies between 51\% in case M07R5m up to 73\% in case M09R3m. Hence, the $\gamma$ terms (the antisymmetric part of the alpha-tensor) account for 49\% in case M07R5m down to 27\% in case M09R3m.

In order to better quantify this relative influence we can compute the  $\alpha_{\rm{P}}/\alpha_{\varphi}$ ratio: 
\begin{equation}
\frac{\alpha_{\rm{P}}}{\alpha_{\varphi}} = \frac{3}{2(r_{top}^3-r_{bcz}^3)} \times \iint dr d\theta r^2 \sin \theta \left| \frac{\langle \BB_{\rm{P}} \rangle \cdot \nabla \times \langle \mathcal{E}' \rangle}{\langle B_{\varphi} \hat{\varphi} \cdot \nabla \times \langle \mathcal{E}'\rangle}\right|.
\end{equation}

\begin{table}[!th]
\begin{center}
\caption{$\alpha-\Omega$ effects from SVD decomposition}\label{alphaomega}
\vspace{0.2cm}
\begin{tabular}{|c|ccc|c|c|}
\cline{2-6}
\multicolumn{1}{c}{} & \multicolumn{3}{|c|}{$\alpha$ tensor} & $\Omega/\alpha_{\varphi}$ & $\alpha_{P}/\alpha_{\varphi}$ \rule[-9pt]{0pt}{24pt}\\ \hline
                     &  0.120                        & 0.092                        & 0.073 &       &       \rule[-7pt]{0pt}{20pt}\\
M07R5m                 & {\it 0.155} & 0.063                       & 0.061 & 19.7  & 12.4  \rule[-7pt]{0pt}{20pt}\\
                     & {\it 0.220} & {\it 0.119} & 0.097 &       &       \rule[-7pt]{0pt}{20pt}\\ \hline
                     & 0.246                        & 0.194                        & 0.088 &       &       \rule[-7pt]{0pt}{20pt}\\ 
M09R3m                 & {\it 0.166} & 0.125                        & 0.056 & 7.0  & 1.59  \rule[-7pt]{0pt}{20pt}\\ 
                     & {\it 0.042} & {\it 0.053} & 0.030 &       &       \rule[-7pt]{0pt}{20pt}\\ \hline
                     & 0.174                        & 0.157                       & 0.087 &       &       \rule[-7pt]{0pt}{20pt}\\ 
M11R3m                & {\it 0.162} & 0.135                        & 0.054 & 5.53  & 3.18  \rule[-7pt]{0pt}{20pt}\\ 
                     & {\it 0.075} & {\it 0.109} & 0.047 &       &       \rule[-7pt]{0pt}{20pt}\\ \hline
                     & 0.209                        & 0.120                        & 0.112 &       &       \rule[-7pt]{0pt}{20pt}\\
M09R1m                & {\it 0.157} & 0.110                        & 0.089 & 1.81  & 4.31  \rule[-7pt]{0pt}{20pt}\\ 
                     & {\it 0.067} & {\it 0.099} & 0.037 &       &       \rule[-7pt]{0pt}{20pt}\\ \hline
\end{tabular}
\end{center}
\textbf{Note:} Results of the mean field SVD dynamo analysis on four representative models (M07R5m, M09R1m, M09R3m, M11R3m) ordered from top to bottom in increasing Rossby number values. 
The first column represents the $\alpha$ tensor with its symmetric: $\alpha_{\rm{s}}$ and antisymmetric: $\gamma$ ({\it italic}) portions (see Eq \ref{eq:alphagamma}). 
The middle column gives the relative importance of the $\Omega$-effect to the $\alpha$-effect for the regeneration of the toroidal field. 
The last column quantifies the ratio of the $\alpha$-effect used for the regeneration of the poloidal magnetic field to the one used for the regeneration of the toroidal field.
\end{table}

Looking at Table \ref{alphaomega} where we report the value of this ratio for all 4 representative models, we note the predominance of the poloidal field regeneration over the toroidal field regeneration for all models as the ratio $\alpha_{\rm{P}}/\alpha_{\varphi}$ is always above 1. This ratio varies from 1.59 in M09R3m up to 12.4 in case M07R5m.

Turning now to the regeneration of the toroidal field, we know from mean-field dynamo theory that it can be due to either the $\alpha$-effect, coming from the fluctuating {\it emf} $\mathcal{E}'$, or from the $\Omega$ effect that acts on the poloidal field through differential rotation. 
In all our models, we note that the regeneration of $\langle B_{\varphi} \rangle$ by the $\alpha$-effect is small, compared to the one of $\BB_{\rm{pol}}$.
Therefore, we now want to measure the relative influence of the $\Omega$-effect to that of the $\alpha$-effect, since the toroidal magnetic field can be regenerated through both effects:
\begin{equation}
\frac{\Omega}{\alpha_{\varphi}} = \frac{3}{2(r_{top}^3-r_{bcz}^3)} \times \iint dr d\theta r^2 \sin \theta \left| \frac{r \sin \theta \langle B_{\varphi} \rangle \langle \BB_{P}\rangle \cdot \nabla \langle \Omega \rangle}{\langle B_{\varphi}\rangle \hat{\varphi} \cdot \nabla \times \langle \mathcal{E}'\rangle}\right|.
\end{equation}

We note that in all models the $\Omega$-effect is much stronger than the $\alpha$-effect in generating the toroidal magnetic field (the ratio $\Omega/\alpha_{\varphi}$ is greater than 5), except for case M09R1m for which it is closer to 1. This confirms that most of the dynamo models considered in this study can be classified as $\alpha$-$\Omega$ dynamos rather than $\alpha^2$-$\Omega$. Statistically steady simulations such as M09R1m on the contrary are closer to be classified as $\alpha^2$-$\Omega$. Of course, this mean field dynamo classification is mostly useful for short magnetic cycle period cases (illustrated in the table with case M07R5m) as they also follow Parker-Yoshimura rule (see \S\ref{sec:MagneticProperties}). For long magnetic cycle period cases such as M09R3m and M11R3m this is less significant, as we observe a complex nonlinear feedback that leads to a different type of cyclic dynamo. Further, we have shown in section 6 and Fig \ref{Ebudget_vs_time} that these dynamo mechanisms are highly variable in time, and can sometimes be quenched while at other times they become dominant. Hence, a mean field classification on such solutions could vary depending on the dynamo phase considered.

\section{Kinetic helicity in solar and anti solar cases}
\label{sec:KH}

In Figure \ref{kinhel} we display several realizations of the horizontally-averaged radial profile of the kinetic helicity $H_k=\vv \cdot \omm$ in our set of convective dynamo models. These profiles have been averaged over the northern hemisphere only.

\begin{figure}[!h]
\centering
\includegraphics[width=0.99\linewidth]{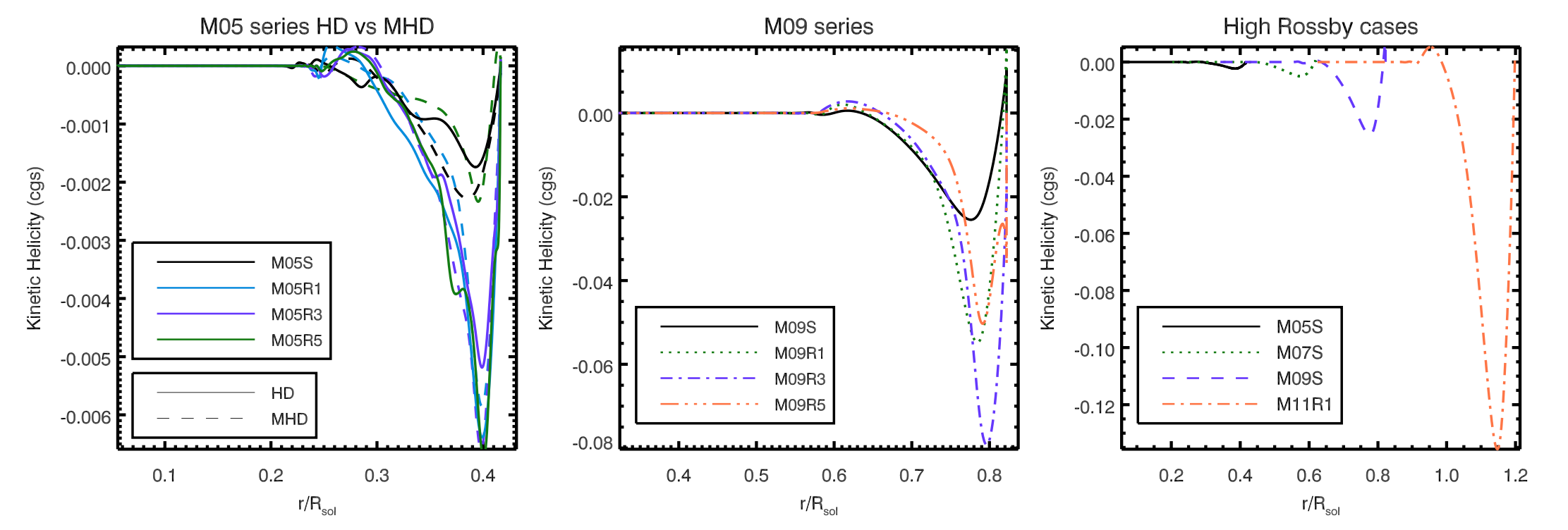}
\caption{Radial kinetic helicity profiles in various models averaged over the northern hemisphere only. Left panel: Comparing kinetic helicity for M05 cases for both the hydrodynamic progenitors and the MHD dynamo runs.  Middle panel: for decreasing Rossby number for $M09$ series. Right panel: For the $Ro_{\rm f} >1$ models spanning the 4 mass bins.}
\end{figure}\label{kinhel}

On the left panel we display the kinetic helicity profiles for the M05m series.
We first note that the kinetic helicity is negative in most of the domain and changes sign at the bottom of the convective envelope and is close to zero in the deep radiative interior below. This sign reversal of $H_k$ is understood by the change of sign of the vorticity field in the downward plumes. As they splash onto the top of the radiative zone (whose realistic stiffness we recall is directly taken from 1-D stellar structure model, see \S\ref{sec:numericalSetup}), they diverge and this yields a change of sign of the local kinetic helicity (see \citealt{2000ApJ...532..593M} for a detailed explanation).
Next, we can study how dynamo-generated magnetic field influences the kinetic helicity content of the convective shell. We do so by comparing the M05 dynamo cases to their hydrodynamic counterpart published in \cite{2017ApJ...836..192B} (dashed vs solid lines). We mostly find that magnetic fields tend to reduce the kinetic helicity content. In some rare cases we find it has little or no influence. In cyclic dynamo cases such as M05R3m we do not see a large influence of the cycle phase on the kinetic helicity content. This confirms that unless magnetic feedbacks are strong on the velocity and vorticity field (via the opposing/drag effect of Maxwell stresses in the converging and cyclonic intersection of downflow lanes), this quantity is not modified much.
In the middle panel, we show how the kinetic helicity evolves with a decreasing Rossby number. We illustrate this by plotting the radial kinetic helicity profiles of the M09m series (other mass bins display similar behavior). We see that as we increase the rotation rate from M09Sm to M09R1m and M09R3m, the peak amplitude near the surface becomes more and more negative (more cyclonic in the northern hemisphere, \textit{i.e}. more right-handed). This seems to stop for case M09R5m. We believe this is due to the strong quenching of the differential rotation and convection state due to the stronger feedback of Maxwell stresses in that case.

Finally, one important question, relevant to $\alpha-\Omega$ dynamo concepts, is how the kinetic helicity behaves in a high Rossby number regime, when the differential rotation harbors an anti-solar rotation profile. Indeed, we already know that in these cases, the gradients of $\Omega$ have a reversed sign. We also know that there is a relationship between the dynamo $\alpha$-effect and kinetic helicity. In the mean field dynamo approach $\alpha = -\frac{\tau}{3} \vv \cdot (\nab \times \vv)$. Hence, knowing if the $\alpha$-effect would change sign or not, can yield interesting information on the dynamo properties (\textit{e.g.} is there or not a breaking of symmetry).
On the right panel of Figure \ref{kinhel}, we display the kinetic helicity radial profiles for the slow rotating cases, those with a high (greater than 1) Rossby number. 
\ASmod{Across the four mass bins,} we see a clear increase in the amplitude of the kinetic helicity in an absolute sense (it becomes more negative near the surface of each model). This is linked to the fact that the velocity amplitude increases by more than one order of magnitude from M05 to M11 series due to the increased stellar luminosity of the more massive cases. Moreover, even though these 4 cases (M05Sm, M07Sm, M09Sm and M11R1m) have anti-solar differential rotation (see Fig \ref{RP}) their kinetic helicity profile is similar to the solar-like cases (negative in the upper layers and positive at the base of the convective zone) as discussed in the two previous panels. This can be understood by the fact that all models still rotate in the same direction when considering their rotating frame.
This means that the mean field $\alpha$-effect is not expected to change sign when the differential rotation ($\Omega$-effect) does.
This conservation of the kinetic helicity sign when changing the Rossby number from greater to lower than 1 is confirmed when displaying the radial vorticity near the surface in two cases M09S and M09R3 (not shown). The vortical nature of the interstices of the downflow lanes (as illustrated with the enstrophy field in Figure \ref{fig:SummaryDynamoM09R3}) is not modified between the two models even though they possess opposite profile of differential rotation. 
We note that there are some debates in the community to include or not a correction from current helicity such that $\alpha_m = -\frac{\tau}{3} (\vv \cdot (\nab \times \vv) - \frac{1}{c\rb} \JJ \cdot \BB)$ (see \citealt{1976JFM....77..321P,2005PhR...417....1B}). So it could be the case that the kinetic helicity does not change sign, but that a correction from the current helicity may. We have assessed this point, and we find that the profile of current helicity is less coherent as a function of depth and does not seem to modify the conclusion of our analysis. 

%\newpage
\bibliography{mybibfile}

\end{document}